\documentclass[11pt]{JHEP3}

\usepackage{latexsym}
\usepackage{epsfig,amssymb,euscript}
\usepackage{amsmath,cancel,slashed}
\DeclareMathAlphabet{\mathpzc}{OT1}{pzc}{m}{it}
\usepackage{array,calc,epsfig,bbm}
\usepackage{multicol}
\usepackage{bbm}
\usepackage{fancybox}

\def\be{\begin{equation}}
\def\ee{\end{equation}}
\newcommand{\eq}[1]{\begin{equation}#1\end{equation}}
\def\bea{\begin{eqnarray}}
\def\eea{\end{eqnarray}}
\def\bseq{\begin{subequations}}
\def\eseq{\end{subequations}}
\newcommand\bbone{\ensuremath{\mathbbm{1}}}
\newcommand{\ul}{\underline}
\newcommand{\nn}{\nonumber}
\newcommand{\beal}{\begin{align}}
\newcommand{\cd}{\mathcal{D}}
\newcommand{\boxedeq}[1]{
\begin{equation}
\fbox{
\rule[0.7cm]{0pt}{0pt}
$#1$
\rule[-0.45cm]{0pt}{0pt}
}
\end{equation}
}
\newcommand{\spl}[1]{\begin{split}#1\end{split}}
\numberwithin{equation}{section} %%
\usepackage[]{graphicx}

\def\d {{\rm d}}

\def\calb         {{\cal B}}

\def\cald         {{\cal D}}
\def\cale         {{\cal E}}
\def\calf         {{\cal F}}
\def\calg         {{\cal G}}

\def\calj         {{\cal J}}
\def\calk         {{\cal K}}
\def\call         {{\cal L}}
\def\calm         {{\cal M}}
\def\caln         {{\cal N}}
\def\calo         {{\cal O}}

\def\calr         {{\cal R}}
\def\cals         {{\cal S}}
\def\calt         {{\cal T}}
\def\calu         {{\cal U}}
\def\calv         {{\cal V}}
\def\calw         {{\cal W}}
\def\calz         {{\cal Z}}

\def\reals        {{\mathbb R}}
\def\zet          {{\mathbb Z}}

\def\tr           {\mathop{\rm Tr}}
\def\Re           {{\rm Re\hskip0.1em}}
\def\Im           {{\rm Im\hskip0.1em}}

\def\sqr#1#2{{\vcenter{\vbox{\hrule height.#2pt
 \hbox{\vrule width.#2pt height#1pt \kern#1pt \vrule width.#2pt}\hrule
 height.#2pt}}}}

\def\T{{\bf T}}

\def\oh{\frac{1}{2}}
\def\a{{\alpha}}
\def\b{{\beta}}
\def\D{{\Delta}}
\def\eps{{\epsilon}}

\def\th{{\theta}}
\def\Lam{{\Lambda}}
\def\lam{{\lambda}}
\def\Om{{\Omega}}
\def\om{{\omega}}
\def\sig{{\sigma}}
\def\G{{\Gamma}}
\def\g{{\gamma}}
\def\p{{\partial}}
\def\raw{\rightarrow}
\def\Raw{\Rightarrow}

\def\slashchar#1{\setbox0=\hbox{$#1$}           % set a box for #1
\dimen0=\wd0                                 % and get its size
\setbox1=\hbox{/} \dimen1=\wd1               % get siste of /
\ifdim\dimen0>\dimen1                        % #1 is bigger
\rlap{\hbox to \dimen0{\hfil/\hfil}}      % so center / in box
#1                                        % and print #1
\else                                        % / is bigger
\rlap{\hbox to \dimen1{\hfil$#1$\hfil}}   % so center #1
/                                         % and print /
\fi}

\title{Generalized non-supersymmetric flux vacua}

\author{Dieter L\"ust${}^{\diamondsuit \clubsuit}$, Fernando Marchesano${}^{\heartsuit}$, Luca Martucci${}^{\clubsuit}$ \& Dimitrios Tsimpis${}^{\clubsuit}$\\

\begin{itemize}

\item  Max-Planck-Institut f\"ur Physik\\
F\"ohringer Ring 6, 80805 M\"unchen, Germany

\item  Arnold-Sommerfeld-Center for Theoretical Physics\\
Department f\"ur Physik, Ludwig-Maximilians-Universit\"at M\"unchen\\
Theresienstra\ss e 37, 80333 M\"unchen, Germany 

\item  CERN PH-TH Division,\\ CH-1211 Geneva 23, Switzerland
 \end{itemize}

\bigskip
 E-mail:
\email{dieter.luest@lmu.de} \& \email{luest@mppmu.mpg.de}, \email{marchesa@cern.ch}, \email{luca.martucci@physik.uni-muenchen.de}, \email{dimitrios.tsimpis@lmu.de}}

\received{\today}               %%
\revised{}
\accepted{}               %% These are for published papers.

\preprint{hep-th\yymmddd}
\preprint{MPP-2008-87\\ LMU-ASC 41/08 \\ CERN-PH-TH/2008-162}

\abstract{We discuss a novel strategy to construct 4D $\caln=0$ stable flux vacua of type II string theory, based on the existence of BPS bounds for probe D-branes in some of these backgrounds. In particular, we consider compactifications where D-branes filling the 4D space-time obey the same BPS bound as they would in an $\caln=1$ compactification, while other D-branes, like those  appearing as domain walls from the 4D perspective, can no longer be BPS. We construct a subfamily of such backgrounds giving rise to 4D $\caln=0$ Minkowski no-scale vacua, generalizing the well-known case of type IIB on a warped Calabi-Yau. We provide several explicit examples of these constructions, and compute quantities of phenomenological interest like flux-induced soft terms on D-branes. Our results have a natural, simple description in the language of Generalized Complex Geometry, and in particular in terms of D-brane generalized calibrations. Finally,  we extend the integrability theorems for 10D supersymmetric type II backgrounds to the $\caln=0$ case and use the results to construct a new class  of $\caln=0$ AdS$_4$ compactifications.}


\begin{document}

\section{Introduction}

Most string compactifications to four space-time dimensions built so far preserve at least $\caln=1$ supersymmetry. The main reason to focus on supersymmetric, also known as BPS, string vacua is twofold. First, supersymmetric string vacua are relatively easy to construct. The underlying supersymmetry/BPS equations are first order differential equations, whose solutions are known in several instances. Well-known examples include Calabi-Yau compactifications, flux compactifications, calibrated brane configurations, supersymmetric, charged black holes and many more. Second, from the phenomenological point of view, supersymmetric string compactifications are a good starting point, since a promising scenario is to assume that space-time sypersymmetry is broken at the TeV scale, much below a string scale or a compactification scale not far from the Planck mass.

On the other hand we know that supersymmetry is eventually broken in nature: the spectrum of elementary particles is non-supersymmetric and observed astrophysical black holes are not supersymmetric or extremal either. Hence, a stringy realization of everyday physics should involve, in some sense, a  non-supersymmetric string vacuum. In contrast to BPS vacua, non-supersymmetric stable string vacua are very difficult to obtain directly, since one has to solve the full string equations of motion. Even in the supergravity approximation,  this implies solving generically cumbersome second order differential equations whose solutions are complicated and to large extent unknown.

In practice, however, one may still hope to break supersymmetry in a controlled, soft way via a `small perturbation' around a certain supersymmetric background, so that most of the nice properties of the latter are kept. This is essentially the strategy we will adopt in this paper in order to construct non-supersymmetric flux vacua. We will keep some of the good properties of known supersymmetric string solutions, and we will introduce some SUSY-breaking parameters in a controllable way. In a sense to be made precise below, we will consider {\em partially-BPS vacua} and, as a result, we will discover that the equations to be solved are essentially still some first order differential equations. 

While our philosophy can be applied to general situations, we will first focus on the construction of non-supersymmetric vacua with vanishing vacuum energy, and potentially leading to $\caln=0$ compactifications of the no-scale type. More specifically, we will consider type II supergravity vacua with broken supersymmetry which generalize to arbitrary supergravity backgrounds the warped CY/F-theory  solutions discussed by Giddings, Kachru and Polchinski (GKP) \cite{gkp}.\footnote{The conditions  characterizing warped  CY solutions with constant dilaton were previously found in \cite{gp1}. See also \cite{drs} for earlier, related (albeit  supersymmetric) solutions.} One obvious motivation is that, in the past few years, the example provided by GKP has generated a lot of activity in the construction of promising string compactification scenarios, like the so-called  KKLT \cite{kklt} and Large Volume Models \cite{LVM}, leading to de Sitter vacua. Note that in both examples one starts with an $\caln=0$ Minkowski no-scale vacuum, which is then modified by the addition  of some stringy ingredient (anti-D3-branes and E3-branes in KKLT and $\a'$ corrections in the Large Volume Models) in order to construct the final dS$_4$ vacuum. It is then reasonable to expect that, by considering generalizations of the GKP construction, one can achieve new ways of constructing de Sitter vacua. These new constructions may have new features, different  from the ones mentioned above. Therefore, in the context of the string Landscape, a basic question is which features and predictions of the KKLT and Large Volume scenarios are robust enough to survive this generalization.

Therefore, in analyzing general $\caln=0$ vacua, we will use as a prototypical example the class of IIB warped CY/F-theory  compactifications discussed in \cite{gkp}. This example is particularly simple since we can identify the source of SUSY-breaking as a non-zero $\calg_3^{(0,3)}$ flux component, that can be turned on without modifying the underlying CY/F-theory structure of the compactification. This nice property of the GKP vacua is not expected to be valid in the generic case of type II vacua considered in the present paper, and thus it cannot be used as the leading principle in our analysis. A first task is then to identify a different characterization of $\caln=0$ GKP vacua which is more suitable for generalizations. 

In this spirit, one may ask how $\caln=0$ GKP vacua fit in the scheme of generalized complex geometry, which has already proven to be an organizing principle in the construction of $\caln=1$ supergravity vacua. This question is answered in Section \ref{sec:vacua}, where the GKP properties are rephrased in terms of the physics of probe D-branes in these backgrounds. Indeed, from \cite{luca} we know that the supersymmetry conditions for a general flux background are equivalent to requiring that certain kinds of probe D-branes obey a BPS bound.\footnote{Let us emphasize that such backgrounds a priori exist without the calibrated probe D-branes on top of them, i.e. they are consistent closed string supergravity backgrounds by themselves. One may then add space-time filling BPS probe D-branes on top of them for model-building applications. In such cases one gets additional consistency relations, like RR tadpole conditions, which for compact examples can be solved by introducing appropriate orientifold geometries. See \cite{Dreview} for recent reviews on this subject.} As we will show, this is only partially true in $\caln=0$ GKP vacua, where a particular set of D-branes, namely some of those that look like domain-walls from the 4D viewpoint, no longer obey this BPS bound, and are thus intrinsically unstable in this background. On the other hand, D-branes that fill the 4D spacetime directions or look like strings in 4D still maintain their BPS properties unchanged with respect to the $\caln=1$ case.

This observation suggests an immediate way to generalize the GKP construction to other settings. Indeed,    instead of considering the whole set of $\caln=0$ supergravity compactifications to 4D Minkowski, we may restrict to those where 4D space-filling and string-like D-branes develop a BPS bound, while 4D domain walls will be lacking  such a `BPSness' property. The analysis of these backgrounds, which we dub  `Domain Wall SUSY-breaking' (DWSB) backgrounds, will be organized as follows:

In Section \ref{sec:dwsb} we translate the DWSB pattern in terms of the usual 10D gravitino and dilatino variations, in order to parameterize the space of DWSB backgrounds. Within this parameter space we single out a particular one-parameter subansatz, which despite its simplicity contains the GKP case, and discuss its main features from the point of view of Generalized Complex Geometry (GCG).\footnote{For applications of GCG to string theory see \cite{Freview} and references therein.} In Section \ref{sec:inteflux} we construct an effective action for compactifications of type II supergravity to 4D whose extremization is equivalent to the 10D equations of motion, and use it to show that our DWSB subansatz corresponds to 10D vacua provided that some first order differential conditions are satisfied. Moreover, we will see that, by imposing some mild assumptions, such 10D vacua will also be stable, tachyon-free vacua, with an effective potential that suggests a no-scale structure. Such intuition will be confirmed in Section \ref{sec:4dint}, where we describe DWSB vacua in a 4D $\caln=1$ language by using the formalism of \cite{lucapaul2}.

As the sections above treat the subject at a rather general and formal level, we proceed to give explicit realizations, as well as phenomenological applications, of DWSB backgrounds. In Section \ref{sec:subcases} we analyze several subfamilies of one-parameter DWSB vacua, some of them related to those obtained in \cite{cg07} from a 4D-like approach. We find good agreement with the results of \cite{cg07}, except for some extra constraints arising from the 10D equations of motion. In Section \ref{sec:examples} we give explicit examples of these subfamilies of vacua, and in Section \ref{sec:fsoft} we perform a soft-term analysis for certain D-branes on them. Finally, in Section \ref{sec:anti} we explore the possibility of adding anti-D-branes in DWSB vacua.

Note that the strategy followed above does not rely on compactifying to a flat 4D space. In fact, thanks to the results of \cite{lucapaul3}, it is straightforward to apply the DWSB concept to compactifications to AdS$_4$, as discussed in Section \ref{sec:ads4}. There we also generalize the DWSB pattern for those backgrounds where probe D-strings do not necessarily have a BPS bound.

Finally, in the last part of the paper we turn to a different, complementary approach to the construction of $\caln =0$ vacua, based on the integrability techniques of \cite{lt,gauntlett,kt}. Although these techniques have so far only been applied to $\caln=1$ backgrounds, it will be shown in Section \ref{sec:inte} that they can be extended to the $\caln=0$ case. As in the case of DWSB backgrounds, it turns out that the presence of a BPS bound is a key ingredient of integrability when localized sources are present in the background. Roughly-speaking, the main idea of this approach is the factorization of the second-order equations of motion into two first-order equations involving spinorial quantities. 
%Schematically, let $\epsilon$ be the ten-dimensional supersymmetry generator, so that $A\epsilon=\psi$, where $\psi$ parameterizes the supersymmetry-breaking and $A$ is a first-order differential operator. The integrability result presented here, amounts to identifying a first-order differential operator $B$ such that, provided the (generalized) Bianchi identities are satisfied, $\epsilon^TB\psi$ is a linear combination of the (second-order) equations of motion. Note that by taking $\epsilon$ to be a Killing spinor, so that $\psi$ vanishes, one reproduces as a corollary the integrability results of \cite{lt,gauntlett,kt} for supersymmetric backgrounds. It follows from the above that an alternative strategy for the construction of general ten-dimensional non-supersymmetric vacua would be to search for backgrounds such that $\psi$ is non-vanishing but nevertheless lies in the kernel of $\epsilon^TB$.
 As an illustration, in section \ref{4+6vacua} we use this method to construct a class  of non-supersymmetric  vacua of the form AdS$_4\times \mathcal{M}_6$, where $\mathcal{M}_6$ can be any nearly-K\"{a}hler manifold.  

Several details and technical computations are relegated to the appendices. In Appendix \ref{ap:conv} we provide the supergravity conventions and definitions used in the main text. In Appendix \ref{ap:sb} we describe in detail how the brane BPSness conditions, expressed in terms of $SO(6,6)$ pure spinors, are related to the more familiar 10D gravitino and dilatino variations. In Appendix \ref{ap:scalarR} we discuss how to express the scalar curvature of a six-dimensional manifold in terms of its pure spinors. Appendix \ref{ap:ng} is dedicated to argue how the results of the paper apply to non-geometric backgrounds as well. Finally, Appendix \ref{inte} contains the details of the derivation of the integrability conditions of Section \ref{sec:inte}, while Appendix \ref{gkpvacua} applies the techniques of that section to prove the integrability of GKP vacua.

%%%%%%%%%%%%%%%%%%%%%%%%%%%%%%
%%%%%%%%%%%%%%%%%%%%%%%%%%%%%%

\section{$\caln = 0$ flux vacua and D-brane stability}\label{sec:vacua}

The purpose of this section is to identify the key features of $\caln=0$ GKP vacua \cite{gkp} that have a natural generalization in the context of type II supergravity. For instance, a basic property of such vacua is that they admit  orientifolds and stable D-branes. As shown below, in the GCG language D-brane stability is rephrased as the existence of certain background calibrations, of the kind discussed in \cite{paul,luca,lucajarah}.\footnote{Differently from more traditional calibrations  \cite{gutowski}, the calibrations of \cite{paul,luca,lucajarah}  allow to consider D-branes  with non-trivial world-volume fluxes and networks thereof.} While the existence of such calibrations is guaranteed for $\caln=1$ backgrounds \cite{luca,lucapaul3}, this is not true for $\caln=0$ vacua. Only a subset of $\caln=0$ vacua contains such D-brane calibrations and, in order to generalize $\caln=0$ GKP vacua, we will restrict to this subset.\footnote{It may seem surprising at first sight that calibrations play such an important role in $\caln=0$ vacua. Note, however, that they are related to the BPS bound for D-branes, which can also be present in absence of supersymmetry. Using calibrations in $\caln=0$ vacua is a novel feature of the present analysis that, to the best of our knowledge, has not been considered before in the string theory literature.}

Having D-brane calibrations in our background is, of course, also a basic requirement for model building, for we need stable D-branes in order to embed the Standard Model in a type II construction. Interestingly, as we will argue in sections \ref{sec:inteflux} and  \ref{sec:inte}, they also constitute a key ingredient to simplify the analysis of the full supergravity equations of motion. Finally, although in this section we will restrict to compactifications to flat space, the same ideas can be applied to $\caln=0$ compactifications to AdS$_4$, a case which we will address in section \ref{sec:ads4}.

%%%%%%%%%%%%%%%%%%%%%%%%%%%%%%

\subsection{GKP vacua as calibrated backgrounds}\label{gkp}

Let us begin our analysis by showing how the $\caln=0$ warped Calabi-Yau/F-theory flux vacua of \cite{gkp} fit in the general picture of D-brane calibrations. In constructing such vacua, one assumes a 10D  spacetime of the form $X_{10} =\mathbb{R}^{1,3}\times_\om \calm_6$. In the string frame, the 10D metric has the form
\bea\label{alph1}
\d s^2_{10}=e^{2A(y)}\d x^\mu\d x_\mu+e^{-2A(y)+\Phi(y)}\hat g_{mn}(y)\d y^m\d y^n
\eea
where $\hat g$ is an F-theory metric on the internal space, with associated K\"ahler form $\hat J$ and normalized holomorphic $(3,0)$-form $\Omega_0 = e^{\Phi/2} \hat \Omega$ satisfying 
\bea\label{alph2}
\frac{1}{3!}\hat J\wedge \hat J\wedge\hat J=-\frac{i}{8}e^{-\Phi}\Omega_0\wedge\overline{\Omega}_0
\eea 
In the limit of constant dilaton, $\hat g$ is a standard CY$_3$ metric. Then, one finds that a class of supergravity backgrounds is given by the following conditions
\bseq\label{wcy}
\begin{align}
\hat{*}_6 \calg_3 &=i\calg_3\ \label{isd}\\
\d(4A-\Phi)&= e^{4A-\Phi}\hat*_{6}\,F_{5}\ \label{warpgkp}\\
\bar\partial\tau&=0\label{holt}
\end{align}
\eseq
where\footnote{We are using conventions which differ from \cite{gkp} in the definition of $\calg_3$ and the Bianchi identities. Our conventions are related to the ones in \cite{gkp} via  $H \raw -H$. } $\calg_3\equiv F_{3}+ie^{-\Phi}H$ and $\tau\equiv C_{0}+ie^{-\Phi}$ is the type IIB axio-dilaton.\footnote{In the warped Calabi-Yau case with constant dilaton, the conditions (\ref{wcy}) were first found in \cite{gp1}.}  Here we express everything in terms of internal RR-field strengths $F_{k}$, which in absence of sources are locally defined by $F=\sum_k F_{k}=\d_H C=(\d+H\wedge)\sum_k C_{k}$. Considering the Bianchi identities in addition to the above equations and assuming the saturation of a BPS bound for sources localized in $\calm_6$
\bea\label{local}
\frac{1}{4}(T^m{}_m-T^\mu{}_\mu)^{\rm loc}\geq T_3\rho_3^{\rm loc}\ 
\eea
one gets a IIB supergravity vacuum. If furthermore 
\bea\label{03cond}
\Omega_0\wedge \calg_3=0\ 
\eea
then the vacuum is supersymmetric. The imaginary-self-duality (ISD) condition (\ref{isd}) is satisfied by 3-forms of type $(0,3)$, primitive $(2,1)$ and non-primitive $(1,2)$.\footnote{That is, those (1,2)-forms that can be written as $\hat J\wedge \alpha$ with $\alpha$ a non-trivial closed $(0,1)$-form.} We can exclude this last possibility by assuming the primitivity condition $\hat J\wedge H=0$, that arises from supersymmetry \cite{gp1}. Finally,  the condition (\ref{03cond}) would also eliminate the $(0,3)$ component, leaving just a $(2,1)$ primitive flux $\calg_3$.

In the following, we would like to characterize this class of vacua from a somewhat different viewpoint: the existence of a BPS/stability bound, analogous to (\ref{local}), for the localized sources. Indeed, in \cite{gkp} the bound (\ref{local}) was argued to be satisfied and saturated by D3-branes and appropriate D7-brane configurations, and one may wonder when an analogous statement can be made for other $\caln=0$ vacua. As we will see, answering this question naturally suggests how to extend the GKP construction to more general settings. 

Let us start by rewriting the the conditions (\ref{wcy}) in a different form, along the lines of
the approach pioneered in \cite{gmpt} to analyze $\caln = 1$ vacua. First, we observe that the total (democratic) 10D RR field-strength $F^{\rm tot}=\sum_k F^{\rm tot}_{k}$ has the form
\bea\label{RRsplit}
F^{\rm tot}=F+\d x^0\wedge\d x^1\wedge\d x^2\wedge\d x^3\wedge e^{4A}\tilde F\ 
\eea  
with the `electric' components given by
\bea\label{dualf}
\tilde F=\tilde*_6 F\ 
\eea
where $\tilde*_6$ is a signed Hodge star operator defined in Appendix \ref{ap:conv} (see eq.(\ref{circstar6})).
Denoting  the internal metric of the background by $g=e^{-2A+\Phi}\hat g$ and  the associated (non-closed) K\"ahler form by $J=e^{-2A+\Phi}\hat J$, the equations (\ref{wcy}) can be rewritten as
\bseq\label{wcy2}
\begin{align}
&\d_H\big[e^{4A-\Phi}\Re \big( e^{iJ}\big)\big] =e^{4A}\tilde F &\hspace{-1.25cm} {\rm gauge\ BPSness} \\
&\d_H\big[e^{2A-\Phi}\Im \big(e^{iJ}\big)\big]=0 &  \text{D-string BPSness} \label{Dterm}
\end{align}
\eseq
where even the assumption that $\hat J=e^{2A-\Phi}J$ is closed follows from the second equation. 

We have dubbed the differential equations above in terms of their 4D physical interpretation. Both equations in (\ref{wcy2}) imply that there is a BPS lower bound for the energy of space-time filling and string-like D-branes, respectively, which is saturated for those D-branes calibrated by (\ref{wcalib}) below. As a result, thanks to (\ref{wcy2}) calibrated D-branes appear in the 4D effective theory as stable gauge theories and string-like defects.\footnote{Condition (\ref{Dterm}) also has a 4D interpretation as D-flatness of the $U(1)$ gauge bosons arising from the bulk \cite{lucapaul2}. So when it is satisfied we have pure F-term SUSY-breaking.}

Let us see how this works, following \cite{paul,luca}. First, note that (\ref{wcy2}) is the kind of condition that a background needs to satisfy in order to contain generalized calibrations for D-branes. In particular, (\ref{wcy2}) are relevant for those D-branes that fill either all four or exactly two dimensions in $\mathbb{R}^{1,3}$, their calibrations being
\bea\label{wcalib}
\omega^{\rm (sf)}&=e^{4A-\Phi}\Re \big( e^{iJ}\big)\quad{\rm and}\quad \omega^{\rm (string)}&=e^{2A-\Phi}\Im \big( e^{iJ}\big)
\eea
respectively. Indeed, one can check that (as in the supersymmetric case of  \cite{luca}) D-branes filling two or four dimensions of $\mathbb{R}^{1,3}$ satisfy the local bound\footnote{For simplicity, we do not indicate possible curvature corrections to the DBI and CS actions.  In BPS saturated expressions, they can be always reintroduced through the substitution $e^{\calf}\rightarrow e^{\calf}\wedge(\text{curv.~corr.})$.}
\bea\label{locbound}
\cale_{DBI}(\Sigma,\calf)\geq \big[ \omega|_\Sigma\wedge e^{\calf}\big]_{\rm top}\ 
\eea
pointwise for any pair $(\Sigma,\calf)$ of an internal $p$-cycle $\Sigma$ with worldvolume flux $\calf$ on it. Here $\cale_{DBI}$ denotes the DBI  energy density
\bea\label{dbi}
\cale_{DBI}(\Sigma,\calf)= e^{qA-\Phi}\sqrt{\det(g|_\Sigma+\calf)}\,\d\sigma\ 
\eea
where $q=4$ for space-filling branes and $q=2$ for D-strings. From this bound, one can identify the potential BPS D-branes as those satisfying
\bea\label{BPScond}
\cale_{DBI}(\Sigma,\calf)_{\rm BPS}= \big[\omega|_{\Sigma}\wedge e^{\calf}\big]_{\rm top}
\eea
Second, and again as in the $\caln=1$ case, one can easily show that D-branes satisfying (\ref{BPScond}) minimize their four-dimensional potential energy
\bea\label{bpsbound}
\calv(\Sigma,\calf)_{\rm BPS} \leq \calv(\Sigma^\prime,\calf^\prime)\ 
\eea
with respect to any configuration $(\Sigma^\prime,\calf^\prime)$  continuously connected to $(\Sigma,\calf)_{\rm BPS}$.\footnote{More precisely, inside its same generalized homology class \cite{lucajarah}.} For this to be true, however, the background has to satisfy the differential equations (\ref{wcy2}), which in terms of  the calibrations (\ref{wcalib}) read
\bea\label{intwcy}
\d_H\omega^{\rm (sf)}=e^{4A}\tilde F\quad \text{and} \quad \d_H\omega^{\rm (string)}=0
\eea
and which lead to the BPS bound (\ref{bpsbound}). For instance, for space-filling/gauge D-branes we have
\bea\label{compustab}
\calv(\Sigma,\calf)_{\rm BPS}&=&\int_\Sigma \big[\cale_{DBI}(\Sigma,\calf)_{\rm BPS}+\cale_{CS}(\Sigma,\calf)_{\rm BPS}\big]=\int_\Sigma (\omega-C^{\rm el})|_{\Sigma}\wedge e^{\calf}\cr &=& \int_{\Sigma^\prime} (\omega-C^{\rm el})|_{\Sigma^\prime}\wedge e^{\calf^\prime} \leq \int_{\Sigma^\prime} \big[\cale_{DBI}(\Sigma^\prime,\calf^\prime)+\cale_{CS}(\Sigma^\prime,\calf^\prime)\big]\cr
&=&\calv(\Sigma^\prime,\calf^\prime)\ 
\eea
where the `electric' RR potential $C^{\rm el}$ is defined by $\d_H C^{\rm el}=e^{4A}\tilde F$ and  (\ref{intwcy}) is used in going from the first to the second line. The case of D-strings is analogous, but without CS contribution.

Although the results above are general, for wCY/F-theory vacua $\omega^{\rm (sf)}$ and $\omega^{\rm (string)}$ take the particular form (\ref{wcalib}). Plugging $\omega^{\rm (sf)}$ into (\ref{BPScond}), one can see that the BPS bound for space-filling D-branes can only be satisfied by D3-branes and D7-branes with appropriate orientation. More precisely, it is always satisfied for D3-branes, while it requires the D7-branes to wrap an internal holomorphic 4-cycle with a primitive $(1,1)$ world-volume flux $\calf$, just like in $\caln=1$ GKP backgrounds \cite{gmm05}.  Similarly, one obtains that  4D BPS D-strings arise from, e.g., a D3-brane wrapping a holomorphic 2-cycle with $\calf=0$. 

To sum up, the GKP conditions (\ref{wcy}) can be restated in the form (\ref{wcy2}) that corresponds to the differential conditions for the existence of calibrations for space-filling and string-like D-branes. These calibrations provide a background structure that automatically ensures the stability of a D-brane or a D-brane  configuration satisfying the bound (\ref{BPScond}). This is completely identical to what happens for BPS D-branes on supersymmetric backgrounds, since all the arguments rely on the same equations. However, in the supersymmetric case we also have the additional condition (\ref{03cond}) and it is natural to ask if it has a similar interpretation in terms of calibrations. 

To arrive at such an interpretation let us, using the ISD property of $\calg_3$, rewrite the SUSY condition (\ref{03cond}) as $H\wedge \Omega_0=0$. We can now make the assumption that $\d\Omega_0=0$ and combine it with $H\wedge \Omega_0=0$ into the condition $\d_H\Omega_0=0$. In this form, the supersymmetry condition (\ref{03cond})  
%is closer to the remaining set of 
%conditions (\ref{wcy2}) and 
can indeed be related to a calibration, now corresponding to D-branes filling 1+2 dimensions in $\mathbb{R}^{1,3}$, i.e. domain walls. The associated calibration in this case is given by \cite{luca}
\bea\label{dwcal}
\omega^{\rm (DW)}=\Re(e^{i\theta}\Omega_0)\ 
\eea 
where $\theta$ is a constant phase specifying the $\caln= 1/2$ preserved by the domain wall. Hence, the presence of $\calg_3^{(0,3)}$ can be equivalently characterized by the supersymmetry breaking condition
\bea\label{gkpdwb}
\d_H\Omega_0= H\wedge\Omega_0\neq 0\, \hspace{2cm} \, \text{DW (non)BPSness}
\eea
We then see that, while $\omega^{\rm (DW)}$ always satisfies (\ref{locbound}) (with $q=3$ in (\ref{dbi})), if the supersymmetry is broken by $\calg_3^{(0,3)}$ then $\d_H\omega^{\rm (DW)} = 0$ will be violated. As a result, for domain-walls the stability argument (\ref{compustab}) above cannot be straightforwardly repeated.

%%%%%%%%%%%%%%%%%%%%%%%%%%%%%%

\subsection{Extensions to generalized settings}\label{extension}

Given the observations above, we would now like to discuss the possible generalizations thereof. 
Consider a generic configuration with 10D space-time $X_{10} = \mathbb{R}^{1,3}\times_\om \calm_6$ and 10D metric 
\bea\label{metricansatz}
\d s^2\, =\, e^{2A(y)}\d x^\mu\d x_\mu+g_{mn}\d y^m\d y^n\ ,
\eea
and RR fields that split as in (\ref{RRsplit}). As in the supersymmetric case, we assume the existence of globally defined (generically non-Killing) spinors that endow the internal space with an $SU(3)\times SU(3)$ structure. This $SU(3)\times SU(3)$ structure can be alternatively characterized by the two internal $SO(6,6)$ pure spinors $\Psi_1$ and $\Psi_2$, which are complex polyforms (see Appendix \ref{ap:conv} for definitions). 
 $\Psi_1$ and $\Psi_2$ define the real polyforms
\bea\label{bcal}
\omega^{\rm (sf)}=e^{4A-\Phi}\Re\Psi_1\quad,\quad \omega^{\rm (string)}=e^{2A-\Phi}\Im\Psi_1\quad,\quad \omega^{\rm (DW)}=e^{3A-\Phi}\Re(e^{i\theta}\Psi_2)
\eea 
that satisfy the algebraic bound (\ref{locbound}) \cite{luca,Martucci06}.

In the supersymmetric case, the bulk supersymmetry conditions can be written completely in terms of $\Psi_1$ and $\Psi_2$ \cite{gmpt} - see eqs.~(\ref{susygeneral}) with $w_0=0$ - and automatically imply that all three polyforms in (\ref{bcal}) are proper  calibrations, satisfying the appropriate differential conditions \cite{luca,Martucci06}.  
So, in order to generalize the GKP example above we will consider backgrounds that, although non-supersymmetric, still admit properly defined calibrations $\omega^{\rm (sf)}$ and $\omega^{\rm (string)}$, as in section \ref{gkp}. This means that the differential conditions
\bseq\label{calgen}
\begin{align}
\d_H\omega^{\rm(sf)}&=\d_H(e^{4A-\Phi}\Re\Psi_1)=e^{4A}\tilde F  &\hspace{-1.25cm} {\rm gauge\ BPSness} \label{cal}\\
\d_H\omega^{\rm (string)}&=\d_H(e^{2A-\Phi}\Im\Psi_1)=0 & \text{D-string BPSness}/\text{D-flatness} 
\label{df}
\end{align}
\eseq
should be satisfied, as in the supersymmetric case. As already indicated in (\ref{df}), the D-string BPSness condition can also be interpreted as a 4D D-flatness condition \cite{lucapaul2}. 

Note  that, since $\tilde*_6^2=-1$ when acting on forms of any degree, the relation (\ref{dualf}) allows to write the gauge BPSness condition (\ref{cal}) as an imaginary anti-self-duality (IASD) condition. Indeed, defining the polyform 
\bea\label{calg}
\calg:=F+ie^{-4A}\d_H\big(e^{4A-\Phi}\Re\Psi_1\big)
\eea  
it is easy to see that
\bea\label{iasd}
\tilde*_6\calg=-i\calg \quad\text{(IASD)} \qquad\Leftrightarrow \qquad  \text{gauge BPSness}
\eea
The IASD condition (\ref{iasd}) relates forms of different degree, with the only exception of the 3-form $\calg_3$ in type IIB, for which (\ref{iasd}) reduces to a more familiar ISD condition 
\bea\label{3isd}
*_6\calg_3=i\calg_3\hspace{3cm}\text{IIB 3-form ISD}
\eea
that incorporates the GKP condition (\ref{isd}) as a special subcase. 

To summarize, in the following we will consider backgrounds satisfying both equations in (\ref{calgen}), which can be merged into
\boxedeq{\label{susypure}
e^{-2A+\Phi}\d_H(e^{2A-\Phi}\Psi_1)=-2\d A\wedge \Re\Psi_1+\tilde*_6 F \quad \quad \text{gauge \& D-string BPSness}}
On the other hand, the supersymmetry breaking will be characterized by
\boxedeq{\label{nonsusypure}
\d_H \omega^{\rm (DW)}\, =\, \d_H(e^{3A-\Phi}\Psi_2)=\{\text{susy-breaking terms}\}\  \hspace{.75cm} \, \text{DW (non)BPSness}}
which implies that $\omega^{\rm (DW)}$ is not a well-defined calibration. We will dub such class of backgrounds as Domain-Wall SUSY-breaking backgrounds, or DWSB backgrounds for short.

Finally, note that everything reduces to the GKP subcase of section \ref{gkp} if we set
\be\label{GKPps}
\Psi^{\rm (GKP)}_1\, =\, e^{iJ}\quad \quad \quad \Psi^{\rm (GKP)}_2\, =\, e^{-3A+\Phi}\Omega_0
\ee
so that (\ref{calgen}) reduces to (\ref{wcy2}), while the supersymmetry breaking (\ref{nonsusypure}) has the specific form (\ref{gkpdwb}).  

%%%%%%%%%%%%%%%%%%%%%%%%%%%%%%
%%%%%%%%%%%%%%%%%%%%%%%%%%%%%%

\section{DWSB backgrounds}\label{sec:dwsb}

From the very definition of DWSB vacua in terms of calibrations, one can easily deduce several common features in their 4D effective theory.\footnote{Our analysis below will be general and, in many aspects, independent of whether our 10D background leads to a 4D effective theory or not. For phenomenological purposes, however, it is useful to assume a 10D $\raw$ 4D compactification.} For instance, the gauge BPSness condition (\ref{cal}) will imply that space-time filling D-branes yield 4D gauge theories without tachyons in the adjoint representation. The D-string BPSness condition (\ref{df}), in turn, forbids fluxes to generate non-vanishing D-terms \cite{lucapaul2}. These features are of course all present in $\caln=0$ GKP vacua\footnote{Strictly speaking, the vanishing of the D-term in GKP vacua is ensured when the internal space is a compact CY. See \cite{anke} for the discussion of a non-compact example with non-vanishing D-term.}, which however display much more specific features like the well-known 4D no-scale structure, a particular pattern of SUSY-breaking soft terms, etc. One may thus wonder whether such specific features can be generalized and, if so, whether they constitute a substantial fraction of the set of DWSB vacua.

In this context, a popular approach to address these questions has been the use of the 4D effective K\"ahler potential and superpotential, that produce an effective potential and a certain soft-term pattern via the usual 4D supergravity formul\ae. While the results of such strategy are quite encouraging, it is important to bear in mind that they neglect key ingredients of flux compactifications like warping effects, that modify non-protected quantities like the K\"ahler potential even in the well-known warped Calabi-Yau case \cite{fp02} (see \cite{giddings,lucapaul2,bcdgmqs06,stud07} for explicit proposals for these corrections, and \cite{fm06} for further evidence). In addition, the effective approach relies on the knowledge of the light fields of the theory, which, beyond the Calabi-Yau approximation, is not usually available.\footnote{An exception is  AdS$_4$-compactifications on nilmanifolds and coset-manifolds, where the warping is constant and the spectrum of light fields can be explicitly worked out \cite{clau}.}

We would then like to generalize the features of GKP $\caln=0$ vacua from a fully 10D perspective, not necessarily tied up to any dimensional reduction scheme. In this spirit, we will first express the DWSB ansatz of the previous section in terms of the more familiar 10D dilatino and gravitino variations, in order to parameterize the space of DWSB backgrounds. As we will see, such parameter space is quite involved and so, in order to achieve interesting physics via a more tractable ansatz, we will single out a subfamily of DWSB backgrounds where the r.h.s. of (\ref{nonsusypure}) depends on a single complex parameter, and where the internal space $\calm_6$ can be understood as a generalized foliation.  Despite this simplification we will show that our one-parameter subansatz contains the set of GKP vacua, as well as many other families of vacua that will be analyzed in Section \ref{sec:subcases}.

%%%%%%%%%%%%%%%%%%%%%%%%%%%%%%

\subsection{Generic DWSB}

In general, an $SU(3)\times SU(3)$ structure is specified by  two internal chiral spinors $\eta_1$ and $\eta_2$, which define a ten dimensional bispinor $\epsilon=(\epsilon_1,\epsilon_2)^T$ via
\be\label{fermsplitmain}
\epsilon_1=\zeta\otimes\eta_1+\ \text{c.c.}\quad\quad\quad\quad \epsilon_2=\zeta\otimes \eta_2+\ \text{c.c.}
\ee
In $\caln=1$ supergravity backgrounds $\eps$ is identified with the ten dimensional supersymmetry generator. In our case, since we are assuming an approximate supersymmetry, such $\eps$ should also exist, although it will of course not satisfy the usual Killing equations. If we restrict ourselves to 4D Poincar\'e invariant backgrounds, we will generically have\footnote{Here $\Delta \eps_i$ is not the usual dilatino variation, but rather the modification defined in (\ref{modified}).}
\bea
\delta\psi^{(1)}_\mu=\frac12e^{A}\hat{\g}_{\mu}\zeta\otimes \calv_1+\ \text{c.c.}\quad& &\quad \delta\psi^{(2)}_\mu=\frac12e^{A}\hat{\g}_{\mu}\zeta\otimes \calv_2+\ \text{c.c.}\cr
\delta\psi^{(1)}_m=\zeta\otimes \calu^1_m+\ \text{c.c.}\quad& &\quad \delta\psi^{(2)}_m=\zeta\otimes \calu^2_m+\ \text{c.c.}\cr
\Delta\epsilon_1=\zeta\otimes \cals_1+\ \text{c.c.}\quad& &\quad \Delta\epsilon_2=\zeta\otimes \cals_2+\ \text{c.c.}
\label{vari}
\eea
where $\calv_{1,2}$, $\calu^{1,2}_m$ and $\cals_{1,2}$ are internal spinors parametrizing the supersymmetry breaking. Their explicit form in terms of $\eta_1$ and $\eta_2$, as well as most of the technical computations performed in this subsection is relegated to Appendix \ref{ap:sb}. 

By restricting to backgrounds satisfying (\ref{susypure}), the SUSY-breaking spinors $\calv_{1,2},\ \cals_{1,2}$, $\calu^{1,2}_m$ will be constrained. The obtained DWSB pattern turns out to be (see Appendix \ref{ap:sb})
\bea\label{calpres}
\calv_{1} =r\eta^*_{1}\quad & &\quad \calv_{2} =r\eta^*_{2}\ \cr
 \cals_{1}=-r\eta^*_{1}+p^2_m\gamma^m\eta_{1}\quad & &\quad \cals_{2}=-r\eta^*_{2}+p^1_m\gamma^m\eta_{2}\ \cr
 \calu^{1}_m=p^{1}_m\eta_{1}+q^{1}_{mn}\gamma^n\eta_{1}^*\quad & &\quad \calu^{2}_m=p^{2}_m\eta_{2}+q^{2}_{mn}\gamma^n\eta_{2}^*\quad 
\eea
with the following extra constraints
\bea\label{sbconst}
\Re p^1_m\,  = & 0 & =\, \Re p^2_m \cr
(1+iJ_1)^k{}_n q^1_{mk}=\, & 0 & =\, (1-iJ_2)^k{}_m q^1_{kn}\cr
(1+iJ_2)^k{}_n q^2_{mk}=\, & 0 & =\, (1-iJ_1)^k{}_m q^2_{kn}
\eea
where $J_1$, $J_2$ are the (almost) complex structures defined by $\eta_1$, $\eta_2$, respectively.

Plugging this into the DW BPSness condition, we obtain that the r.h.s. of (\ref{nonsusypure}) has the form
\bea\label{dwbroken}
e^{-3A+\Phi}\d_H(e^{3A-\Phi}\Psi_2) &=&  ir(-)^{|\Psi_2|}\Im\Psi_1+ \frac12 q^1_{mn}\gamma^n\Psi^*_1\gamma^m-\frac12 q^2_{mn}\gamma^m\Psi_1\gamma^n\cr &&+[(p^2-p^1)^m\iota_m+(p^1+p^2)_m\d y^m\wedge]\Psi_2  
\eea
where $|\Psi_2|$ is the degree mod 2 of the polyform $\Psi_2$. Note  that the generalized complex structure defined by $\Psi_2$ is integrable if and only if the susy-breaking parameters $r$ and $q^{1,2}_{mn}$ are vanishing \cite{gualtieri}.

%%%%%%%%%%%%%%%%%%%%%%%%%%%%%%

\subsection{One-parameter DWSB}\label{1param}

It is useful to further restrict the generic DWSB ansatz above to a simpler and  more tractable one. In the following, we will consider a simple subfamily of DWSB backgrounds, which still contains the GKP case, where the SUSY-breaking ansatz (\ref{calpres}) depends on just one parameter, $r$. As usual, this choice of 10D subansatz will be reflected in the 4D effective theory, and more precisely in the structure of 4D F-terms. We will come back to the 4D interpretation of this ansatz in Section \ref{sec:4dint}, where we will show that these one-parameter backgrounds are, in fact, no-scale vacua.

Let us then constrain the above ansatz (\ref{calpres}) by first setting $p^i_m  = 0$, and then by expressing the complex parameters $q^i_{mn}$ in terms of the complex parameter $r$ as
\bea\label{restr}
\calv_{1} =-\cals_{1}=r\eta^*_{1}\quad & &\quad \calv_{2} =-\cals_{2}=r\eta^*_{2}\  \cr
 \calu^{1}_m=-\frac{1}2\, r\, \Lambda^n{}_m\gamma_n\eta_{1}^*\quad & &\quad \calu^{2}_m=-\frac{1}2\, r\, \Lambda_m{}^n\gamma_n\eta_{2}^*\quad 
\eea
 where $\Lam$ is an $O(6)$ rotation matrix. This DWSB subansatz thus reads, in terms of the DW (non)BPSness equation  (\ref{nonsusypure}):
\bea\label{finansatz}
\d_H(e^{3A-\Phi}\Psi_2) &=&i\, r\, e^{3A-\Phi}\big[(-)^{|\Psi_2|}\Im\Psi_1+\frac12 \Lambda_{mn}\gamma^m(\Im\Psi_1)\gamma^n\big]
\eea 

While in principle $\Lam$ depends on 15 real parameters, we will specify them in terms of the background, so that $r$ is the only remaining parameter. Indeed, recall that given an $SU(3) \times SU(3)$ structure specified by (\ref{fermsplitmain}), the internal spinors are related by
\be\label{spinrel}
\eta_1\, =\, i U \eta_2
\ee
where $U$ is a unitary, in general point-dependent operator acting on six-dimensional spinors\footnote{With our Clifford algebra conventions of Appendix \ref{ap:conv} we can choose $U$ to be a real $8\times 8$ matrix.} so that
\be
U\gamma_{(6)}\, =\, \mp\gamma_{(6)}U\quad\text{in IIA/IIB}\ .
\ee
This implies that in IIB $U$ defines an element of $SU(4)$, while in IIA we can choose any 6D real vector $v$ so that it is now $\slashed{v}\gamma_{(6)}U$ that gives an element of $SU(4)$. In the vector representation, such rotations will be described by an $O(6)$ matrix $\Lambda^m{}_n$ satisfying
\be\label{localor}
U \gamma_m U^{-1}\, =\, \Lambda^n{}_m\gamma_n
\ee
Hence $\Lambda^m{}_n$ can be understood as an element of $SO(6)$ in IIB, while in IIA $\det\Lambda=-1$ and $\Lambda$ can be identified with an element of $SO(6)$ up to a reflection in an arbitrary direction.\footnote{For an explicit construction of $\Lambda$ in terms of $U$, see e.g. section 3 of the second paper in \cite{hassan}.}   

Therefore an obvious choice for $\Lambda$ in (\ref{restr}) is the one given by the local $O(6)$ rotation in (\ref{localor}). As a first consistency check, one can see that this choice is compatible with the constraints (\ref{sbconst}) of the general DWSB ansatz (\ref{calpres}). Indeed, the subansatz (\ref{restr}) implies that
\be
q^{1}_{mn}=-\frac14\, r(1-iJ_1)^k{}_n\Lambda_{km}\quad \text{and}  \quad q^{2}_{mn}=-\frac14\, r(1-iJ_2)^k{}_n\Lambda_{mk}
\label{qrestr}
\ee 
In addition,  from (\ref{spinrel}) and (\ref{localor}) there follows the identity
\be
(J_1)^m{}_k\Lambda^k{}_n\, =\, \Lambda^m{}_k(J_2)^k{}_n
\ee
and thus $q^{i}_{mn}$ in (\ref{qrestr}) satisfy 
\be
 (1-iJ_2)^k{}_m q^1_{kn}\, =\, 0\quad \quad (1-iJ_1)^k{}_m q^2_{kn}=0
\ee
in agreement with (\ref{sbconst}).

It is however easy to see that (\ref{spinrel}) does not fully specify the rotation $U$. Indeed, since $\eta_1$ and $\eta_2$ have the same norm, we can rewrite (\ref{spinrel}) as $\eta_1 = i U_1^{-1} U_\eta U_2 \eta_2$, where $U_i \eta_i = \eta$ for some fixed spinor $\eta$. Any element $U_\eta$ belonging to the stabilizer subgroup of $\eta$, $SU(4)_\eta \simeq SU(3)$ will then satisfy this relation. Hence, since $\dim_{\mathbb{R}} SU(4)_\eta =8$, eight real parameters  still need to be specified in the ansatz (\ref{restr}).\footnote{In fact, these would be eight  real functions, since $U$ is in general point-dependent.} In the following we will consider background geometries that naturally do the job.

Some intuition on how to construct such geometries can be obtained from our knowledge of fermionic D-brane actions. Indeed, unitary operators with the properties of $U$ above are naturally found when constructing the $\kappa$-symmetry operator associated with D-branes in a  certain background \cite{bkop97}: To each kind of  D-brane we can associate a different rotation $U$, and vice versa. From this point of view, the fact that $SU(4)_\eta$ is non-trivial only means that locally there is more than one kind of BPS D-brane in a given $\caln=1$ (or almost $\caln=1$) background, that is, for a given choice of $\eta_1$ and $\eta_2$. This also suggests that $\Lam$ will be fixed by the background if, at least locally, a D-brane is singled out over the whole family of possibilities.

Indeed, one can easily do so as follows. Given the metric ansatz (\ref{metricansatz}), let us first suppose that we can split the internal space tangent bundle as $T_{\calm_6} =T_\Pi\oplus T^\perp_\Pi$, where $T_\Pi$ is a subbundle of odd/even dimension $n$ in IIA/IIB, and $T^\perp_\Pi$ is its orthogonal completion. Furthermore, we consider a real two-form $R\in \Lambda^2T_\Pi^*$, and construct the operator $U$ as
\be
U\, =\, \gamma^{n}_{(6)}\sum_k\frac{\epsilon^{\alpha_1\ldots\alpha_{n-2k}\beta_1\ldots\beta_{2k}}}{(n-2k)!k!2^k\sqrt{\det(g|_\Pi+R)}}\, \gamma_{\alpha_1\ldots\alpha_{n-2k}}R_{\beta_1\beta_2}\cdots R_{\beta_{2k-1}\beta_{2k}}
\ee 
where the indices $\alpha,\beta$ correspond to some basis $e_\alpha$ of $T_\Pi$ and we denote with $|_\Pi$ the pull-back to $T_\Pi$. The associated  $O(6)$ transformation $\Lambda$ has the following explicit form
\be\label{lambda}
\Lambda\,=\,\bbone_\perp-(g|_{\Pi}+R)^{-1}(g|_\Pi-R)\ , 
\ee
where $\bbone_\perp$ denotes the projection along $T^\perp_\Pi$. Since we are also assuming that $U$ satisfies the condition (\ref{spinrel}), $(T_\Pi,R)$ is  a (generalized) subbundle calibrated by $\Re\Psi_1$. 

Finally, using this parameterization of $U$ in terms of $T_\Pi$ and $R$, we can write the DW (non)BPSness equation (\ref{finansatz}) as follows:\footnote{Recall that $\sigma$ is the operator that reverses the order of the indices of a form.}
\be\label{finansatz2}
\d_H(e^{3A-\Phi}\Psi_2) \, =\, 4i\, r\,(-)^{|\Psi_2|} e^{3A-\Phi}\frac{\sqrt{\det g|_\Pi}}{\sqrt{\det(g|_\Pi+R)}} e^{-R} \wedge \sigma(\d \text{Vol}_\perp)
\ee
where $\d \text{Vol}_\perp$ is the volume form of $T^\perp_\Pi$, such that $\d \text{Vol}_6=\d \text{Vol}_\Pi\wedge \d \text{Vol}_\perp$. This equation strongly constrains the choice of $T_\Pi$ and $R$, until now just restricted by algebraic conditions.  Indeed, the r.h.s. of (\ref{finansatz2}) is $\d_H$-exact and so $\d_H$-closed.  By Frobenius' theorem, it is not difficult to realize that this implies that $T_\Pi$ must be integrable, foliating $\calm_6$ into leaves $\Pi$, and furthermore that $\d R=H|_\Pi$. Thus, this construction applies to internal spaces $\calm_6$ that  can be foliated by calibrated generalized submanifolds $(\Pi,R)$.  This geometry can be physically probed by space-filling D-branes wrapping $\Pi$ with $\calf = R$, that can move around and span the entire internal space $\calm_6$. We will refer to such calibrated D-branes as {\it aligned} with the background.

In addition, (\ref{finansatz2}) constrains the choices of $r$ which, up to now, could be taken to be an arbitrary complex function of $\calm_6$. However, given a choice of generalized foliation $(\Pi,R)$, $r$ must be chosen such that the r.h.s. of (\ref{finansatz2}) is d$_H$-exact. This will still be true if $r$ is multiplied by an overall complex constant, but not if changed by an arbitrary complex function. Hence, we see that our ansatz truly depends on a single complex parameter,  $r$.

Furthermore, we see that the SUSY-breaking term appearing on the right-hand side of (\ref{finansatz}) can be interpreted as a polyform which is Poincar\'e dual to the generalized submanifold  $(\Pi,R)$. The minimal degree of the polyform appearing in this SUSY-breaking term corresponds to the codimension of $\Pi$, while higher degree contributions are induced by a non-vanishing $R$. Recall that the {\em type} of a pure spinor is defined as its lowest degree. Then  (\ref{finansatz}) can be satisfied only if
\be\label{typeconst}
\text{type}(\Psi_2)\, \leq\,  \text{codim}(\Pi)-1\ .
\ee
For example, in IIB $SU(3)$-structure backgrounds $\Psi_2$ is a three-form, hence $\Psi_2$ is of type three and the above construction applies only if $\Pi$ corresponds to D3 or D5 branes.  In the next subsection we will discuss in more detail the properties of these configurations from the viewpoint of generalized complex geometry. 

Let us now check that the GKP case of subsection \ref{gkp} fits into this one-parameter subfamily of backgrounds. Recall that in this particular case we are in type IIB theory and our pure spinors/polyforms are given by  (\ref{GKPps}). Also, supersymmetry  is broken via a non vanishing $H^{(0,3)}=-ie^\Phi F^{(0,3)}$ flux. Using the warped CY/F-theory relation $\eta_1=i\eta_2\equiv\eta$ we obtain that the SUSY-breaking spinorial parameters are
\bea
\cals_1=-i\cals_2=-\calv_1=i\calv_2=\frac14 \slashed{H}^{(0,3)}\eta\quad & \quad  &\quad \calu^1_m=-i\calu_m^2=\frac14 \slashed{H}_m^{(0,3)}\eta
\eea
which indeed fit into the ansatz (\ref{restr}), with
\bea
\label{gkpparam}
r=-\frac14 e^{-3A+\Phi}\Omega_0\cdot H^{(0,3)}\quad &\quad & \quad \Lambda=\bbone
\eea
where the contraction $\cdot$ is defined in (\ref{cdot}). In this case, the DW (non)BPSness equation (\ref{gkpdwb}) reads
\be
\d_H(e^{3A-\Phi}\Psi_2)\, =\, H\wedge\Omega_0\, =\, 4i\,r\, e^{3A-\Phi}\d \text{Vol}_6
\ee
so clearly $\Lam = \bbone$ corresponds to choosing the leaves $\Pi$ of the above description as the points of $\calm_6$, singling out D3-branes of all other space-filling BPS D-branes. This is not surprising, since in warped Calabi-Yau compactifications with ISD fluxes D3-branes at any point are automatically BPS and, at least at the classical level that we are working, do not feel any effect of the background fluxes, much in contrast to D7-branes. 

 In the same spirit as GKP, we can interpret each family of these one-parameter DWSB vacua as a deformation of a class of $\caln=1$ vacua where (\ref{spinrel}) is also true. Since this relation between 6D internal spinors is related to the spectrum of localized BPS sources like space-filling D-branes and O-planes, turning on the one-parameter deformation will not change the definition of a BPS gauge D-brane, even if for $r\neq 0$ our theory is no longer supersymmetric. Note  that this nicely matches the fact that for DWSB vacua the gauge BPSness condition (\ref{cal}) is unchanged with respect to the supersymmetric case. 

Finally, let us point out that the GKP SUSY-breaking parameters in (\ref{gkpparam}) have a well-defined interpretation from the 4D effective physics viewpoint. On the one hand, $r$ is related to the vev of the auxiliary field $T$, the complexified overall K\"ahler modulus and, since this is the modulus which dominates SUSY breaking in the GKP scenario, to the four-dimensional gravitino mass. On the other hand, $\Lambda$ is related to the structure of soft-terms felt by D7-branes in this kind of backgrounds \cite{gmm05,Martucci06,bcdgmqs06}. As we will see in sections \ref{sec:4dint} and \ref{sec:fsoft}, similar statements apply to the more general family of one-parameter DWSB vacua obtained from (\ref{restr}).

%%%%%%%%%%%%%%%%%%%%%%%%%%%%%%

\subsection{Generalized geometry of DWSB backgrounds}\label{gcsec}

We would now like to discuss some of the geometrical features of the above class of DWSB backgrounds, and again the GCG language \cite{gualtieri} turns out to be the most natural for this purpose.
First of all, let us rename the main objects of our discussion, by defining  $\calz:=e^{3A-\Phi}\Psi_2$ and  
\be\label{jtilde}
\tilde\jmath_{(\Pi,R)}\, :=\, 4(-)^{|\Psi_2|} e^{3A-\Phi}\frac{\sqrt{\det g|_\Pi}}{\sqrt{\det(g|_\Pi+R)}} e^{-R} \wedge \sigma(\d \text{Vol}_\perp)
\ee
We have used the letter $\tilde\jmath$ intentionally, since $\tilde\jmath_{(\Pi,R)}$ can be thought of as a smeared version of the source current $j$ associated with a D-brane wrapping a generalized submanifold $(\Pi,R)$. In particular, since $(\Pi,R)$ are calibrated, $\tilde\jmath_{(\Pi,R)}$ is normalized so that
\be
\langle \Re\Psi_1,\tilde\jmath_{(\Pi,R)}\rangle\, =\, 4(-)^{|\Psi_2|} e^{3A-\Phi}\d{\rm Vol}_6
\ee 

The DWSB equation (\ref{finansatz2}) can then be rewritten in the more concise form
\be\label{intviol}
\d_H\calz\, =\, i\,r\,\tilde\jmath_{(\Pi,R)}
\ee
$\calz$ is a {\rm complex} pure spinor, that defines an {\rm almost} generalized complex structure $\calj$ that is {\em not} integrable. The obstruction to its integrability is given exactly by the non-vanishing $r\tilde\jmath_{(\Pi,R)}$ term that appears on the r.h.s. of (\ref{intviol}). This absence of integrability implies that either we have no natural complex/symplectic coordinates defined by $\calz$ on the internal space, or the $H$-twist does not respect them. For instance, in the GKP case $\calj$ defines and ordinary complex structure which is still integrable and the (generalized) non-integrability of $\calj$ reduces to the presence of a non-vanishing $H^{0,3}$ flux. 

On the other hand, $\tilde\jmath_{(\Pi,R)}$ defines a Dirac structure (see e.g. \cite{gualtieri}) that is integrable precisely because of (\ref{intviol}). Indeed, as we have seen in the previous subsection, eq.(\ref{intviol}) implies that $T_\Pi$ can be integrated into the foliation $\Pi$ and that $\d R=H|_\Pi$. The maximal isotropic sub-bundle\footnote{We recall that the Clifford action of a generalized vector $\mathbb{X}=X+\xi\in T_{\calm_6} \oplus T^*_{\calm_6}$ on a $SO(6,6)$ spinor  (i.e. a polyform) $\alpha$ is given by $\mathbb{X}\cdot \alpha:=\iota_X\alpha+\xi\wedge \alpha$.} 
\bea
T_{(\Pi,R)}&=&\{\mathbb{X}\in T_{\calm_6}\oplus T^*_{\calm_6}\ :\ \mathbb{X}\cdot  \tilde\jmath_{(\Pi,R)}=0\}=\cr
&=&\{X+\xi\in T_\Pi\oplus T^*_{\calm_6}\ :\ \xi|_\Pi=\iota_X R\}\ ,
\eea
then corresponds to the generalized tangent bundle of the foliation $(\Pi,R)$. The integrability of the Dirac structure, i.e. the existence of the foliation  $(\Pi,R)$,  is then equivalent to requiring that $T_{(\Pi,R)}$ is involutive under the twisted  Courant bracket $[\cdot,\cdot]^H_C$.\footnote{Let us recall that, if $\mathbb{X}=X+\xi$ and $\mathbb{Y}=Y+\chi$, their twisted Courant bracket is given by $[\mathbb{X},\mathbb{Y}]^H_C=[X,Y]+\call_X\chi-\call_Y\xi-\frac12\d(\iota_X\chi-\iota_Y\xi)+\iota_X\iota_YH$.} 

For completeness, let us give two additional characterizations of $T_{(\Pi,R)}$. First, it can be directly defined in terms of the $O(6)$ matrix $\Lambda$ in (\ref{lambda}) as follows
\be
X+\xi\in T_{(\Pi,R)}\quad\Leftrightarrow \quad (1+\Lambda)X+(1-\Lambda)g^{-1}\cdot\xi=0
\ee
Second, it can also be defined in terms of the following linear operator acting on $T_{\calm_6}\oplus T^*_{\calm_6}$
\be
\calr_{(\Pi,R)}\, =\, \left( \begin{array}{cc} P & 0 \\ R P + P^T R & -P^T \end{array}  \right)\ ,
\ee
where $P=\bbone_\|-\bbone_\perp$ is the canonical product structure associated with $T_\Pi$.  One can think of $\calr_{(\Pi,R)}$ as the generalized product structure  associated with $T_{(\Pi,R)}$, since $\calr^2_{(\Pi,R)}=\bbone$ and $T_{(\Pi,R)}=\{\mathbb{X}\in T_{\calm_6}\oplus T^*_{\calm_6}\ :\  \calr_{(\Pi,R)}\mathbb{X}=\mathbb{X}\}$. Of course, the integrability of the Dirac structure constrains $\calr$ and $\Lambda$, that must satisfy appropriate differential conditions. 

In any case, the fact that the Dirac structure is integrable has a direct consequence, namely that the Courant bracket defines a Lie algebroid structure\footnote{See, e.g., \cite{gualtieri} for definitions and properties of Lie algebroids and Dirac structures in the GCG context.} on  $T_{(\Pi,R)}$ and furthermore that one can define a differential $\d_{(\Pi,R)}$ acting on the graded complex $\bigoplus^6_{k=0}\Lambda^kT^*_{(\Pi,R)}$. 

Let us now come back to the generalized almost complex structure $\calj$, and analyze it from this latter point of view. We can indeed think of $\calj$ as a almost Dirac structure associated with the maximal isotropic sub-bundle $L\subset (T_{\calm_6}\oplus T^*_{\calm_6})\otimes \mathbb{C}$ defined as the annihilator of $\calz$ (i.e. $L\cdot\calz=0$).  The fact that $\calj$ is not integrable means that $L$ is not involutive under the Courant bracket, and so we can define neither an associated Lie algebroid structure on $L$, nor the corresponding differential  $\bar\partial_\calj$  acting on $\bigoplus^6_{k=0}\Lambda^kT^*_{(\Pi,R)}$. The differential $\bar\partial_\calj$ would be the generalization of the usual Dolbeaut differential of complex geometry, and its existence is equivalent to the integrability of the generalized complex structure. 

The foliation $(\Pi,R)$, however, is calibrated and thus an almost generalized complex foliation, in the sense that $T_{(\Pi,R)}$ is stable under $\calj$.\footnote{The fact that $(\Pi,R)$ is an almost generalized complex foliation could be taken as the only assumption made for this construction, since (\ref{intviol}) automatically implies that $\langle \Im\Psi_2,\tilde\jmath_{(\Pi,R)}\rangle=0$, and this provides the additional D-flatness condition that implies that $(\Pi,R)$ is calibrated \cite{luca,Martucci06}.} Then, we can consider the following complex bundle
\be
L_{(\Pi,R)}\,:=\,L\cap T_{(\Pi,R)}\,=\, \{\mathbb{X}\in (T_{\calm_6}\oplus T^*_{\calm_6})\otimes \mathbb{C}\ :\  \mathbb{X}\cdot\calz=0\ \text{and}\ \mathbb{X}\cdot\tilde\jmath_{(\Pi,R)}=0\}
\ee
The key point is that $L_{(\Pi,R)}$ is involutive under the Courant bracket thanks to (\ref{intviol}), since for any $\mathbb{X},\mathbb{Y}\in L_{(\Pi,R)}$ we obviously have that   $[\mathbb{X},\mathbb{Y}]^H_C\in T_{(\Pi,R)}$ and furthermore
\be
 [\mathbb{X},\mathbb{Y}]^H_C\cdot\calz\, =\, -\mathbb{Y}\cdot\mathbb{X}\cdot\d_H\calz=-r\mathbb{Y}\cdot\mathbb{X}\cdot\tilde\jmath_{(\Pi,R)}=0
\ee
This implies that the Courant bracket defines a Lie algebroid structure on $L_{(\Pi,R)}$. We will denote with $\bar\partial_{(\Pi,R)}$ the associated differential acting on the graded complex $\bigoplus^3_{k=0} \Lambda^k L^*_{(\Pi,R)}$.

To summarize, we have seen that $\calj$, although non integrable, allows to define a `holomorphic' differential $\bar\partial_{(\Pi,R)}$ with respect to the foliation $(\Pi,R)$. This is a remarkable property of the class of one-parameter DWSB configurations that could also hold for more general DWSB backgrounds, and could also have interesting implications for the associated 4D effective theories. For example, in the case of supersymmetric compactifications, it turns out that the first cohomology group $H^1_{\bar\partial}$ of this Lie algebroid defines the local moduli space of BPS D-branes \cite{lucapaul1}, and the fact that it comes with natural holomorphic structure is related to the fact that this moduli space is a complex manifold. It is thus natural to speculate that $H^1_{\bar\partial}(\Pi,R)$ could define some holomorphic structure on the space of calibrated leaves, and so on the moduli space of aligned D-branes. For instance, in the GKP case $H^1_{\bar\partial}(\Pi,R)=L^*_{(\Pi,R)}=T_{\calm_6}^{1,0}$, and so the space of all leaves (that in this case are just points) corresponds to $\calm_6$. The holomorphic structure of the space of leaves corresponds, in turn, to the holomorphic structure of $\calm_6$ defined by the non-integrable (in the $H$-twisted sense) pure spinor $\calz\equiv \Omega_0$. This matches the fact that the moduli space of D3-branes does not change with respect to the supersymmetric case, and it is still a complex manifold $\calm_6$. It would be interesting to see if an analogous statement applies for other aligned D-branes in more general DWSB vacua.

Finally, it is also tempting to speculate that the non-integrability of $\calj$ may be related to a closed string moduli space that is not a complex manifold. Indeed, again looking at the GKP example, is easy to see that as long that we break supersymmetry via a non-vanishing $H^{3,0}$,  the moduli space of complex structures and complex axio-dilaton is no longer a complex manifold. This is easy to see in the toroidal GKP example of Section \ref{sec:examples}, where the moduli space is given by the solution to the non-holomorphic equations (\ref{nonholmoduli}). In more general $\caln=0$ GKP compactifications, the same holds because the ISD condition (\ref{isd}) is not holomorphic. Finally, the constraints that the equations of motion impose on DWSB backgrounds  (see next section) are also of non-holomorphic nature, so one would expect the same statement to apply. A complete analysis of all these interesting problems is beyond the scope of the present work, but we hope to come back to them in the near future.

%%%%%%%%%%%%%%%%%%%%%%%%%%%%%%
%%%%%%%%%%%%%%%%%%%%%%%%%%%%%%

\section{The effective potential of type II flux vacua}\label{sec:inteflux}

In the last section we have specified a set of non-supersymmetric backgrounds that generalize the well-known $\caln=0$ GKP vacua. The next main question to be addressed is whether these generalized backgrounds are stable vacua of the theory. As discussed in the introduction, even at the classical supergravity level that we are working this is not an easy question to address. Not only need we to show that our backgrounds satisfy the equations of motion of type II supergravity, but also that they do not contain any closed string tachyons.

The purpose of this section is to address both questions and to argue that, indeed, the DWSB backgrounds discussed in section \ref{1param} are tachyon-free vacua. Our strategy will be to construct the effective action for type II compactifications to 4D directly from the 10D type II supergravity action,\footnote{In some particular cases one may hope to construct such potential from a 4D effective K\"ahler potential and superpotential, a practice largely followed in the literature. However, as argued above this strategy is not fully reliable in general situations.} and to show that the extremization of the action is equivalent to satisfying the type II equations of motion. We will also address the absence of closed string tachyons for our compactifications to flat space, that will be guaranteed if the effective potential $\calv_{\rm eff}$ entering this action is positive semi-definite. In order to present a simple, clear derivation of our results we will, in subsection \ref{fulleffpot}, rewrite the type II effective potential in terms of the pure spinors $\Psi_1$ and $\Psi_2$, which played a key role in classifying $\caln=0$ vacua in Section \ref{sec:vacua}. Then, in subsection \ref{effdwsb}, we will analyze $\calv_{\text{eff}}$ for the particular case of the one-parameter DWSB backgrounds of section \ref{1param}. We will see that, by truncating the full potential in a way naturally suggested by the geometry of these configurations, we obtain a positive semi-definite potential  $\calv_{\rm eff}$. This analysis will be essential in proving the no-scale properties of these vacua, a subject that we will address in the next section.

Note  that, in some sense, our philosophy in analyzing $\calv_{\rm eff}$ is quite similar to the one in \cite{ccdl03}, where 4D $\caln=1$ heterotic vacua were analyzed directly from the 10D action. The present approach is however more general since  we do not restrict to $\caln=1$ but also analyze $\caln=0$ vacua of the theory. This has a direct consequence, namely that while in \cite{ccdl03} the equations of motion do not impose any additional constraint to the supersymmetry equations and Bianchi identities, we do find additional constraints to the DWSB conditions (\ref{susypure}) and (\ref{nonsusypure}). We will compute explicitly such extra constraints in the case of the one-parameter DWSB backgrounds of section \ref{1param}, and give a 4D interpretation of them in Section \ref{sec:4dint}. While these extra constraints are automatically satisfied in GKP vacua, they could be non-trivial in other DWSB compactifications, so in Section \ref{sec:subcases} we will analyze them for several families of DWSB vacua.

Finally, let us point out that besides computing the effective potential, another quite powerful approach  to prove that $\caln=0$ backgrounds are indeed vacua is based on the integrability results of \cite{lt,gauntlett,kt}. While these integrability  techniques have so far only been applied to $\caln=1$ backgrounds, we will show in Section \ref{sec:inte} that they can  be extended  to $\caln=0$ backgrounds, and we will use them to construct novel $\caln=0$ vacua.

%%%%%%%%%%%%%%%%%%%%%%%%%%%%%%

\subsection{Effective potential and equations of motion}

Let us begin our discussion by introducing an effective 4D action that gives the complete set of 10D equations of motion for backgrounds with metric of the form (\ref{metricansatz}). In fact, we will be more general and consider the 10D metric ansatz
\be
\d s^2_{10}\, =\, e^{2A(y)}\d s^2_{X_4}+g_{mn}\d y^m\d y^n
\label{10dansatz}
\ee
where $X_4$ is a general 4D space whose metric $g_4$ only depends on the 4D coordinates $x^\mu$, and all the other fields (warping included) depend only on the internal coordinates $y^m$. The  `effective' 4D action for these configurations is\footnote{In order to simplify the expressions to follow, we work in units of $2\pi\sqrt{\alpha^\prime}=1$, so that all D-brane tensions are equal. The dimensionful expressions can be easily obtained by reinstating $2\pi\sqrt{\alpha^\prime}$, as in Section \ref{sec:examples}. Furthermore,  we are neglecting anomalous curvature-like corrections to the D-brane and O-plane contribution not to clutter the notation. They can be easily added without affecting the results of the discussion.} 
\be\label{4daction}
S_{\rm eff}\, =\, \int_{X_4}\d^4 x\sqrt{-g_4}\Big(\frac12\,\caln R_{4}-2\pi\calv_{\rm eff}\Big)
\ee
where $R_4$ is the 4D scalar curvature,
\be\label{ckp}
\caln \, =\, 4\pi\int_{\calm_6}e^{2A-2\Phi} \d \text{Vol}_6
\ee
is the warped-volume of the internal space (and gives the conformal K\"ahler potential of the 4D description - see \cite{lucapaul2} and section \ref{sec:4dint}), and
\bea\label{4Deffpot}
\calv_{\rm eff}&=&\int_{\calm_6}{\rm d}\text{Vol}_6\, e^{4A}\Big\{e^{-2\Phi}[-\calr+\frac12 H^2-4(\d\Phi)^2+8\nabla^2A+20(\d A)^2]-\frac12\tilde F^2\Big\}\cr
&&+\sum_{i\in\text{loc. sources}}\tau_i \Big(\int_{\Sigma_i}e^{4A-\Phi}\sqrt{\det(g|_{\Sigma_i}+\calf_i)}-\int_{\Sigma_i}C^{\rm el}|_{\Sigma_i}\wedge e^{\calf_i}\Big)
\eea
is the type II effective potential density. In (\ref{4Deffpot}), the first line (where $\calr$ refers to the 6D scalar curvature) contains the closed string sector, while the second line contains the contribution from localized sources, labeled by the index $i$. We only consider D-branes and O-planes as localized sources, with $\tau_{\text{D$p$}}=1$ and $\tau_{\text{O$q$}}=-2^{q-5}$ respectively.\footnote{For O-planes one should set $\calf=0$ and, despite their contribution to the supergravity action, they should not be seen as dynamical objects of the compactification.} They both couple to the electric RR-field $C^{\rm el}$, defined by $e^{4A}\tilde F=\d_H C^{\rm el}$. Note  that here we are using the `electric' frame to describe the RR-degrees of freedom, instead of using the `magnetic' one where the fundamental field is  $C$, that is (locally) defined by $\d_H C =F=-\tilde*_6 \tilde F$. In any case in this effective formulation, in either frame, the usual problems related to the self-duality conditions  on the total RR-field strengths are not present. 

Let us now to compare the 10D equations of motion for the above ansatz with the  equations  obtained by extremizing the 4D effective action (\ref{4daction}).\footnote{In (\ref{4daction}) we are omitting kinetic terms for the internal fields since we are only  considering  configurations where they are constant along the 4D directions.} One can check that, by varying (\ref{4daction}) with respect to the dilaton $\Phi$, $B$-field, internal metric $g$, electric RR-potentials $C^{\rm el}$ and open-string degrees of freedom, one gets the same set of equations as obtained by first varying the full 10D supergravity plus D-brane action with respect to the same fields and then restricting to our class of configurations. Furthermore, by varying (\ref{4daction}) with respect to the the warping, one gets the trace of the external Einstein equations. Finally, from the variation with respect to $g_4$ one gets that the 4D space is Einstein, clearly in agreement with the 10D picture, with $R_4=8\pi\calv_{\rm eff}/\caln$. This same equation can indeed be obtained by integrating over the internal space the trace of the 10D external Einstein equations, multiplied by an appropriate factor. Thus, we conclude that the effective action (\ref{4daction}) reproduces {\em the full set} of equations of motion that must be imposed on the most generic flux compactification to an Einstein 4D space.

In particular, the `electric' RR equations of motion reproduce the ordinary (`magnetic') Bianchi identities
\be\label{intBI}
\d_H F\, =\, -\d_H \tilde*_6 F\, =\, -j_{\rm tot}\, :=\, -\sum_i\tau_i j_i
\ee
where the source current $j$ for a D-brane wrapping an internal cycle $\Sigma$ is defined by
\be
\int_\Sigma\chi|_\Sigma\wedge e^\calf\, =\, \int_{\calm_6}\langle \chi, j\rangle
\ee
for any polyform $\chi$, where $\langle\cdot,\cdot\rangle$ is the six-dimensional Mukai pairing (see (\ref{mukain})). Also, for later convenience, let us observe that the following combined variation of 
(\ref{4daction}) reproduces the external components of modified  Einstein equations (\ref{einstein2}): 
\bea\label{warpdilvar}
\frac{\delta S_{\rm eff}}{\delta A}+ 2\frac{\delta S_{\rm eff}}{\delta \Phi}\, =\, 0& \quad\Leftrightarrow \quad& \text{modified external Einstein eqs.}
\eea
%

%%%%%%%%%%%%%%%%%%%%%%%%%%%%%%

\subsection{Effective potential, pure spinors and calibrations}
\label{fulleffpot}

To proceed, let us rewrite the effective potential $\calv_{\rm eff}$ in terms of the generalized calibrations of Section \ref{sec:vacua}. Consider first the contribution coming from the closed string RR sector + localized sources, i.e. the second line plus the last term in the first line of (\ref{4Deffpot}). This piece of $\calv_{\rm eff}$ can be rewritten as\footnote{The square of a polyform is the sum of the squares of its components of definite degree, as defined in appendix \ref{ap:conv}, before eq.~(\ref{cdot}). The same rule applies to the absolute value squared $|\ldots|^2$, defined after eq.~(\ref{cdot}), that will be used below.}
\begin{align}\label{1part}
\frac12\int_{\calm_6}\d\text{Vol}_6&\,e^{4A}\big[\tilde F-e^{-4A}\d_H(e^{4A-\Phi}\Re\Psi_1)\big]^2-\frac12\int_{\calm_6}\d\text{Vol}_6\,e^{-4A}\big[\d_H(e^{4A-\Phi}\Re\Psi_1)\big]^2 \nn \\ &
+\sum_{i\in \text{loc. sources}}\!\!\!\!\tau_i\int_{\calm_6}e^{4A-\Phi}\Big(\d\text{Vol}_6\,\rho^{\rm loc}_i-\langle \Re\Psi_1,j_i\rangle  \Big)
\nn\\ & +\int_{\calm_6}\langle  e^{4A-\Phi}\Re\Psi_1-C^{\rm el},\d_H F+j_{\rm tot}\rangle 
\end{align}
where  the Born-Infeld density $\rho^{\rm loc}$  associated with a D-brane/O-plane wrapping $(\Sigma,\calf)$ is defined  as
\be
\rho^{\rm loc}\, =\,  \frac{\sqrt{\det (g|_\Sigma+\calf)}}{\sqrt{\det g}}\,\delta(\Sigma)
\ee
Note  that, in terms of $\rho^{\rm loc}$ the algebraic inequality (\ref{locbound}) takes the form
\be\label{densbound}
\rho^{\rm loc}\, \geq\, \frac{\langle e^{-\Phi}\Re\Psi_1,j\rangle}{\d {\rm Vol}_6}
\ee
where by $1/{\d {\rm Vol}_6}$ we mean that we remove the ${\d {\rm Vol}_6}$ factor in the numerator. 
As boundary condition, we impose that the orientifolds be calibrated, so that for them the above inequality is saturated. Thus, the term in the second line of (\ref{1part}) is always positive.

We now face the challenging problem of expressing the six-dimensional scalar curvature $\calr$ in terms of the pure spinors $\Psi_1$ and $\Psi_2$, that indeed contain the full information about the metric. Recently, a formula was found in \cite{cassani2}, that solves this problem provided that some restrictions on the form of $\d_H\Psi_{1,2}$ are satisfied (see eqs.~(4.19-20) in \cite{cassani2}). Unfortunately, our backgrounds do not satisfy these restrictions and, furthermore, in \cite{cassani2} the warp-factor is considered constant. Thus, the  result  therein needs to be  appropriately generalized. By going through the derivation in \cite{cassani2}, it is possible to obtain such a generalization. We carry out this somewhat technical discussion in Appendix \ref{ap:scalarR}.  Applying (\ref{gencur2}), one can easily compute the terms in (\ref{4Deffpot}) that complete (\ref{1part}).  The resulting full potential (\ref{4Deffpot}) takes the form
\bea\label{finpot0}
\calv_{\rm eff}&=&\frac12\int_{\calm_6}\d\text{Vol}_6\,e^{4A}\big[\tilde F-e^{-4A}\d_H(e^{4A-\Phi}\Re\Psi_1)\big]^2\cr
&& +\frac12\int_{\calm_6}\d\text{Vol}_6\,\big[\d_H(e^{2A-\Phi}\Im\Psi_1)\big]^2+\frac12\int_{\calm_6} \d\text{Vol}_6 e^{-2A}\, 
\big|\d_H(e^{3A-\Phi}\Psi_2)\big|^2\cr
&&+\sum_{i\in \text{D-branes}}\tau_i\int_{\calm_6}e^{4A-\Phi}\Big(\d\text{Vol}_6\,\rho^{\rm loc}_i-\langle \Re\Psi_1,j_i\rangle  \Big)\cr
&&-\frac14\int_{\calm_6} e^{-2A}\Big\{\frac{|\langle \Psi_1,\d_H(e^{3A-\Phi}\Psi_2)\rangle |^2}{\d\text{Vol}_6}+\frac{|\langle \bar\Psi_1,\d_H(e^{3A-\Phi}\Psi_2)\rangle |^2}{\d\text{Vol}_6}\Big\}\cr
&&-4\int_{\calm_6} \d\text{Vol}_6\,e^{4A-2\Phi}\big[(u_{\rm R}^1)^2+(u_{\rm R}^2)^2\big]+\int_{\calm_6}\langle  e^{4A-\Phi}\Re\Psi_1-C^{\rm el},\d_H F+j_{\rm tot}\rangle\cr &&
\eea
where $u_{\rm R}^{1,2}:=(u^{1,2}_m+u^{*1,2}_m)\d y^m$ are the real extension (and contain the same information) of the vector-like SUSY-breaking terms entering the expansion (\ref{fexp}).  The corresponding contribution to $\calv_{\rm eff}$  can also be expressed in terms of $\Psi_1$ and $\Psi_2$ by using (\ref{ups}).

In this new form the effective potential $\calv_{\rm eff}$ depends explicitly on the warping, dilaton, $H$-fields and RR-fields $\tilde F$. As a consequence, the corresponding field equations  in pure spinor form can be directly derived by varying (\ref{finpot0}) with respect to these fields. On the other hand,  (\ref{finpot0}) depends only implicitly on the internal metric through the pure spinors $\Psi_1$ and $\Psi_2$, the volume form $\d \text{Vol}_6=(i/8)\langle\Psi_1,\bar\Psi_1\rangle=(i/8)\langle\Psi_2,\bar\Psi_2\rangle$ and the Hodge star operator $\tilde*_6$. Nevertheless, the variation of these objects under metric deformations can be easily described. Indeed, under a deformation $\delta g^{mn}$, the volume element gives $\delta\sqrt{\det g}=-(1/2)\delta g^{mn}g_{mn}\sqrt{\det g} $ as usual, the Hodge star gives
\be
\delta \langle \tilde*_6\chi_1,\chi_2\rangle\, =\, \delta g^{mn}\big[ \langle \tilde*_6\iota_m\chi_1,\iota_n\chi_2\rangle-\frac12g_{mn}\langle \tilde*_6\chi_1,\chi_2\rangle\big]
\ee
for any pair of polyforms $\chi_1$ and $\chi_2$, and finally the pure spinors $\Psi_1$ and $\Psi_2$ give 
\be
\delta\Psi_i\, =\, -\frac12\delta g^{mn}\, g_{k(m}\d y^k\wedge \iota_{n)}\Psi_i \quad \quad i=1,2
\ee
Thus, by using these simple rules, one can vary (\ref{finpot0}) inside (\ref{4daction}) with respect to the internal metric and straightforwardly obtain the internal Einstein equations in pure spinor form.\footnote{By that we mean a form where the derivatives act only on the pure spinors. In fact, the residual dependence on the metric contained in the Hodge star can be eliminated by decomposing the forms in the generalized Hodge decomposition defined by the pure spinors: a decomposition in $\pm i$-eigenspaces of $\tilde*_6$ (see, e.g., the Appendix A of \cite{lucapaul2}). }

As simple check, let us see how (\ref{finpot0}) behaves in the case of supersymmetric compactifications. If we consider compactifications to AdS$_4$, we can plug in the supersymmetric conditions (\ref{susygeneral}) into (\ref{finpot0}) and obtain from (\ref{4daction}) that the 4D cosmological constant $2\pi\calv_{\rm eff}/\caln$ is given by $-3|w_0|^2$, in agreement with the 10D result. If we now take $w_0 = 0$ and restrict to compactifications to flat space, then the extremization of $S_{\rm eff}$ amounts to that of $\calv_{\rm eff}$, and it is easy to see that $\caln=1$ backgrounds solve the equations of motion. Indeed, the first four terms in (\ref{finpot0}) are positive definite and vanish exactly when the configuration is supersymmetric (i.e. when (\ref{calgen}) is satisfied and the r.h.s. of (\ref{nonsusypure})  vanishes). Of the remaining terms, the first two are negative definite but still quadratic in the SUSY-breaking terms, while the last term can be seen to be the product of two quantities that vanish because of the gauge BPSness condition  and the Bianchi identity (\ref{intBI}) respectively. Thus, clearly any supersymmetric compactification to flat space extremizes the potential (\ref{finpot}). This provides an alternative proof that all supersymmetric flux compactifications (with calibrated sources) are solutions of equations of motion, already demonstrated by integrability arguments in \cite{kt}.

As a further check, let us again focus on compactifications to $\reals^{1,3}$ and analyze their modified external Einstein equations. As explained above, they can be obtained by considering the combined variation (\ref{warpdilvar}) of $\calv_{\rm eff}$. Now, it is easy to see that, if we vary (\ref{finpot0}) around a configuration that satisfies the gauge and string BPSness conditions (\ref{calgen}),  the terms containing $\Psi_2$ do not contribute at all. Thus, the gauge and string BPSness, the Bianchi identities and the calibration condition for the sources are sufficient to prove that the modified external Einstein equations are satisfied. This can also be verified directly as follows. First note that the external components of the modified Einstein equation (\ref{einstein2}) reduce to
\be\label{exteinstein}
\nabla^m(e^{-2\Phi} \nabla_m e^{4A})\, =\,  e^{4A}\tilde F\cdot \tilde F+e^{4A-\Phi}\!\!\!\!\!\!\!\!\sum_{i\in\text{loc. sources}}\!\!\!\!\tau_i\rho_i^{\rm loc}
\ee
Using the Bianchi identity (\ref{intBI}) we can rewrite this equation as
\bea\label{first}\nonumber
-\d(e^{-2\Phi}*_6 \d e^{4A})&=& \langle\tilde*_6 \tilde F, e^{4A}\tilde F\rangle- \langle\d_H\tilde*_6\tilde F, e^{4A-\Phi}\Re\Psi_1\rangle \\ && 
+e^{4A-\Phi}\!\!\!\!\!\!\!\!\sum_{i\in\text{loc. sources}}\!\!\!\!\tau_i\Big[\rho_i^{\rm loc}\d\text{Vol}_6-\langle \Re\Psi_1,j_i\rangle\Big]
\eea
In addition, using the string-BPSness/D-flatness condition (\ref{df}) one can prove that 
\be\label{remaining}
\d(e^{-2\Phi}*_6 \d e^{4A})\, =\, \d\langle\tilde*_6\d_H(e^{4A-\Phi}\Re\Psi_1), e^{-\Phi}\Re\Psi_1\rangle_5
\ee
and so the modified external Einstein equations can be restated in the form
\bea\label{second}
\langle e^{4A}\tilde F-\d_H(e^{4A-\Phi}\Re\Psi_1), F+\d_H\tilde*_6(e^{-\Phi}\Re\Psi_1)\rangle \cr
+\, e^{4A-\Phi}\!\!\!\!\!\!\!\!\sum_{i\in\text{loc.~sources}}\!\!\!\!\tau_i\Big[\rho_i^{\rm loc}\d\text{Vol}_6-\langle \Re\Psi_1,j_i\rangle\Big] & = &0
\eea
The term in the first line vanishes after imposing  (\ref{cal}) while the term in the second line vanishes if we assume the saturation of the bound (\ref{densbound}). Note  that in this derivation we have only assumed that we have a DWSB background, and not any specific subansatz.

%%%%%%%%%%%%%%%%%%%%%%%%%%%%%%

\subsection{Effective potential for DWSB backgrounds}
\label{effdwsb}

Let us now use $\calv_{\rm eff}$ to compute which additional constraints need to be imposed on the DWSB backgrounds of section \ref{1param} to promote them to DWSB vacua. One can immediately see that the only terms in (\ref{finpot0}) that can give a non-trivial contribution to these equations are
\bea\label{restrvdw}
\calv_{\rm eff} & \supset &
\frac12\int_{\calm_6} e^{-2A} \langle\tilde*_6
\big[\d_H(e^{3A-\Phi}\Psi_2)\big],\d_H(e^{3A-\Phi}\bar\Psi_2)\rangle\cr
&  & -\frac14\int_{\calm_6} e^{-2A}\Big\{\frac{|\langle \Psi_1,\d_H(e^{3A-\Phi}\Psi_2)\rangle |^2}{\d\text{Vol}_6}+\frac{|\langle \bar\Psi_1,\d_H(e^{3A-\Phi}\Psi_2)\rangle |^2}{\d\text{Vol}_6}\Big\}
\eea
since the rest are quadratic in quantities vanishing in the backgrounds under consideration. As follows from the discussion at the end of the previous subsection, the variation of (\ref{restrvdw}) with respect to the combination (\ref{warpdilvar}) vanishes on our configurations. The same is true for the variation of (\ref{restrvdw}) with respect to the dilaton $\Phi$, although for this one has to use the particular form (\ref{finansatz2}) of the DWSB condition. Thus, dilaton and external Einstein equations are satisfied in our class of backgrounds. 

On the other hand, by varying (\ref{restrvdw}) with respect to the $B$-field and taking (\ref{finansatz}) into account, we obtain the equation
\be\label{Bps}
\d\Big[ e^{4A-2\Phi}\langle \Im(r^*\Psi_2), 3\Re\Psi_1 +\frac12(-)^{|\Psi_2|}\Lambda^{mn}\gamma_m\Re\Psi_1\gamma_n \rangle_3  \Big]\, =\, 0
\ee
where the explicit $r$ dependence can be eliminated via $r=2(-)^{|\Psi_1|}\langle \Re\Psi_1,\d_H\Psi_2\rangle/\langle \Psi_1,\bar\Psi_1\rangle$. For example in the GKP case, one can easily see that (\ref{Bps}) is satisfied by using (\ref{GKPps}) and $\Lambda^{\rm(GKP)}=\bbone$.  Of course, in the supersymmetric case  (\ref{Bps}) is trivially satisfied since $r=0$.

Finally, one can obtain the non-trivial contributions to the internal Einstein equations by varying (\ref{restrvdw}) with respect to the internal metric, using the rules described in subsection \ref{fulleffpot}. Massaging the resulting equation and taking our restriction (\ref{finansatz}) into account, we arrive at the following equation\footnote{ To show (\ref{intmps}) one should take the following
identity into account: $\Omega^{(2)}_{mnp}=
-\Lambda^{s}{}_{m}\Lambda^{q}{}_{n}\Lambda^{r}{}_{p}\Omega^{(1)}_{sqr}$,
which is a consequence of (\ref{spinrel}) and (\ref{localor}).    }
\be\label{intmps}
\Im\Big\{\big\langle g_{k(m}\d y^k\wedge \iota_{n)}\Psi_2,\d_H\big[e^{A-\Phi}r^*\big(3\Re\Psi_1+\frac12 (-)^{|\Psi_2|}\Lambda^{kr}\gamma_k\Re\Psi_1\gamma_r\big)\big]\big\rangle\Big\}\, =\,0
\ee
Again, (\ref{intmps}) is satisfied in the supersymmetric case $r=0$. As a less trivial check, one can use again  (\ref{GKPps})  and $\Lambda^{\rm(GKP)}=\bbone$ to easily see that GKP vacua satisfy (\ref{intmps}) as well. Thus, our formalism reproduces the known fact that GKP configurations  solve the full set of equations of motion. 

Let us summarize our results. Provided that the Bianchi identities (\ref{intBI}) are satisfied,  the DWSB backgrounds described in section \ref{1param}, with calibrated localized sources, automatically solve the dilaton and  external Einstein equations. On the other hand, the $B$-field and internal Einstein equations reduce to (\ref{Bps}) and (\ref{intmps}) respectively. Thus, inside this DWSB subansatz, we need only impose  (\ref{Bps}) and (\ref{intmps}) to get a true vacuum.

Note  that both equations (\ref{Bps}) and (\ref{intmps}) can be unified in the following integrated condition
\be\label{bplusg}
\int_{\calm_6} e^{A-\Phi}\Im\big\{ r^*\big\langle \delta_{g,B}\big[\d_H(e^{3A-\Phi}\Psi_2)\big], 3\Re\Psi_1+\frac12(-)^{|\Psi_2|}\Lambda^{mn}\gamma_m\Re\Psi_1\gamma_n \big\rangle  \big\}\, =\, 0
\ee
for any deformation $\delta_{g,B}$ of internal metric and $B$-field. Since in our DWSB vacua we have
\be\label{dwid}
\big\langle \d_H(e^{3A-\Phi}\Psi_2), 3\Re\Psi_1+\frac12(-)^{|\Psi_2|}\Lambda^{mn}\gamma_m\Re\Psi_1\gamma_n \big\rangle\, \equiv\, 0
\ee
the condition (\ref{bplusg}) may be seen as a requirement of stability of (\ref{dwid}) under deformations of $\d_H(e^{3A-\Phi}\Psi_2)$. We will give a 4D interpretation of (\ref{dwid}) and (\ref{bplusg}) in the next section  (see the discussion following (\ref{falpha})). 

A very interesting property of this DWSB subansatz is that, by slightly constraining the fields entering (\ref{finpot0}), the effective potential becomes positive semi-definite.  In order to see this, one must first impose the Bianchi identities (\ref{intBI}), so that the last term in (\ref{finpot0}) vanishes. Second, the possible modified dilatino variations should be restricted so that $u^{1,2}_m=0$ identically. In other words we impose that, even off-shell, $\cals_i$ in (\ref{vari}) are only described by singlets under the $SU(3)\times SU(3)$ structure group. From (\ref{ups}), (\ref{psgen}) and (\ref{psexp}) we see that this condition is equivalent to excluding some vector-like components in the violation of string and DW BPSness (\ref{df}) and (\ref{nonsusypure}). 
Note that such a restriction affects only the NS-sector and involves vector-like modes under the $SU(3)\times SU(3)$ structure group, that are anyway expected  not to give rise to light - possibly tachyonic - zero-modes.\footnote{For example, in the $SU(3)$-structure GKP case with underlying Calabi-Yau geometry, vector-like zero-modes are excluded.}  Finally, let us consider a particular setting where the internal space is foliated by generalized leaves $(\Pi,R)$ of the kind described in section \ref{1param}, but without imposing any particular condition on them.  For this generalized foliation the allowed DWSB condition is given by
\be\label{psdwsb}
\d_H(e^{3A-\Phi}\Psi_2)\, =\, ir\tilde \jmath_{(\Pi,R)}
\ee
as in subsection \ref{gcsec}, where $r$ will eventually correspond to our SUSY-breaking parameter. We stress that $\tilde\jmath_{(\Pi,R)}$, while necessarily compatible with (\ref{psdwsb}), is otherwise a quite generic real pure spinor, associated with the generic foliation $(\Pi,R)$, and need not obey any particular calibration condition. In other words, $\tilde\jmath_{(\Pi,R)}$ is `off-shell'. For example, in the GKP case this restriction is much milder than the assumption that $\Omega_0$  in (\ref{alph2}) is closed, that is indeed assumed in the derivations of (warped) effective potentials in the literature \cite{gkp,giddings,stud07}.
 
Applying this `truncation' to the potential (\ref{finpot0}), we get 
\bea\label{finpot}
\calv^{\text{DWSB}}_{\rm eff}&=&\frac12\int_{\calm_6}\d\text{Vol}_6\,e^{4A}\big[\tilde F-e^{-4A}\d_H(e^{4A-\Phi}\Re\Psi_1)\big]^2\cr
&& +\, \frac12\int_{\calm_6}\d\text{Vol}_6\,\big[\d_H(e^{2A-\Phi}\Im\Psi_1)\big]^2\cr
&&+\, \frac12\int_{\calm_6} e^{-2A}|r|^2\Big[\langle\tilde*_6\tilde\jmath_{(\Pi,R)},\tilde\jmath_{(\Pi,R)}\rangle-\frac{|\langle \Psi_1,\tilde\jmath_{(\Pi,R)}\rangle |^2}{\d\text{Vol}_6}\Big]\cr
&&+\, \sum_{i\in \text{D-branes}}\tau_i\int_{\calm_6}e^{4A-2\Phi}\Big(\d\text{Vol}_6\,\rho^{\rm loc}_i-\langle \Re\Psi_1,j_i\rangle  \Big)
\eea
where we have used $\langle\tilde*_6\tilde\jmath_{(\Pi,R)},\tilde\jmath_{(\Pi,R)}\rangle=16\,e^{6A-2\Phi}\d \text{Vol}_6$. All of the four terms appearing in the above four lines are positive definite: the first two trivially; the last one because of (\ref{densbound}), and the third one because of an analogous local bound
\be\label{folbound}
\langle\tilde*_6\tilde\jmath_{(\Pi,R)},\tilde\jmath_{(\Pi,R)}\rangle\, \geq\, \frac{|\langle \Psi_1,\tilde\jmath_{(\Pi,R)}\rangle |^2}{\d\text{Vol}_6}
\ee
that again follows from an algebraic inequality similar to (\ref{locbound}).

Thus, upon some mild restrictions, we end up with a positive semi-definite warped potential (\ref{finpot}) that vanishes precisely for our foliated DWSB vacua. The vanishing of the first and second terms imposes the gauge and string-BPSness  conditions (\ref{calgen}), while the vanishing of the last term implies that all D-branes must be calibrated. Finally, the vanishing of the third term implies that the bound (\ref{folbound}) must be saturated and this, together with  (\ref{psdwsb}), is equivalent to the requirement that the fibers of the generalized foliation $(\Pi,R)$ must be  calibrated by $\omega^{\rm(sf)}$. Furthermore, we explicitly see how, once the generalized foliation $(\Pi,R)$ is chosen to be calibrated -- which, in the GKP case, is guaranteed by $\d\Omega_0 = 0$ -- the dependence of $\calv_{\rm eff}$ on the SUSY-breaking parameter $r$ disappears. This encodes the fundamental no-scale structure of these class of vacua, a point to which we will return in section \ref{sec:4dint}.

As a simple check, we can compute $\calv^{\text{DWSB}}_{\rm eff}$ in the GKP case, once we plug in the usual restrictions considered in, e.g.,  \cite{gkp,giddings,stud07}. Namely, let us impose from the beginning the relation between warping and RR five-form (\ref{warpgkp}), the underlying CY-structure and the fact that we have calibrated sources. The only surviving contribution in (\ref{finpot}) comes from the second line, that gives the following effective potential
\bea\label{gkppot}
\calv^{\rm GKP}_{\rm eff}&=&\frac12\int_{\calm_6} \d\text{Vol}_6\,e^{4A}[*_6F+e^{-\Phi}H]^2\cr
&=&\frac14\int_{\calm_6} \d\text{Vol}_6\,e^{4A}\big|(1+i*_6)\calg_3\big|^2\ .
\eea
Once translated into Einstein frame variables (where $A^{\rm E}=A-\Phi/4$), this indeed reproduces the warped potential found in \cite{giddings,stud07}.

%%%%%%%%%%%%%%%%%%%%%%%%%%%%%%
%%%%%%%%%%%%%%%%%%%%%%%%%%%%%%

\section{4D structure of DWSB vacua}\label{sec:4dint}

After constructing a stable set of (4+6)D vacua, the obvious question to address is which kind of effective 4D theories (if any) they give rise to. We have already mentioned several features of the DWSB vacua of section \ref{sec:vacua} that follow from their definition, like 4D Poincar\'e invariance, D-flatness and the absence of open string tachyons.\footnote{We will generalize the first two features in Section \ref{sec:ads4}.} We have, in addition, considered a subfamily of models where SUSY-breaking depends on a single parameter $r$, on which the scalar potential density $\calv_{\rm eff}$ does not depend at its minima. 

All this, of course, points to a  4D theory with pure F-term breaking and a no-scale structure,  as in the well-known case of \cite{gkp}. We would then like to recast our intuition in terms of ordinary 4D $\caln=1$ terminology, in order to answer basic questions such as what is the relation between the DWSB parameter $r$ and the 4D gravitino mass.
 
This is not such an easy task because, as mentioned before, in order to have a full 4D description one needs an appropriate prescription to truncate the theory to a finite number of 4D modes, and such a prescription is not always available. We can, nevertheless, use the formalism and results of \cite{lucapaul2} to describe our DWSB configurations in a simple 4D $\caln=1$ language, so that most of the features of the 4D effective theory become manifest. By exploiting such an idea we will be able to describe the 4D F-term structure of our DWSB vacua and, in particular, to rewrite the effective potential of the previous section as a 4D F-term potential. This will allow us, in turn, to unveil the no-scale structure of one-parameter DWSB vacua from a 4D viewpoint.

To proceed let us define as in \cite{lucapaul2} the rescaled pure spinors
\be
t\,:=\, e^{-\Phi}\Psi_1  \quad \quad \calz\,:=\, e^{3A-\Phi}\Psi_2
\label{rescpure}
\ee 
where $\calz$ was  already introduced in section \ref{gcsec}. In terms of these rescaled spinors the gauge and string BPSness conditions (\ref{calgen}) become
\bseq\label{susygent}
\begin{align}
\d_H\big(e^{4A}\Re t\big) &=\,  e^{4A}\tilde*_6 F \label{cal1}\\
\d_H\big(e^{2A}\Im t\big) &=\, 0 \label{cal2}
\end{align}
\eseq
while the DWSB condition for one-parameter backgrounds (\ref{finansatz2}) reads
\be\label{susygenz}
\d_H\calz\, =\, i\,r\,\tilde\jmath_{(\Pi,R)}
\ee
with $\tilde\jmath_{(\Pi,R)}$ defined as in (\ref{jtilde}). 

The importance of the definitions (\ref{rescpure}) is that $t$ and $\calz$ contain the full NS sector information of our supergravity background,\footnote{Strictly speaking this is not true, as $t$ and $\calz$ do not contain the $B$-field information. The full NS sector information is contained in the twisted version of $t$ and $\calz$, denoted $t^{\rm tw}$ and $\calz^{\rm tw}$. By twisting of a polyform we mean the operation $\omega\rightarrow e^{-B}\omega$, so that the $\d_H$ differential becomes the ordinary exterior derivative. Also, it is $\calz^{\rm tw}$ and the complexification of $\Re t^{\rm tw}$ that should be considered as 4D chiral fields, and not their untwisted counterparts. These untwisted spinors have however the nice property of being globally well-defined, so in the following we will express everything in terms of them. One may express everything in terms of $t^{\rm tw}$ and $\calz^{\rm tw}$ at any time by also replacing $\d_H$ by $\d$. We refer the reader to \cite{lucapaul2} for further details on the 4D relevance of twisted spinors.\label{fttwist}} and that one can consider $\calz$ as a chiral field of the associated  4D $\caln=1$ description, while a second chiral field $\calt$ contains the information about $t$ and the RR-sector. More explicitly, one can show that the independent degrees of freedom contained in $t$ can be described by $\Re t$ alone \cite{hitchin}. Then, one defines 
\be\label{calt}
\calt\, :=\, \Re t-iC
\ee
where $C$ is the RR gauge potential locally defined by $F=F^{\text{bg}}+\d_H C$, and $F^{\text{bg}}$ is some fixed background non-trivial flux. Thus, $\calt$ and $\calz$ are chiral fields describing the complete flux compactification. Finally, it is useful to define the complex polyform
\be\label{gform}
G\, :=\, F+i\d_H\Re t
\ee
that may be considered as the field strength of $\calt$.\footnote{The polyform $G$ in (\ref{gform}) is related to $\calg$ defined in (\ref{calg}) by $\calg=G+4i\d A\wedge \Re t$ and thus in  (\ref{denssup}) and (\ref{suppot}) below one may substitute $G$ with $\calg$. Furthermore, notice that $\calg$ and $G$ coincide in the approximation of constant warping.}  

As discussed in \cite{lucapaul2}, warped flux compactifications are naturally described by a 4D $\caln=1$ Weyl invariant formulation. The usual Einstein frame formulation can then be recovered by gauge fixing the Weyl symmetry, as we will briefly recall at the end of this section. Using the 4D formalism of \cite{toineconf},  one can then introduce a superpotential $\calw$ and conformal K\"ahler potential $\caln$ as functions of $\calz$ and $\calt$ \cite{lucapaul2}. In order to do so, one first defines the following superpotential  and  conformal K\"ahler potential  densities
\be\label{denssup}
W\, =\, \pi(-)^{|\calz|+1}\langle \calz,G\rangle\quad\quad N\, =\, \frac{i\pi}{2}\langle\calz,\bar\calz\rangle^{1/3}\langle t,\bar t\rangle^{2/3}
\ee
where, here and in the following, $t$ should be considered as a function of $\calt$. Then the superpotential and conformal K\"ahler potential are given by 
\be\label{suppot}
\calw \, =\, \int_{\calm_6} W\, =\, \pi (-)^{|\calz|+1}\int_{\calm_6} \langle \calz, G\rangle
\ee
and 
\be\label{kahler}
\caln\, =\, \int_{\calm_6} N\, =\, \frac{\pi i}{2}\int_{\calm_6} \langle t,\bar t\rangle^{2/3}\langle\calz,\bar\calz\rangle^{1/3}
\ee
matching the previous definition (\ref{ckp}). Note  that $\calw$ and $\caln$ can be seen as the `zero modes' of the densities $W$ and $N$, that can in turn be seen as defining a local $\caln=1$ structure pointwise in the internal space $\calm_6$.\footnote{This local  structure may be seen as a warped $\caln=1$ version of the local $\caln=2$ structure discussed in \cite{grana} for flux compactifications with approximatly constant warp-factor.} 

Given these definitions, let us consider the problem of relating our DWSB parameter $r$ and the gravitino mass.  We can do this directly, starting with the 10D gravitino $\psi_M=(\psi^{(1)}_M,\psi^{(2)}_M)^T$ and dilatino $\lambda=(\lambda^{(1)},\lambda^{(2)})^T$. Recall that in compactifications to four dimensions one can define the following combination with diagonal 4D kinetic term
\be\label{diag10d}
\tilde\psi_\mu\, :=\, \psi_\mu+\frac12\Gamma_\mu(\Gamma^m\psi_m-\lambda)
\ee 
where $\mu,\nu,\ldots$ denote 4D indices and $m,n,\dots$ are internal 6D indices. Then, using the underlying $SU(3)\times SU(3)$ structure, one can give a natural definition of the 4D gravitino density $\psi^{\rm 4D}_\mu$ as follows
\be\label{4Dgravitino}
\tilde\psi^{(1)}_\mu\,=\frac12\,\psi^{\rm 4D}_\mu\otimes \eta^*_1+\text{\ c.c.}\quad\quad \tilde\psi^{(2)}_\mu\,=\frac12\,\psi^{\rm 4D}_\mu\otimes \eta^*_2+\text{\ c.c.}
\ee
The associated 4D kinetic term is then given by
\be\label{lockin}
\frac{i}{2}\int_{\calm_6} N (\bar\psi^{\rm 4D}_\mu\hat\gamma^{\mu\nu\rho}\nabla_\nu\psi^{\rm 4D}_\rho)
\ee
and the standard 4D gravitino is given just by the zero mode $\psi^{{\rm 4D}}_{(0)\mu}$ of $\psi^{{\rm 4D}}_\mu$, that is constant in the internal space. Indeed, it has the correct kinetic term \cite{toineconf}
\be\label{4dgravi}
\frac{i}{2}\,\caln\, \bar\psi^{{\rm 4D}}_{(0)\mu}\hat\gamma^{\mu\nu\rho}\nabla_\nu\psi^{{\rm 4D}}_{(0)\rho}\
\ee
that is essentially defined by the conformal K\"ahler potential $\caln$. On the other hand, $\psi^{{\rm 4D}}_\mu$ contains information about higher order KK modes in the reduction of the 10D fermions and (\ref{lockin}) corresponds to the generalization of the 4D formula  (\ref{4dgravi}) that includes higher KK gravitini. 

From this, it follows that the 4D gravitino mass is given by  $m_{3/2}=\calw/\caln$ \cite{toineconf}. Analogously, we can define the gravitino mass density as
\be\label{grm1}
{\mathfrak m}_{3/2}\,=\, W/N
\ee
and so we can think of the ordinary gravitino mass $m_{3/2}$ as the average value of ${\mathfrak m}_{3/2}$ evaluated with respect to the warped volume: 
\be\label{gmi}
m_{3/2}\, =\, \frac{\int_{\calm_6} {\mathfrak m}_{3/2}\, e^{2A-2\Phi}\d\text{Vol}_6}{\int_{\calm_6} e^{2A-2\Phi}\d\text{Vol}_6}
\ee

In the particular case of one-parameter DWSB vacua, it is not difficult to show that the above definitions, together with (\ref{susygent}) and (\ref{susygenz}), lead to the  identification
\be\label{rmid}
{\mathfrak m}_{3/2}\, =\, -2e^A r
\ee
that, up to a warp factor, identifies the SUSY-breaking parameter $r$ with the gravitino mass density. One can reach the same conclusion in a different way, starting from the following 4D formula for the  SUSY transformation of the 4D gravitino density 
\be\label{4Dsusytransf}
\delta\psi^{\rm 4D}_\mu=\nabla_\mu\zeta^*+\frac{W}{2 N}\gamma_\mu\zeta\ 
\ee
that is obtained by extending the transformation for the ordinary 4D gravitino \cite{lucapaul2}. Then, by using (\ref{diag10d}), (\ref{4Dgravitino}) and the 10D gravitino and dilatino variations for DWSB vacua, one can directly compute $\delta\psi^{\rm 4D}_\mu$ in terms of $r$. Indeed, comparing the result with (\ref{4Dsusytransf}) and (\ref{grm1}), one gets the same identification (\ref{rmid}).

For a generic supersymmetric background, equations (\ref{susygent}) and (\ref{susygenz}) (with $r=0$), as well as their extension to AdS$_4$ compactifications (\ref{susygeneral}), can be understood as F- and D-flatness conditions for $\calw$ and $\caln$ \cite{lucapaul2}.\footnote{See also \cite{cassani1} for a similar analysis that uses  the ${\cal N}=2$ formalism of \cite{grana}, assuming approximately constant warping.} Hence, one may try to obtain a 4D interpretation of DWSB  by expressing (\ref{susygenz}) in 4D $\caln=1$ terms.  Since imposing (\ref{cal2}) amounts to requiring D-flatness in our vacuum \cite{lucapaul2}, the source of  the SUSY-breaking must derive from a non-vanishing F-term. In the following, we would like to understand more precisely such a pattern of F-term SUSY-breaking. 

Let us start by observing that, by using the D-flatness condition (\ref{cal2}), it is possible to rewrite (\ref{cal1}) as
\be\label{genselfd}
G_{(-1)}\,=\,0 \quad\text{and}\quad G_{(3)}\,=\,0
\ee
where  ${}_{(k)}$ denotes the $k$-th components in the the generalized Hodge decomposition $\Lambda^\bullet T^*_{\calm_6}=\bigoplus V_{(k)}$ induced by $\calz$ (see, e.g., appendix A of \cite{lucapaul2}). On the other hand eq.~(\ref{susygenz}), encoding the DWSB pattern, decomposes as follows under the generalized Hodge decomposition
\be\label{decdwb}
(\d_H\calz)_{(2)}\,=\,0\quad\quad (\d_H\calz)_{(0)}\,=\,i\,r\,\tilde\jmath_{(\Pi,R)}
\ee

The generic holomorphic variation of $\calz$ takes values in $V_{(3)}\oplus V_{(1)}$ while  the holomorphic fluctuations of $\calt$ take value in $V_{(0)}\oplus V_{(-2)}$ \cite{lucapaul2}. One can immediately see that the first conditions in (\ref{genselfd}) and (\ref{decdwb}) are equivalent to the vanishing of the F-terms $\cald \calw:=\partial \calw-3\calw\, \partial \log \caln$ associated with the holomorphic variations $\delta\calz_{(1)}$ and $\delta\calt_{(-2)}$ respectively  \cite{lucapaul2}, that is
\be\label{ft1}
\cald_{\calz_{(1)}} \calw\, =\, 0\quad\quad \cald_{\calt_{(-2)}} \calw\, =\, 0
\ee 

On the other hand, we can analogously consider the F-term densities $\cald W:=\partial W-3W\,\partial \log N$ associated with the densities $W$ and $N$.\footnote{The reader should not be confused here. We name $\cald W$ `F-term density' not because it is a density for the F-term $\cald \calw$, but because it is defined in terms of the densities $W$ and $N$.} Generically one has $\cald_{\calz_{(3)}} W\equiv 0$ and
\be\label{nonvt}
\cald_{\calt_{(0)}} W\, =\, \pi i [(-)^{|\calz|}(\d_H\calz)_{(0)}+2i{\mathfrak m}_{3/2} e^{2A}\Im t ]\ .
\ee
From (\ref{susygenz}) we then conclude that, if ${\mathfrak m}_{3/2}=-2e^Ar\neq 0$, then $\cald_{\calt_{(0)}} W \neq 0$. The reason is that $\tilde\jmath_{(\Pi,R)}$ is a pure spinor while $\Im t$  is a `stable' $SO(6,6)$ spinor \cite{hitchin} not pure by definition, and so the two terms in (\ref{nonvt}) cannot cancel each other out. More precisely, using (\ref{finansatz}) we get the following non-vanishing F-term density
\be\label{tft}
\cald_{\calt_{(0)}} W\, =\, -\frac12\pi e^{2A}{\mathfrak m}_{3/2}\Big[{3}\Im t+  \frac12 (-)^{|t|} \Lambda_{mn}\gamma^m\Im t\gamma^n\Big]\ .
\ee
Interestingly, this particular form for $\cald_{\calt_{(0)}} W$ automatically implies the second condition in (\ref{genselfd}), $G_{(3)}=0$. Indeed, let us introduce the fluctuation of $\calt$ with respect to the complex scalar field $\a$
\be\label{fluct}
\delta_\alpha\calt\, =\, \alpha\, \tilde*_6\tilde \jmath_{(\Pi,R)}
\ee
Then, the property $G_{(3)}=0$ is equivalent to 
\be\label{falpha}
\cald_{\alpha} W\, :=\, \langle\tilde*_6\tilde \jmath_{(\Pi,R)},\cald_{\calt_{(0)}} W \rangle\, =\, 0
\ee
that directly follows from (\ref{dwid}). On the other hand, if we denote by $\delta\calt^\perp$ the other fluctuations of $\calt$ orthogonal to $\delta_\alpha\calt$ with respect to the `pointwise' metric $\langle\tilde*_6 ~.~,~.~\rangle$, then generically $\cald_{\calt^\perp_{(0)}} W\neq 0$.

Note  that (\ref{falpha}) suggests an interpretation of the constraints (\ref{Bps}) and (\ref{intmps}) that  need to be imposed to have DWSB vacua. Indeed, the equivalent constraint (\ref{bplusg}) coming from the field equations can be written as
\be
\int_{\calm_6} e^{-2A}\Im\langle \tilde*_6 \delta_{g,B}\big(\d_H\bar \calz\big),\cald_{\calt_{(0)}} W \rangle\, =\, 0 
\ee
Recalling that the DWSB condition takes the form (\ref{susygenz}) for these backgrounds, the field equations could be seen as a  stability condition of the F-flatness (\ref{falpha}) under deformations of $\d_H\calz$. 

We have then concluded that, in one-parameter DWSB vacua, the supersymmetry-breaking is characterized by the non-vanishing F-term density $\cald_{\calt^\perp_{(0)}} W$. Let us however stress that in order to have $\caln=1$ vacua one has to require, in  addition to (\ref{ft1}),  that the F-terms $\cald_{\calt_{(0)}} \calw$ and $\cald_{\calz_{(3)}} \calw$ vanish \cite{lucapaul2}. Of course, in generic supersymmetric vacua (including compactifications to AdS$_4$) one has a constant ${\mathfrak m}_{3/2}\equiv m_{3/2}$ and thus $\cald_{\calt_{(0)}} \calw =\cald_{\calz_{(3)}} \calw =0$ implies $\cald_{\calt_{(0)}} W =0$ (while $\cald_{\calz_{(3)}} W \equiv 0$ identically). The converse is not true, and for example in our case $\cald_{\calt_{(0)}} W=0$ would not be enough to assure supersymmetry: the condition $\cald_{\calz_{(3)}} \calw = 0$ has to be considered in addition, and leads to the the condition $G_{(-3)}=0$

On the other hand, in the non-supersymmetric case, if $\cald_{\calt_{(0)}} W \neq 0$, then we clearly have $\cald_{\calt_{(0)}} \calw \neq 0$, but also $\cald_{\calz_{(3)}} \calw \neq 0$ even if $\cald_{\calz_{(3)}} W \equiv 0$. Indeed, assuming $\cald_{\calt_{(0)}} W\neq 0$ we obtain
\be
\cald_{\calz_{(3)}} \calw \, =\, \frac{\pi i}{2}\,  ({\mathfrak m}_{3/2}-m_{3/2})\, e^{-4A}\bar\calz\ ,
\ee
 and thus $\cald_{\calz_{(3)}} \calw=0$ only if ${\mathfrak m}_{3/2}=const.$, which is generically not the case. For instance, in the warped CY case one gets
\be
\partial(e^{-4A}{\mathfrak m}_{3/2} )\, =\, 0
\ee
and thus a constant ${\mathfrak m}_{3/2}$ would require a constant warping, which cannot happen  if the fluxes are non-vanishing. If we nevertheless take the constant warping/dilaton approximation, as often assumed in the literature, we get  $\cald_{\calz_{(3)}} \calw \simeq 0$ and we are thus led to identify the warping as the origin of the non-vanishing F-term $\cald_{\calz_{(3)}} \calw$. This is in agreement with the interpretation of $\delta\calz_{(3)}$ as describing fluctuations of the warping \cite{lucapaul2}.   

To summarize, we have shown that in order to get the conditions (\ref{susygent})-(\ref{susygenz})  in the 4D formulation of \cite{lucapaul2}, we need to impose D-flatness (that corresponds to (\ref{cal2})), partial F-flatness (\ref{ft1}), and that SUSY-breaking is described by the non-vanishing F-term density (\ref{tft}). The latter in turn implies that the F-term  $\cald_{\calt_{(0)}} \calw$ is non-vanishing, as well as $\cald_{\calz_{(3)}} \calw$ that does not vanish due to warping effects. Since in the constant warp-factor approximation $\calt$ and $\calz$ may be identified with the chiral fields of hyper- and vector-multiplets of an  underlying $\caln=2$ structure \cite{grana,grimm}, we see that one-parameter DWSB may be seen as a quaternionic SUSY-breaking (using the terminology of \cite{cg07}), up to `warping mediation' to the vector multiplet sector.\footnote{When using this $\caln=2$ terminology, we are being inevitably sloppy. As already mentioned, it can really be used only in the constant warp-factor approximation. For example, the $\delta\calz_{(3)}$ fluctuation is unphysical in that limit \cite{grana} and does not really correspond to any vector multiplet.}

We can now show that the potential (\ref{finpot}) can be understood as a no-scale potential in the case of our DWSB vacua. In order to do so, we have to restrict from the beginning to SUSY-breaking configurations that satisfy (\ref{psdwsb}) with a calibrated associated foliation $(\Pi,R)$. This implies that the superpotential (\ref{suppot}) depends only on the fluctuation $\delta_\alpha\calt$ defined in (\ref{fluct}) and on the generic $\calz$, while it does not depend on $\delta\calt^\perp$. We also impose the D-flatness condition (\ref{cal2}). Then, the potential (\ref{finpot}) takes the following form
\be\label{poteffns}
\calv^{\rm no-scale}_{\rm eff}=\int_{\calm_6} \d\text{Vol}_6\, e^{4A}\Big( |G_{(-1)}|^2+|G_{(3)}|^2\Big)
\ee
It is not difficult to see that this potential has indeed the features of no-scale superpotentials. Namely, it is positive semi-definite and is the sum of the following F-term (densities) associated with the chiral fields $\delta\calz$ and $\delta_\alpha \calt$ that the superpotential density depends on:
\be\label{vanft}
\cald_\calz W \sim G_{(-1)}\quad\quad \cald_{\alpha} W \sim G_{(3)}
\ee

It is useful to keep in mind the GKP subcase as an example, whose no-scale structure was discussed e.g. in \cite{gkp,giddings}. In this case, the restriction on calibrated foliations $(\Pi,R)$ boils down to requiring that the associated $(3,0)$-form $\Omega_0$ be closed and thus the $H$-field is the only source of the SUSY-breaking. Then, the above fluctuation (\ref{fluct}) that affects the superpotential corresponds to the axion-dilaton fluctuations and indeed the usual Gukov-Vafa-Witten superpotential \cite{gvw} (which is what (\ref{suppot}) reduces to by the usual CY truncation) depends on it. The D-flatness corresponds to restricting to K\"ahler spaces and primitive $H$-fields and both conditions are automatic for warped Calabi-Yau's. Then the F-terms (\ref{vanft}) entering the potential (\ref{gkppot}) in its form (\ref{poteffns})  correspond to the F-terms associated with the complex structure moduli and the axion-dilaton. 

Finally, let us end this section by briefly recalling how to obtain the more usual Einstein frame formulation \cite{toineconf,lucapaul2}. One must first isolate the compensator $Y$ by the rescaling 
\be
\calz\quad\rightarrow\quad Y^3 \calz\ ,
\ee
that corresponds to the complexification of the rescaling $e^{A}\rightarrow |Y|e^A$ of the warping. Then the Einstein frame quantities $\calw_{\rm E}$ and $\calk_{\rm E}$ are defined by
\be
\calw \quad\rightarrow\quad\frac{Y^3}{M^3_{\rm P}}\calw_{\rm E}\quad\quad\quad\quad \caln\quad\rightarrow\quad |Y|^2e^{-\calk_{\rm E}/3}\ ,
\ee
with analogous expressions for the densities $W_{\rm E}$ and $e^{-K_{\rm E}/3}$. One can then go to the Einstein frame by gauge-fixing $Y\equiv M_{\rm P}e^{\calk_{\rm E}/6}$. For example, in this frame the gravitino mass takes the usual form
\be\label{grm3}
m_{3/2}\, =\, \frac{1}{M^2_{\rm P}}\,e^{\calk_{\rm E}/2}\calw_{\rm E}\ .
\ee
while the gravitino mass density reads
\be\label{grm2}
{\mathfrak m}_{3/2}=\frac{1}{M^2_{\rm P}}\,e^{\calk_{\rm E}/2}e^{(K_{\rm E}-\calk_{\rm E})/3} W_{\rm E}\ .
\ee 

%%%%%%%%%%%%%%%%%%%%%%%%%%%%%%
%%%%%%%%%%%%%%%%%%%%%%%%%%%%%%

\section{Some subcases in detail}\label{sec:subcases}

So far our discussion has remained at a rather general, even abstract, level. It is then reasonable to wonder to which set of supergravity vacua it applies and, in particular, if the landscape of DWSB vacua is non-trivial. In fact, most of our results apply to the subfamily of backgrounds constructed in section \ref{1param},  where we restricted the DWSB ansatz to depend on a single complex parameter. As we have shown, this subfamily contains a non-trivial set of non-SUSY flux vacua, since it contains the GKP construction. One may still wonder, however, if  it really includes any interesting class of $\caln=0$ vacua beyond the GKP case.

The purpose of this section is to show that this is indeed the case and, in addition, to give a flavor of what DWSB vacua look like. Again, we will focus on the DWSB backgrounds of section \ref{1param}, 
which can always be related to $\caln=1$ backgrounds by taking the SUSY-breaking parameter $r \rightarrow 0$. From the discussion of section \ref{1param}, we are led  to consider backgrounds involving fibered\footnote{In this section we assume that the foliations  characterizing the one-parameter DWSB vacua of section \ref{1param} are fibrations of the internal space.} internal spaces of the form   
\be\label{fibrationpi}
\Pi\quad \hookrightarrow\quad \calm_6\quad\rightarrow\quad {\cal B}
\ee
with a base $\calb$ and  fibers $\Pi$. Also, from section \ref{sec:4dint} we know that for these backgrounds to be vacua the fibers need to be calibrated by $\omega^{\rm (sf)}$. In particular, this implies that in some cases  this fibration must be supplemented by a two-form $R$ on the fibers $\Pi$, such that $\d R=H|_\Pi$, which together with $\Pi$ specifies a generalized fibration $(\Pi, R)$ that specifies our DWSB ansatz.

So, in the following, we will spell out several classes of backgrounds inside the DWSB subansatz of section \ref{1param}, giving explicit formul\ae{} relating fluxes, geometrical structures and SUSY-breaking parameters, and leaving the construction of explicit examples for the next section. As we will see, some of these backgrounds have been already identified in \cite{cg07} as no-scale $\caln = 0$ vacua via the use of effective K\"ahler and superpotentials derived in the GCG literature.\footnote{We will not reproduce all of the constructions of \cite{cg07}, since several of them do not fit into the DWSB ansatz of Section \ref{sec:vacua}. Indeed, for some cases in \cite{cg07} tachyonic adjoint fields are found in the D-brane worldvolumes, something that is forbidden from the very beginning by our gauge BPSness condition (\ref{cal}).} Here we will revisit these constructions from a full 10D viewpoint, generalizing them and giving the additional conditions needed to satisfy the 10D equations of motion. Finally, we will also construct a novel set of $\caln=0$ vacua not considered before. In particular, we will consider backgrounds where the generalized foliation $(\Pi, R)$ contains a non-trivial two-form $R$. While these kind of backgrounds (dubbed magnetized backgrounds henceforth) are not very common in the literature, we will show that simple examples of them can be obtained via a $\beta$-deformation \cite{lm,alessandrobeta,quiverbeta} of usual GKP vacua. Further examples can also be obtained via non-geometric backgrounds, whose analysis is relegated to Appendix \ref{ap:ng}.

%%%%%%%%%%%%%%%%%%%%%%%%%%%%%%

\subsection{$SU(3)$-structure backgrounds}
\label{subcases}

Let us first consider backgrounds whose structure group is $SU(3)$. As recalled in appendix \ref{4dstructure}, these are defined by the fact that the internal 6D spinors $\eta_1$ and $\eta_2$ are parallel. Then the $SU(3)$ structure can be characterized by the pair $(J,\Omega)$ defined in (\ref{jomega}), where $J$ is a K\"ahler\footnote{By K\"ahler form we mean the (1,1)-form associated with the almost complex structure  $J^m{}_n$ as in (\ref{jomega}), and which may or may not be closed. While we denote both objects (two-form and complex structure) by the same symbol, it should be clear to which one we refer from the context.} and $\Omega$ is a nowhere vanishing $(3,0)$ form. Locally, we can choose a vielbein $e^a$  such that
\bea\label{su(3)vielbein}
J&=& -(e^1\wedge e^4+e^2\wedge e^5+e^3\wedge e^6)\cr
\Omega&=& (e^1+ie^4)\wedge(e^2+ie^5)\wedge (e^3+ie^6)
\eea

These $SU(3)$-structure backgrounds are morally closer to Calabi-Yau vacua in the sense that there is (at least formally) a weak flux limit where the deviation from a Calabi-Yau metric is small compared to the KK scale. A useful way to describe such deviation is given by the five torsion classes $W_i$, defined as
\bea\label{tc}
\d J &=&-\frac32\Im(\overline{W}_1\Omega)+W_4\wedge J+W_3\cr
\d\Omega &=& W_1 J\wedge J+W_2\wedge J+\overline{W}_5\wedge \Omega
\eea
where $W_1$ is a complex scalar, $W_2$ is $(1,1)$ and primitive, $W_3$ is real $(2,1)+(1,2)$ and primitive, $W_4$ is a real one-form, $W_5$ is a $(1,0)$-form. So, as long as any of these torsion classes does not vanish we will be away from the CY case $\d J = \d \Om = 0$.

Following this general philosophy, we would like to investigate $\caln=1 \raw \caln =0$ deformations of $SU(3)$-structure backgrounds that fit into our DWSB ansatz, and see how the torsion classes depend on the SUSY-breaking parameter $r$. This will simplify the comparison of our approach with the one taken in \cite{cg07}, where similar $SU(3)$-structure $\caln=0$ backgrounds were considered and the discussion was presented from the point of view of torsion classes (see also \cite{lawrence}). In order to do a more direct comparison, we will restrict to type IIB backgrounds with calibrated D5-branes and type IIA backgrounds with calibrated D6-branes.

%%%%%%%%%%%%%%%%%%%%%%%%%%%%%%

\subsubsection{IIB vacua with aligned D5-branes}\label{IIBD5}

Type IIB $SU(3)$-structure backgrounds have the following pure spinors (see appendix \ref{4dstructure})
\be\label{su3}
\Psi_1\,=\,e^{i\theta}e^{iJ}\quad \quad \quad \quad \quad\Psi_2\,=\,e^{-i\theta}\Omega
\ee
Let us now consider backgrounds that admit BPS (i.e., calibrated) D5-branes wrapped on a two-cycle $\Sigma$ with a worldvolume flux $\calf=0$. Then, certain restrictions should be applied to $\Psi_1$ and $\Psi_2$. Indeed, following the arguments of \cite{luca,Martucci06} one may see that such a D5-brane may develop F and/or D-terms in its 4D gauge theory if non-BPS. F-flatness requires that $\Sigma$ be almost-complex with respect to the almost-complex structure defined by $\Omega$, while the D-flatness condition fixes
\be\label{d5theta}
\theta_{\text{D5}}\, =\, -\pi/2 
\ee
on top of $\Sigma$. To make contact with the setup of \cite{cg07}, we will also require that $\theta$ be constant. In that case, the background conditions for string and gauge BPSness (\ref{calgen}) translate to
\bea
H=F_{1}=F_{5}=0\quad&\quad e^{2A-\Phi}=\text{const.}\quad& \quad \d J\wedge J=0\cr
& *_6 F_{3}=-e^{-2\Phi}\d (e^\Phi J)&
\label{DWSBD5}
\eea
Note  that the last condition is equivalent to the ISD condition (\ref{3isd}) since using (\ref{calg}), (\ref{d5theta}) and the second condition in (\ref{DWSBD5}) we can write  $\calg_3= F_{3}+ie^{-2\Phi}\d(e^{\Phi} J)$. Thus
\be
 \calg_3= F_{3}+ie^{-2\Phi}\d(e^{\Phi} J)\quad\quad \text{is ISD}
\label{ISDD5}
\ee
which can be decomposed as\footnote{Recall that the ISD condition on a three-form implies that it only has pieces $(0,3)$, $(2,1)_{\text{prim}}$ and $(1,2)$ of the form $\eta^{0,1}\wedge J$.} 
\be
\begin{array}{lcl}
\,[F_{3}+i\d(e^{-\Phi} J)]^{3,0}\,=\,0 & & [F_{3}+i\d(e^{-\Phi} J)]^{1,2}\,=\,0\\
\,[F_{3}+e^{-2\Phi}i\d(e^{\Phi} J)]^{2,1}\quad  \text{is} & \text{primitive}
\end{array}
\ee
This is the analogue of the ISD condition eq.(\ref{isd}) in the GKP case, as discussed in general in subsection \ref{extension}.

Let us now induce supersymmetry breaking via our DWSB ansatz of section \ref{1param}, while still preserving the $SU(3)$-structure above. We need an internal space fibered as in (\ref{fibrationpi}), with calibrated generalized fibers $(\Pi,R)$ that, because of (\ref{su3}) and (\ref{typeconst}), must satisfy $\text{codim}(\Pi)\geq 4$. Thus the only possibility is that the fibers $\Pi$ are almost-complex two-cycles (i.e. such that $\Omega|_\Pi=0$) with $R=0$.  Note  that, from this general approach, one obtains the setting that was assumed as the starting point in \cite{cg07}. 

Since $\Pi$ is almost complex, in the basis entering (\ref{su(3)vielbein}) we can choose $e_1$ and $e_4$ as tangent to $\Pi$. As a consequence, the K\"ahler form $J$ splits as follows
\be\label{splitJ}
J\, =\, J_{\calb_4}+J_{\Pi_2}
\ee
with $J_{\calb_4}=-(e^2\wedge e^5+e^3\wedge e^6)$ and $J_{\Pi_2}=-e^1\wedge e^4$. Then the DWSB ansatz (\ref{finansatz2}) takes the form
\be\label{tors}
\d(e^{A}\Omega)\, =\, -2e^Ar\,J_{\calb_4} \wedge J_{\calb_4}
\ee
It is not difficult then to identify the source of SUSY-breaking with the non-vanishing $(0,3)$ component of $F_{3}+i\d(e^{-\Phi} J)$. This is clear from the general formula (\ref{rmid}) relating $r$ to the gravitino mass density  (\ref{grm1}), that in this case is given by
\be
{\mathfrak m}_{3/2}\, =\, -2e^A r\, =\, 2e^{A+\Phi} \frac{\Omega\wedge \left(F_{3}+i\d(e^{-\Phi} J)\right)}{\Omega\wedge\bar\Omega}
\ee

Finally, we have to impose the field equations (\ref{Bps}) and (\ref{intmps}) to get a true   vacuum. The $B$-field equation  (\ref{Bps}) is automatically satisfied while the internal Einstein equation  (\ref{intmps}) reduces to\footnote{Here and in the following examples with $R=0$, it is useful to note that  $3\Re\Psi_1-\frac12\Lambda^{mn}\gamma_m\Re\Psi_1\gamma_n=4(\Re\Psi_1-\d
\text{Vol}_\Pi)$ when $R=0$, as follows from (\ref{finansatz}) and (\ref{finansatz2}). }

\be\label{d5ein}
\Re\langle g_{k(m}\d y^k\wedge \iota_{n)}\Omega, \d(e^{-A}r^* J_{\calb_4})\rangle\, =\, 0
\ee
Since $g_{k(m}\d y^k\wedge \iota_{n)}\Omega$ is either a $(3,0)$ or a primitive $(2,1)$ form, it is sufficient to impose
\be\label{d5einstein}
[\d(e^{-A}r J_{\calb_4})]^{3,0}=[\d(e^{-A}r J_{\calb_4})]_{\rm prim}^{2,1}=0
\ee
to guarantee (\ref{d5ein}). For example, if $A$ and $r$ are constant along the fiber $\Pi$ (as is typical when localized sources are wrapped along the fibers) and  $\d J_{\calb_4}=\d f\wedge J_{\calb_4}$ for some real function defined on the base $\calb_4$, then the conditions (\ref{d5einstein}) amount to $[\d (e^{-A}r f)]^{1,0}=0$. We will construct simple examples in the next section where the latter is satisfied.

Let us now write the above conditions in terms of the torsion classes defined in (\ref{tc}). First of all, from (\ref{tors}) we have a direct relation between our susy-breaking parameter $r$ and the first torsion class:
\be
W_1=-\frac23 r
\ee
In addition, we have
\bea
W_2\,=\,2W_1(J_\calb-2J_\Pi)\quad &&\quad W_3\,=\,-ie^\Phi (F_{3})^{2,1}_{\rm prim}+\text{ c.c.}\quad\cr
\quad W_4\,=\,0\quad &&\quad W_5\,=\,-(\d A)^{1,0}
\eea
Furthermore, there are some additional relations coming from the BPSness conditions (\ref{calgen}) that we repeat here for completeness
\be
\d(2A-\Phi)\,=\,0\quad\quad H\,=\,F_{1}\,=\,F_{5}=0\quad \quad (F_{3})^{2,1}_{\rm non-prim}\,=\,-i e^{-\Phi}(\d\Phi)^{1,0}\wedge J
\ee

All the above conditions are in perfect agreement with the conditions proposed in \cite{cg07} for $SU(3)$ vacua with D5/D9-branes. In \cite{cg07} this set of relations were first obtained in the constant warp-factor/dilaton approximation, and later completed to include non-trivial dilaton and warp-factor. In this second step, the prescription was imposed that any non-singlet condition under the $SU(3)$-structure group should remain unchanged with respect to the supersymmetric case. It is quite satisfactory to rederive the same set of conditions from our 10D approach. Note, however, that we also find an extra condition, namely (\ref{d5ein}), that must be satisfied by any true vacuum. It would be interesting to see which backgrounds  this extra condition excludes as true vacua.

%%%%%%%%%%%%%%%%%%%%%%%%%%%%%%

\subsubsection{IIA vacua with aligned D6-branes}\label{IIAD6}

We now turn to IIA backgrounds with $SU(3)$ structure and D6-branes. The pure spinors have the following form 
\be\label{d6ps}
\Psi_1\,=\,\Omega\quad\quad\Psi_2\,=\,e^{-i\theta}e^{iJ}\ ,
\ee
where we have reabsorbed in $\Omega$ the phase appearing in (\ref{su3ps}). The BPSness conditions (\ref{calgen}) now become
\bea\label{d6cal}
&F_{4\, }=F_{6}=0\quad\quad \d(e^{2A-\Phi}\Im\Omega)\,=\,0\quad\quad H\wedge \Im\Omega\,=\,0&\cr
&*_6F_{0}\,=\,e^{-\Phi}H\wedge \Re\Omega\quad\quad*_6 F_{2}\,=\,-e^{-4A}\d(e^{4A-\Phi}\Re\Omega)\ .&
\eea

Following the same argument as for D5-branes, if we want to have BPS D6-branes, they should satisfy the  F-flatness and D-flatness conditions derived in \cite{luca,Martucci06}. F-flatness now implies that the internal three-cycle $\Sigma$ wrapped by the D6-brane should be Lagrangian with respect to $J$ (i.e. $J|_\Sigma=0$) and that $\calf=0$, while the D-flatness requires that $\Im\Omega|_\Sigma=0$. 

Let us then assume that the fibration (\ref{fibrationpi}) characterizing the SUSY-breaking has special Lagrangian three-cycles as fibers\footnote{Another possibility would be given by coisotropic five-dimensional fibers} and thus  $R=0$ by the above arguments. In the vielbein basis introduced in  (\ref{su(3)vielbein}) we can choose $e_1,e_2,e_3$ as tangent to the fibers $\Pi_3$. The resulting pure spinor expression (\ref{finansatz}) is then given by
\be\label{d6sb}
\d_H(e^{3A-\Phi-i\theta}e^{iJ})\, =\, 4ire^{3A-\Phi}\Im\Omega_{\calb_3}
\ee
where $\Im\Omega_{\calb_3}=-e^4\wedge e^5\wedge e^6$. This equation in turn implies that $\theta$ and $e^{3A-\Phi}$  are constant and thus the following relations
\be\label{d62}
H+i\d J\, =\, 4ie^{i\theta}r\,\Im\Omega_{\calb_3}\quad\quad J\wedge\d J=0\quad\quad J\wedge H\,=\,0 
\ee 
must be satisfied. Note  that the first condition in (\ref{d62}) implies the other two, as well as the condition $H\wedge \Im\Omega=0$ in (\ref{d6cal}). In this case the gravitino mass can be written as
\be
{\mathfrak m}_{3/2}\, =\, -2e^A r\, =\, 3i\, e^{A-i\theta}\,\frac{(H+i\d J)\wedge \Re\Omega}{J\wedge J\wedge J}
\ee

To summarize, for IIA backgrounds with aligned D6-branes wrapping $\Pi_3$ we must impose the following relations
\bea\label{d6fin}
&F_{4}\,=\,F_{6}\,=\,0\quad\quad\quad e^{3A-\Phi}\,=\,\text{const.}\quad\quad\quad \d(e^{-A}\Im\Omega)\,=\,0&\cr &e^{\Phi}*_6 F_{2}\,=\,-e^{-A}\d(e^{A}\Re\Omega)\quad\quad\quad e^{\Phi} F_{0}\,=\,-4\Im (e^{i\theta}r)&\cr 
&H\,=\,  -4\Im(e^{i\theta} r)\,\Im\Omega_{\calb_3}\quad\quad\quad \d J\,=\,4\Re(e^{i\theta} r)\,\Im\Omega_{\calb_3}&
\eea
in addition to which we must consider the constraints coming from the equations of motion (\ref{Bps}) and (\ref{intmps}). In the present context they  reduce to
\be\label{d6eom}
\d\big[e^{-2A}r\,(\Re\Omega-\Re\Omega_{\Pi_3})\big]\,=\,0
\ee
where $\Re\Omega_{\Pi_3}=e^1\wedge e^2\wedge e^3$.

Let us now translate the above conditions in terms of torsion classes. Again, we have a direct relation between the first torsion class and $r$:
\be
W_1\, =\, -\frac23\,\Re (e^{i\theta}r)
\ee 
while the other torsion classes are given by
\bea
&W_2\,=\,e^\Phi(F_{2})^{1,1}_{\rm prim}\quad\quad\quad W_3\,=\,-\frac32 W_1(4\Im \Omega_{\calb_3}-\Im\Omega)\ ,&\cr
&W_4\,=\,0\quad\quad\quad W_5\,=\,(\d A)^{1,0}\ .
\eea
The remaining conditions not encoded in the torsion classes are 
\bea
&F_{4}=F_{6}\,=\,0\quad\quad\quad \d(3A-\Phi)\,=\,0\quad\quad\quad \Im(e^{i\theta}r)\,=\,-\frac14 e^{\Phi}F_{0}&\cr 
&H\,=\,-\frac{1}{6W_1}e^\Phi F_{0}\d J\quad\quad\quad e^\Phi F_{2}\,=\,2\d A\cdot \Im\Omega-2W_1 J+ W_2&
\eea
as well as the equation of motion (\ref{d6eom}). Again, one can check that such conditions reproduce the ones found in \cite{cg07} via a 4D approach and some educated guessing, while the relation (\ref{d6eom}) is new.

%%%%%%%%%%%%%%%%%%%%%%%%%%%%%%

\subsection{IIB vacua with static $SU(2)$-structure}
\label{staticsec}

Let us now consider classes of backgrounds beyond the ones in \cite{cg07}, but to which our 10D approach applies equally well. We will first consider static $SU(2)$-structure backgrounds, defined by the condition that the two internal spinors $\eta_1$ and $\eta_2$ are everywhere  orthogonal. In the case of IIB, this means that we can introduce a one-form $\theta=\theta_m\d y^m$ such that
\be
\eta_2\,=\,-\frac{i}{2}\theta_m\gamma^m\eta^*_1
\ee
Note  that, by definition,
\be
\theta^m\bar\theta_m\,=\,2\quad\quad\quad (1+iJ_1)^n{}_m\theta_n\,=\,(1+iJ_2)^n{}_m\theta_n=0
\ee
We can now split the two $SU(3)$-structures defined by $\eta_1$ and $\eta_2$ in components tangent and orthogonal to $\theta$ as follows (see e.g. \cite{mpz})
\be
\left\{\begin{array}{l} J_1\,=\,-\frac{i}{2}\theta\wedge\bar\theta +\mathpzc{j}\\ 
J_2\,=\,-\frac{i}{2}\theta\wedge\bar\theta -\mathpzc{j}\end{array}\right.
\quad\quad\quad
\left\{\begin{array}{l} \Omega_1\,=\,-\theta\wedge\mathpzc{w}\\ 
\Omega_2\,=\,\theta\wedge\bar{\mathpzc{w}}\end{array}\right. 
\ee
with $\iota_\theta \mathpzc{j}=\iota_{\bar\theta} \mathpzc{j}=\iota_\theta \mathpzc{w}=\iota_{\bar\theta} \mathpzc{w}=0$.  The resulting pure spinors have the form
\bea
\Psi_1&=& \mathpzc{w}\wedge e^{\frac12\theta\wedge\bar\theta} \cr 
\Psi_2&=&\theta\wedge e^{i\mathpzc{j}}
\eea

Let us now consider a D5-brane wrapping a two-cycle $\Sigma$. Its BPS conditions read \cite{Martucci06}
\be\label{su2d5}
\begin{array}{lcl}
\theta|_\Sigma=0\quad\quad\quad \mathpzc{j}|_\Sigma=0\quad\quad\quad \calf=0\quad&\quad&\quad{\rm F-flatness}\\
\Im\mathpzc{w}|_\Sigma=0 \quad&\quad&\quad{\rm D-flatness}
\end{array}
\ee
Again, if we choose the two-dimensional fibers $(\Pi,R)$ in the generalized fibration associated with the  SUSY-breaking, they have to obey the same conditions. Thus $R=0$ and the right-hand side of (\ref{finansatz2}) is a four-form, implying that
\be\label{hol1form}
\d(e^{3A-\Phi}\theta)\,=\,0
\ee
This means that we can locally introduce a complex coordinate $z$ such that $\d z=e^{3A-\Phi}\theta$. The hypersurfaces $D$ defined by $z=\text{constant}$ then admit an $SU(2)$ structure defined by the pair $(\mathpzc{j}|_D,\mathpzc{w}|_D)$. Finally, (\ref{su2d5}) reduces to the statement that calibrated two-cycles are SLag cycles inside the leaves $D$, and hence  the fibers $\Pi$ in the fibration (\ref{fibrationpi}) define a SLag fibration of the leaves $D$. From (\ref{finansatz2}), this also implies that
\be\label{jH}
\d \mathpzc{j}|_D=0\quad\quad\quad H|_D=0
\ee
so that $\mathpzc{j}|_D$ is a symplectic form on the leaves $D$.

We can select a local basis $e_a$ such that $e_1,e_2,e_4,e_5$ are tangent to $D$ and 
\bea
\mathpzc{j}&=&-(e^1\wedge e^4+e^2\wedge e^5)\cr
\mathpzc{w}&=&(e^1+ie^4)\wedge (e^2+ie^5)\cr
\theta&=&e^3+ie^6\
\eea
Furthermore we can take $e_1$ and $e_2$ tangent to $\Pi$. With choice of basis, let us decompose $\d \mathpzc{j}$ and $ H$ as follows
\be\label{decjH}
\d \mathpzc{j}\,=\,(f\wedge \bar\theta+\,\text{c.c.}) + \frac{i}2 u\wedge\theta\wedge \bar\theta\quad\quad\quad H\,=\,(g\wedge \bar\theta+\,\text{c.c.})+
\frac{i}2 h\wedge\theta\wedge \bar\theta
\ee
with $f,g$ complex two-forms and $u,h$ real one-forms that can be expanded in the basis  $e^1,e^2,e^4,e^5$. Then (\ref{finansatz2})  reduces to the condition
\be\label{sbsu2}
g+if\,=\,2r\, e^4\wedge e^5
\ee
that identifies  the origin of SUSY-breaking in the component of $H+i\d \mathpzc{j}$ along $e^4 \wedge e^5 \wedge \bar\theta$.  

Further constraints come form the string BPSness condition (\ref{df}), that  gives
\be\label{su2half}
\d(e^{2A-\Phi}\Im\mathpzc{w})\,=\,0\quad \quad\quad h\wedge \Im \mathpzc{w}\,=\,e^{4A-\Phi}[\d(e^{\Phi-4A}\Re\mathpzc{w})]_D
\ee
where ${}_D$ indicates that, in the expansion in the vielbein basis $e^a$, we pick up only terms containing $e^1,e^2,e^4,e^5$. The first condition in (\ref{su2half}) may be rephrased by saying that the $SU(2)$ structure $(\mathpzc{j}|_D,\mathpzc{w}|_D)$ on $D$ is `half-flat'. In addition, from the gauge BPSness condition (\ref{cal}) we get
\bea\label{su2RR}
&F_{5}\,=\,0\quad\quad\quad *_6F_{3}\,=\,-e^{-4A}\d(e^{4A-\Phi} \Re\mathpzc{w})&\cr
&e^{\Phi}*_6 F_{1}\,=\,H\wedge \Re\mathpzc{w}-i(2\d A-\d\Phi)\wedge\Im\mathpzc{w}\wedge \theta\wedge\bar\theta &
\eea
The second condition can be restated  as
\bea
\calg_3=F_3+ie^{-4A}\d(e^{4A-\Phi} \Re\mathpzc{w})\hspace{1cm}\text{is ISD}
\eea
as discussed in general in subsection \ref{extension}.

It remains to discuss the possible constraints coming from the equations of motion (\ref{Bps}) and (\ref{intmps}). For both, we need first to compute
\be
3\Re\Psi_1-\frac12\Lambda^{mn}\gamma_m\Re\Psi_1\gamma_n\, =\, -4 e^4\wedge e^5+2i\theta\wedge \bar\theta\wedge(e^1\wedge e^5+e^4\wedge e^2)
\ee
Then, the $B$-field equations (\ref{Bps}) reduce to
\be
\Im\langle\theta,\d(e^{A-\Phi}r^*\, e^4\wedge e^5)\rangle_4\,=\,0
\ee
This is solved if for example we impose that
\be\label{simpres}
\d(e^{A-\Phi}r^*\, e^4\wedge e^5)\,=\,0
\ee
Note  that if we assume that (\ref{simpres}) is indeed satisfied, the internal Einstein equations (\ref{intmps}) get simplified too, reducing to the following conditions
\be\label{einsu(2)}
\begin{array}{ccc}
e_1(e^{-7A+2\Phi}r)\,=\,e_2(e^{-7A+2\Phi}r)\,=\,0 &\quad \quad & e_4(e^{-7A+2\Phi}r)\,=\,-e^{-7A+2\Phi}r\,\iota_{e_1}h \\
g\wedge e^4\wedge e^5\,=\,0 & \quad \quad& e_5(e^{-7A+2\Phi}r)\,=\,-e^{-7A+2\Phi}r\,\iota_{e_2}h 
\end{array}
\ee
The more general case where (\ref{simpres}) is not satisfied can be worked out straightforwardly and for simplicity we refrain from discussing it explicitly.

%%%%%%%%%%%%%%%%%%%%%%%%%%%%%%

\subsection{Magnetized vacua from $\beta$-deformations}
\label{msb}

So far we have considered DWSB vacua described by generalized fibrations $(\Pi,R)$ that are  ordinary ones, with  no `magnetic flux' $R$ supplementing (\ref{fibrationpi}). The purpose of this subsection is to describe a simple way to construct magnetized DWSB vacua.

Indeed, suppose that we have a GKP vacuum that admits a $U(1)\times U(1)$ isometry group, and thus defines a ${\bf T}^2$ fibration of the internal space $\calm_6$. Then, we can obtain a new (dual) background by performing a $\beta$-deformation, which is a particular transformation of the $SO(2,2)$ extended T-duality group associated with the $U(1)\times U(1)$ symmetry. Recently, this trick was used in \cite{lm} to find the supergravity background dual to the field theory $\beta$-deformation \cite{beta} of the $\caln=4$ SYM theory. In \cite{alessandrobeta} it was realized that generalized geometry provides a very natural way to describe $\beta$-deformations and \cite{quiverbeta} followed this approach to find new classes of backgrounds dual to $\beta$-deformed $\caln=1$ superconformal quiver gauge theories. Here we will apply the procedure described in \cite{quiverbeta} to compact non-SUSY vacua. 

To proceed, let us call $y^1,y^2$ the coordinates along ${\bf T}^2$ (with $y^{1,2}\simeq y^{1,2}+1$) and $y^3,\ldots,y^6$ the remaining coordinates. Then a $\beta$ deformation can be described as a $O(6,6)$ T-duality  transformation, acting on the $H$-twisted extension bundle $E$ defined in (\ref{extbundle}) (which is locally isomorphic to $T_M\oplus T^*_M$), of the form
\be
\calo_\beta\,=\,\left( \begin{array}{cc} \bbone & \beta \\ 0 & \bbone \end{array}\right)
\ee
where $\beta=\gamma\, \partial_{y^1}\wedge\partial_{y^2}$, with $\gamma$ constant. The action on the twisted pure spinors defined in footnote \ref{fttwist} and (\ref{pstduality}) is given by
\be
\calo_\beta\cdot\,=\, e^\beta\cdot=1+\beta\cdot\quad\quad\text{with }\quad \quad \beta\cdot\, :=\,\gamma \,\iota_{y^1}\iota_{y^2}
\ee
By assumption, the background fields and the pure spinors are invariant under the $U(1)\times U(1)$ group describing ${\bf T}^2$ and thus it is easy to see that $\calo_\beta\cdot$ commutes with the exterior derivative $\d$ appearing in the twisted version of (\ref{calgen}) and (\ref{finansatz2}). This explicitly shows that, starting with a GKP vacuum (that has $\tilde\jmath_{\rm GKP}\sim \d y^1\wedge\ldots\wedge \d y^6$),  after a $\beta$-deformation we get a DWSB vacuum with
\be
\tilde \jmath^{\rm tw}_{({\bf T}^2,R)}\, \sim\, (\gamma-\d y^1\wedge\d y^2\wedge)\d y^3\wedge\ldots\wedge \d y^6
\ee
Clearly, $\tilde \jmath_{({\bf T}^2,R)}=e^{-B}\wedge\tilde \jmath^{\rm tw}_{({\bf T}^2,R)}$ describes a generalized fibration with non-zero $R$, up to non-generic cancellations generated by the $B$-field twist. By the very same argument, one can also see that the new $\Psi_2$ contains a point-dependent one-form and thus generically this background corresponds to a non-static $SU(2)$-structure.

Let us be more specific and consider a simplified setting, where the original GKP vacuum has constant dilaton $e^\Phi\equiv g_s$ and the internal space has factorized structure $\calm_6 \simeq {\bf T}^2\times D$ with metric
\be
g^{\rm GKP}\,=\,e^{-2A}\hat g\,=\,e^{-2A} (\hat g_{\bf T^2}+\hat g_D)
\ee
where $\hat g=\hat g_{\bf T^2}+\hat g_D$ is the unwarped CY metric. Furthermore, we assume that $B^{\rm GKP}$ has no legs along $y^1,y^2$.\footnote{See \cite{lm,quiverbeta} for examples with more complicated geometries.} Then, by smearing the background D3-branes and O3-plane along ${\bf T}^2$ and performing the $\beta$-deformation one gets a NS background
\bea\label{betaNS}
\d s^2 &=&e^{-2A} \left( g_s^{-2} e^{2\Phi} \d s^2_{\bf T^2}+ \d s^2_D\right) \cr
B&=&B^{\rm GKP}+\Delta B\, := \,B^{\rm GKP}- \g \, g_s^{-2} e^{2\Phi-4A} \, \det \hat g_{\bf T^2}\, \d y^1\wedge \d y^2 \cr
e^\Phi&=&g_s(1+\gamma^2 e^{-4A}\det \hat g_{\bf T^2})^{-1/2}
\eea
as well as RR background fields given by
\bea\label{betaRR}
F\, =\, e^{-\Delta B}\wedge (e^\beta\cdot F^{\rm GKP})
\eea
This implies that the ISD 3-form of this background is
\be
\calg_3\, =\, F_3 + i e^{-\Phi} \Re \left(e^{i\a} \left[H - i e^{-2A}\d(e^{2A} J) \right] \right)
\ee
where $J = g_s^{-2}e^{-2A} (e^{2\Phi} J_{\T^2} + J_D)$ and
\be
\text{cos}\, \a\, =\, g_s^{-1} e^{\Phi}\quad \quad \text{and} \quad \quad \text{sin}\, \a \, = \, g_s^{-1} e^{\Phi} \g \,e^{-2A} (\text{det} \hat{g}_{\T^2})^{1/2}
\ee
so that $\calg_3$ interpolates between the expressions $\calg_3 = F_3 + i e^{-\Phi}H$ and (\ref{ISDD5}).\footnote{Such interpolating ISD form also appears in the supergravity description of non-geometric backgrounds \cite{ms07}, which as discussed in Appendix \ref{ap:ng} are another example of magnetized backgrounds, and in interpolating $SU(3)$ solutions in the sense of \cite{fg03}.}

The pure spinors $e^{-\Phi}\Psi_{1,2}$ and $\tilde \jmath_{(\bf{T}^2,R)}$ also transform as the RR fields in (\ref{betaRR}) and so comparing the end result with the DWSB ansatz (\ref{finansatz2}) one gets
\be
R\,=\, \frac{1}{\gamma} g_s^{-2} e^{2\Phi}\,\d y^1\wedge \d y^2\, =\, - \g^{-2} e^{4A} \D B
\ee
The $\beta$-transform does, of course, also act on D-branes. Instead of the GKP D3-brane we will have D5-branes wrapping ${\bf T}^2$ at different points in $D$. We can split the gauge invariant world-volume field-strength into $\calf=B|_{{\bf T}^2}+{\rm F}/2\pi$, where F is a pure $U(1)$ field-strength. As in \cite{quiverbeta}, the  precise value of F generated by the $\beta$-deformation can be easily computed to be
\be
{\rm F}=\frac{2\pi}{\gamma}\,\d y^1\wedge \d y^2
\ee
If $\gamma=m/n$, the $\beta$-deformation maps $n$ coinciding D3-branes to $m$ coinciding D5-branes on which F is quantized (see e.g. \cite{quiverbeta}). However, it is important to observe that $\calf=0$ on the resulting D5-branes since on them $e^A\rightarrow 0$ and so $B|_{{\bf T}^2}\rightarrow -(1/\gamma)\d y^1\wedge \d y^2$. Hence these D5-branes are unmagnetized and no D3-brane charge is induced on them. The same conclusion can be reached by directly looking at the calibration condition for D5-branes wrapped on $\T^2$.

%This is consistent with the fact that the same $\calf$ should be  present on orientifolds too and they do not generically allow this to be non-trivial. The same conclusion is reached by directly looking at the calibration condition for the D5-branes and O5-planes and indeed we consistently have $R\rightarrow 0$ on the sources. Thus, since only $\calf$ enters the equations of motion, we can consider a generic $\gamma$ and for example set F$=0$ by turning on a Wilson line that shifts the $B$-field to $B^{\rm new}=B^{\rm old}+(1/\gamma)\d y^1\wedge \d y^2$. However, $\gamma$ will be anyway generically constraint by the quantization of the RR-fluxes, as shown explicitly in the next section.

%%%%%%%%%%%%%%%%%%%%%%%%%%%%%%
%%%%%%%%%%%%%%%%%%%%%%%%%%%%%%

\section{Simple examples}\label{sec:examples}

Let us now provide explicit constructions of the DWSB subcases that were analyzed in the previous section, in order to illustrate the physics of DWSB vacua in a more concrete way.  As our ansatz involves the fibration (\ref{fibrationpi}), the natural laboratory to build such vacua are twisted tori and 
 other toroidal-like geometries, which is the kind of geometric backgrounds that we will consider here. This will allow us to further compare the results of our 10D approach with those of \cite{cg07} where detailed soft term computations were performed for the particular case of twisted tori. 
 
 Note  that constructing explicit backgrounds not only allows to materialize the general ideas and computations carried above, but also to push the calculation of quantities of phenomenological interest, like the spectrum of D-brane soft terms that we will discuss in the next section. Again, since such computations will be carried out for twisted tori and close relatives, it will be straightforward to compare with the soft term results of \cite{cg07}.

%%%%%%%%%%%%%%%%%%%%%%%%%%%%%%

\subsection{Type IIB on $SU(3)$-structure twisted tori}\label{ex1}

To construct an explicit type IIB $SU(3)$-structure compactification let us, following \cite{Schulz04}, first consider the following NSNS background\footnote{In this section we recover the $2\pi \sqrt{\a'}$ factors neglected in the rest of the paper.}
\bseq\label{twD5}
\begin{align}
\label{ex1fullmetric}
\d s^2 & =  e^{2A} \d s^2_{\reals^{1,3}} +  \d s^2_{\calm_6}  \\
\label{ex1metric}
\d s^2_{\calm_6} & =  \a' (2\pi)^2 \Big\{e^{2A} \big[ {R_1^2}(\eta^1)^2 +  R_4^2(\eta^4)^2 \big]  + e^{-2A} \sum_{j=2,3,5,6} R_j^2 (\d y^j)^2\Big\} \\
\label{ex1H}
H & =  0
\end{align}
\eseq
where the warp factor $A$ only depends on $\calb_4 = \{y^2, y^3, y^5, y^6\}$ and $\eta^1$,  $\eta^4$ are non-closed one-forms satisfying 
\be
\d\eta^a\, =\, \oh f^a_{jk}\, \d y^j \wedge \d y^k, \quad a = 1,4,\quad j,k = 2,3,5,6
\label{ex1twisteda}
\ee
and so, in the limit of constant warp factor, $\calm_6$ is nothing but a twisted six-torus. In terms of the $SU(3)$-structure backgrounds discussed in section \ref{IIBD5} we have that the $SU(3)$-invariant forms are
\bseq\label{JOtwD5}
\begin{align}
\label{ex1J}
J & =  -\a'(2\pi)^2 \left[ e^{2A} R_1R_4\, \eta^1 \wedge \eta^4  + e^{-2A}\left(R_2R_5\, \d y^2 \wedge \d y^5 + R_3R_6\, \d y^3 \wedge \d y^6\right) \right] \\
\Om & =  \a'^{3/2} (2\pi)^3 e^{-A} \left(R_1\eta^1 + i R_4\eta^4 \right) \wedge \left( R_2\d y^2 + i R_5 \d y^5\right) \wedge \left( R_3 \d y^3 + i R_6 \d y^6\right) 
\label{ex1Om}
\end{align}
\eseq

In order to ensure that  $\calm_6$ is compact, we will demand that the structure constants $f^a_{ij}$ be integer constants. Besides that, in principle they can take any discrete value. Similarly, the six radii $R_a,R_i$ can take any real value and the dilaton $\Phi$ may be an arbitrary function.\footnote{We could also consider a richer NS ansatz by giving a more complicated metric to ${\calb_4}$, by turning on a closed $B$-field, etc. but we are ignoring such possibilities for the sake of simplicity.} Imposing D-string and gauge BPSness, namely eqs.(\ref{DWSBD5}), implies that
\bseq
\begin{align}
\label{ex1rel1}
e^{\Phi-2A} & \equiv  g_s \, = \, const.\\
\label{ex1rel3}
\d J \wedge J & =  0 \quad \Raw \quad f^a_{25} R_3R_6 + f^a_{36} R_2R_5\, =\, 0, \quad a =1,4
\end{align}
\eseq
as well as
\be
g_s *_{\calm_6} F_3 \, = \,\a' (2\pi)^2 R_1R_4 \left[ 4\d A \wedge \eta^1 \wedge \eta^4 +  \d \left(\eta^1 \wedge \eta^4\right) \right]
\label{ex1rel2}
\ee
which can also be seen as constraints arising from the equations of motion \cite{Schulz04}. This last condition can be rewritten as
\be
g_s\, F_3 \, = \,  *_{\calb_4} \d e^{-4A}  - \a' (2\pi)^2   \left(R_1^2\, \eta^1 \wedge *_{\calb_4} \d \eta^1 + R_4^2\, \eta^4 \wedge *_{\calb_4} \d \eta^4 \right)
\label{ex1rel2b}
\ee
or
\be
F_3 \, = \,  g_s^{-1} *_{\calb_4} \d e^{-4A}  - \a' (2\pi)^2   \tilde{g}_s^{-1} \left(\frac{R_1}{R_4} \, \eta^1 \wedge *_{\calb_4} \d \eta^1 + \frac{R_4}{R_1}\, \eta^4 \wedge *_{\calb_4} \d \eta^4 \right)
\label{ex1rel2c}
\ee
where we have defined $\tilde{g}_s = g_s / R_1R_4$ as the vev of the 4D dilaton field, as in \cite{Schulz04}. Here, $*_{\calb_4}$ stands for Hodge duality in the unwarped coordinates $\{y^2, y^5, y^3, y^6\}$.%\footnote{Note  that $\calb_4$ is a set of coordinates that form a non-closed 4-chain, and hence they are not really part of a $\T^4$. For most of our considerations below, however, we can still think of them as such.} 

The first component of (\ref{ex1rel2b}) arises from the backreaction of D5-branes and O5-planes transverse to $\calb_4$, and vanishes in the limit of constant warp factor. On the other hand, the second terms survives this limit, and should be thought as a properly quantized background flux. Indeed, one should be able to write (\ref{ex1rel2b}) as
\be
F_3 \, = \, g_s^{-1} *_{\calb_4} \d e^{-4A}  - \a' (2\pi)^2 \left(\eta^1 \wedge F^1_{\calb_4} + \eta^4 \wedge F^4_{\calb_4} \right)\, \equiv\,  g_s^{-1}*_{\calb_4} \d e^{-4A} + F_3^{\text{bg}}
\label{ex1rel2d}
\ee
where $F^a_{\calb_4}$ are primitive, integer-valued two-forms of $\calb_4$, just like $\d\eta^a$. In practice, the fact that all these forms are integer-valued and the equality between (\ref{ex1rel2c}) and (\ref{ex1rel2d}) will select particular values for the radii $R_i$ and $g_s$. In some cases, this process can be described via an effective potential lifting these would-be moduli. For instance, note that if we further impose the constraint
\be
\d\left(e^A\Om\right) \, =\, 0
\label{ex1relDW}
\ee
then $\d\eta^a$ must be $(1,1)$ primitive forms of $\calb_4$. This implies that they are integer anti-self-dual forms of $\calb_4$. This indeed fixes the $\calb_4$ complex structure moduli, since not every $\calb_4$ admits ASD integer 2-forms. In addition, since then $F^1_{\calb_4} = - \tilde{g}_s^{-1} R_1/R_4\, \d\eta^1$ and $F^4_{\calb_4} = - \tilde{g}_s^{-1} R_4/R_1\, \d\eta^4$, this puts constraints on $R_1/R_4$ and on $\tilde{g}_s$.

However, from the discussion in section \ref{IIBD5} (see also \cite{Schulz04}) we know that (\ref{ex1relDW}) is only imposed by supersymmetry, and that it will not be satisfied for DWSB vacua. Indeed, in general we have
\be
\d\left(e^A\Om\right) \, =\, \a'^{3/2} (2\pi)^3 \left[ (\d\eta^1)^{SD} + i (\d\eta^4)^{SD}\right] \wedge \Om_{\calb_4} 
\label{ex1relDWSB}
\ee
where $SD$ selects the self-dual component of $\d\eta^a$ and 
\be
\Om_{\calb_4} \, = \, \left( R_2\d y^2 + i R_5 \d y^5\right) \wedge \left( R_3 \d y^3 + i R_6 \d y^6\right) 
\ee
One can rewrite this expression more explicitly by defining
\be
J_{\calb_4}\, =\, -\a'(2\pi)^2 e^{-2A}\left(R_2R_5\, \d y^2 \wedge \d y^5 + R_3R_6\, \d y^3 \wedge \d y^6\right)
\ee
using the splitting $J = J_{\calb_4} + J_{\Pi_2}$ in (\ref{splitJ}). Then
\be
\d\left(e^A\Om \right)  = \frac{e^{4A}}{2\pi\sqrt{\a'}} \left[ \frac{f^1_{23} + i f^4_{23}}{R_2R_3} -   \frac{f^1_{56} + i f^4_{56}}{R_5R_6} + i \left( \frac{f^1_{26} + i f^4_{26}}{R_2R_6} -  \frac{f^1_{53} + i f^4_{53}}{R_5R_3} \right)  \right] \oh J_{\calb_4} \wedge J_{\calb_4} 
\label{ex1relDWSBb}
\ee
from which one can extract the value of $r$ in (\ref{tors}): $e^{3A}$ times a complex constant. This in turn implies that $\d ( e^{-A} r J_{\calb_4}) = 0$, so that (\ref{d5einstein}) is automatically satisfied and no background relations beyond (\ref{DWSBD5}) and (\ref{tors}) arise from the equations of motion. Finally, the only non-trivial Bianchi identity is that of $F_3$, which reads
\be
g_s\, \d F_3 \, = \, -  \nabla^2_{\T^4} e^{-4A} \d{\rm vol}_{\T^4}  - \a' (2\pi)^2  \left(R_1^2\, \d\eta^1 \wedge *_{\T^4} \d\eta^1 + R_4^2\, \d\eta^4 \wedge *_{\T^4} \d\eta^4 \right)
\label{ex1BIF3}
\ee

Let us now compute the dilatino and gravitino variations for this background. First, let us note that one can rewrite $F_3^{\text{bg}}$ as
\be
e^{\Phi}\left(F_3 - F_3^{\text{bg}} \right)\, =\, 2  *_{\calm_6} (\d\Phi \wedge J_{\Pi_2})
\label{ex1defbg}
\ee
which will be a useful expression when computing the dilatino and gravitino operators. Indeed, the dilatino operator reduces to
\be
\label{ex1dilatino}
\calo\, =\, \slashed{\p}\Phi - \oh e^{\Phi} \slashed{F}_3\sig_1\, =\, - \oh e^{\Phi} \slashed{F}_3^{\text{bg}} \sig_1 + 2\slashed{\p}\Phi P_-^{\Pi_2}
\ee
where we have used (\ref{ex1defbg}) and defined the fermionic projectors\footnote{Such kind of projectors arise in the fermionic action of D-branes, as discussed in next section. In particular, note that the projection condition $P_-^{\Pi_2} \eps = 0$ amounts to
\be\nonumber
\eps_1\, =\, i \slashed{J}_{\Pi_2} \g_{(6)} \eps_2\, =\, i \G^{41} \g_{(6)} \eps_2  \, =\, \G_{O5,\Pi_2}\, \eps_2\,
\label{ex1projb}
\ee
where $\G_{O5,\Pi_2}$ is the action of an O5-plane wrapping $\Pi_2$ on bulk spinors.}
\be
P_\pm^{\Pi_2}\, =\, \oh \left(1 \pm i\slashed{J}_{\Pi_2} \g_{(6)} \sig_1 \right) 
\label{ex1proj}
\ee
Hence,  if one considers a background spinor  $\eps$ projected out by $P_-^{\Pi_2}$, then the corresponding dilatino variation will only depend on $F_3^{\text{bg}}$. If in addition one defines the antisymmetrized geometric flux
\be
f_{mnp}\, =\, 3\,  g_{a[m}f^a_{np]}
\label{ex1metricf}
\ee
and uses the identity  $e^{\Phi} \slashed{F}_3^{\text{bg}} = i \slashed{f} \slashed{J}_{\Pi_2} \g_{(6)} = i \slashed{J}_{\Pi_2} \g_{(6)}  \slashed{f}$, one has
\be
\calo\eps \, = \,  - \oh e^{\Phi} \slashed{F}_3^{\text{bg}} \sig_1 \eps + 2\slashed{\p}\Phi P_-^{\Pi_2} \eps\, =\,  - \oh \slashed{f} \eps
\label{ex1dilatinob}
\ee
Similarly, one can compute the external gravitino variation to be
\be
\cald_\mu \, =\, \p_\mu + \frac{1}{4} \G_\mu \slashed{\p} \Phi  - \frac{1}{8}  \G_\mu e^{\Phi} \slashed{F}_3 \sig_1\, =\, \p_\mu - \frac{1}{8}\G_{\mu} e^\Phi \slashed{F}_3^{\text{bg}} \sig_1+ \frac{1}{2} \G_\mu \slashed{\p}\Phi P_-^{\Pi_2}
\label{ex1gravext}
\ee
and the internal gravitino variation as 
\be
\cald_m \, =\, \nabla_m  + \frac{1}{8} e^{\Phi} \slashed{F}_3 \G_m \sig_1
\label{ex1gravint}
\ee
where
\be
\nabla_{m} \, = \, \p_{m} + \frac{1}{2} \Lambda_m{}^{n} \left(\slashed{\p}A \G_{n}  -  \p_nA + \oh \slashed{f}_{n} \right)
\label{ex1spinc}
\ee
and
\be
\Lambda_{mn}\, =\, g_{mn} -2 e^{{a}}_m e_{{a}n},\quad \quad a \in \Pi_2
\label{ex1Lam}
\ee
where $a \in \Pi_2$ means that the flat index $a$ only runs over the $\Pi_2$ fiber coordinates. Hence one finds an internal gravitino variation of the form
\bseq
\begin{align}
\label{ex1gravintf}
\cald_{a}  & =  \p_{a}  - \frac{1}{8} \G_{a} \slashed{f}  - \frac{1}{4} \left(\slashed{f} +  2\slashed{\p}\Phi \right)\G_{a} P_-^{\Pi_2} \quad \quad a \in \Pi_2 \\
\label{ex1gravintb}
\cald_{j}  & =   \left( \p_{j} - \frac{1}{4} \p_{j} \Phi\right) + \frac{1}{8} \Lam_j{}^m\G_m  \slashed{f}  + \frac{1}{4} \left( \slashed{f}+ 2\slashed{\p}\Phi \right) \Lam_j{}^m\G_m  P_-^{\Pi_2} \quad \quad j \in \calb_4
\end{align}
\eseq
where we have separated the internal index $m$ into a fiber index $a$ and a base index $j$.

To summarize, if we consider a spinor $\eps = e^{A/2} \eps'$ such that $\eps'$ is constant and satisfies $P_-^{\Pi_2} \eps' = 0$, then the supersymmetry variations amount to
\bseq
\begin{align}
\calo\eps & =  - \oh \slashed{f} \eps \\
\cald_\mu\eps & =   - \frac{1}{8}\G_{\mu} \slashed{f} \eps \quad \quad \quad \quad \quad \quad \quad \quad  \mu \in \reals^{1,3} \\
\cald_{a} \eps & =   - \frac{1}{8} \G_{a} \slashed{f}  \eps \, =\, \frac{1}{8} \Lam_{a}{}^n \G_n \slashed{f} \quad \quad \ a \in \Pi_2 \\
\cald_{j} \eps & =  \frac{1}{8} \Lam_{j}{}^n \G_n \slashed{f} \quad \quad \quad \quad \quad \quad \quad \ \ j \in \calb_4
\end{align}
\eseq

Comparing with the general DWSB ansatz (\ref{calpres}), we have that
\be
\calv_1 \, = \, - \frac{1}{4} \slashed{f} \eta_1 \quad \quad \text{and} \quad \quad\calu^i_m \, =\, \frac{1}{8} \Lam_{mn} \G^n \slashed{f} \eta_i
\label{ex1susyparam}
\ee
In addition, the modified dilatino variation now reads
\be
\Delta\eps\, =\, \left(\G^M\cald_M - \calo\right)\eps\, =\, \frac{1}{8} \Lambda_{mn} \G^m \G^n\, \slashed{f} \eps\, =\, \frac{1}{4} \slashed{f} \eps
\label{ex1dilat}
\ee
And so we recover the ansatz (\ref{restr}), with
\be
r\eta_i^*\, =\, -\frac{1}{4} \slashed{f}^{(0,3)}\eta_i
\label{ex1r}
\ee
in agreement with (\ref{ex1relDWSBb}).

%%%%%%%%%%%%%%%%%%%%%%%%%%%%%%

\subsection{Type IIA on $SU(3)$-structure twisted tori}\label{ex2}

Let us now turn to type IIA background with the following NSNS ansatz
\bseq
\begin{align}
\label{ex2fullmetric}
\d s^2 & =  e^{2A} \d s^2_{\reals^{1,3}} +  \d s^2_{\calm_6} \\
\label{ex2metric}
\d s^2_{\calm_6} & =  \a' (2\pi)^2 \Big[e^{2A} \sum_{a=1,2,3} {R_a^2}(\eta^a)^2  + e^{-2A} \sum_{j=4,5,6} R_j^2 (\d y^j)^2\Big] \\
\label{ex2H}
H & =  \a' (2\pi)^2 \Big(N \d y^4\wedge \d y^5 \wedge \d y^6 + \sum_{a=1,2,3} B_a\, \d\eta^a \wedge \d y^{a+3} \Big)
\end{align}
\eseq
where $N \in \zet$, $B_a \in \reals$, and the warp factor now depends on $\{y^4, y^5, y^6\}$. The one-forms $\eta^1$, $\eta^2$, $\eta^3$ now satisfy (\ref{ex1twisteda}) for $a =1,2,3$ and  $j,k = 4,5,6$ and so again, in the limit of constant warp factor, we recover a twisted six-torus. In terms of the notation of section \ref{IIAD6}, the $SU(3)$-invariant forms are
\bseq
\begin{align}
\label{ex2J}
J & =  -\a'(2\pi)^2 \left[ R_1R_4\, \eta^1 \wedge \d y^4  + R_2R_5\, \eta^2 \wedge \d y^5 + R_3R_6\, \eta^3 \wedge \d y^6 \right] \\ 
\Om & =  \a'^{3/2} (2\pi)^3  \left(e^A R_1\eta^1 + i e^{-A} R_4 \d y^4 \right) \wedge \nonumber\\
 &\qquad\qquad\left( e^{A} R_2 \eta^2 + i e^{-A} R_5 \d y^5\right) \wedge \left( e^{A} R_3 \eta^3 + i e^{-A} R_6 \d y^6\right) 
\label{ex2Om}
\end{align}
\eseq
where $J$ can be complexified as
\be
J_c\, =\, - \a' (2\pi)^2\, i \sum_{a=1,2,3} T_a\, \eta^a \wedge \d y^{a+3}, \quad \quad T_a = R_aR_{a+3} + i B_a
\label{ex2Jc}
\ee
and the $T_a$ encode the light K\"ahler-like moduli of the compactification (see e.g. \cite{camara2005}). Note  that, from this point of view, $H = H^{\text{bg}} + \d\Re J_c$, where $H^{\text{bg}}  = \a' (2\pi)^2 \, N \d y^4\wedge \d y^5 \wedge \d y^6$ is some `background' component of $H$. From the point of view of the supergravity such splitting is somewhat arbitrary, and so the physical quantity can only be $H^{\text{bg}} + \d J_c$.

Imposing string and gauge BPSness via (\ref{d6cal}) restricts the rest of the background as
\bseq
\begin{align}
\label{ex2rel1}
e^{\Phi-3A} & \equiv  g_s \, = \, const.\\ \label{ex2rel3}
\d (e^{-A} \Im \Om) & =  0 \quad \Raw \quad
\left\{
\begin{array}{l}
f^2_{45} R_2R_6 + f^3_{64} R_3R_5\, =\, 0 \\ 
f^3_{56} R_3R_4 + f^1_{45} R_1R_6 \, =\, 0 \\ 
f^1_{64} R_1R_5 + f^2_{56} R_2R_4\, =\, 0
\end{array}
\right.
\end{align}
\eseq
and
\bseq
\begin{align}
g_s *_{\calm_6} F_2 & = - \a'^{\, 3/2} (2\pi)^3  R_1R_2 R_3 \left[ 4\d A \wedge \eta^1 \wedge \eta^2 \wedge \eta^3 +  \d \left(\eta^1 \wedge \eta^2 \wedge \eta^3 \right) \right]
\label{ex2rel2}\\
g_s *_{\calm_6} F_0 & = - \a'^{-1/2} (2\pi)^{-1}  \big(N + B_1f^1_{56} + B_2f^2_{64} + B_3f^3_{45} \big) \frac{\d{\rm vol}_{\calm_6}}{R_4R_5R_6}
\label{ex2rel4}
\end{align}
\eseq
while $F_4 = F_6 = 0$. Also note that $\d J^2 = 0$ and so, in the limit of constant warp factor, we have a half-flat internal manifold.

Again, these last conditions can be rewritten as
\bea
F_2 & = &  g_s^{-1} *_{\calb_3} \d e^{-4A}  - \a' (2\pi)^2 g_s^{-1} \sum_{a=1,2,3} R_a^2\, \eta^a \wedge *_{\calb_3} \d \eta^a
\label{ex2rel2b}\\
& = &  g_s^{-1} *_{\calb_3} \d e^{-4A}  - \a'^{1/2} (2\pi)\, \tilde{g}_s^{-1} \sum_{a, j, k}  \eta^a \wedge \d y^i  \eps_{ijk}f^a_{jk} \frac{R_a^2}{R_j^2R_k^2} 
\label{ex2rel2c}\\
& = & g_s^{-1}*_{\calb^3} \d e^{-4A}  - \a'^{1/2} (2\pi)  \sum_{a=1,2,3} \eta^a \wedge F^a_{\calb_3}\, \equiv \,  g_s^{-1} *_{\calb_3} \d e^{-4A} + F_2^{\text{bg}}
\label{ex2rel2d}\\
F_0 & = &  - \a'^{-1/2} (2\pi)^{-1} (\tilde{g}_s {\rm Vol}(\calm_6))^{-1} \left[N + B_1f^1_{56} + B_2f^2_{64} + B_3f^3_{45} \right]\, \equiv\, F_0^{\text{bg}}
\label{ex2rel4b}
\eea
where now $\tilde{g}_s = g_s /R_1R_2R_3$, $\calb_3$ stands for the unwarped flat metric along $\{y^4, y^5, y^6\}$, ${\rm Vol}(\calm_6)$ is normalized in $\a'$ units and $F^a_{\calb_3}$ is an integer-valued one-form in $\calb_3$. As before, the fact that $F_2^{\text{bg}}$ and $F_0^{\text{bg}}$ are quantized is what selects particular values of the (now K\"ahler) moduli, or else the equations of motion that depend on (\ref{ex2rel2}) are not satisfied. 
Following \ref{IIAD6} we can identify
\be
\Im \Om_{\calb_3} \, = \, - \a'^{3/2} (2\pi)^3 e^{-3A} R_4R_5R_6\, \d y^4\wedge \d y^5 \wedge \d y^6
\ee
and so from 
\be
\bar{H} + \d J_c\, =\, - \frac{e^{3A} \a'^{-1/2} (2\pi)^{-1}}{R_4R_5R_6}  \left[N - i \left(T_1f^1_{56} + T_2f^2_{64} + T_3f^3_{45}\right) \right] \, \Im \Om_{\calb_3}
\label{ex2relDWSB}
\ee
and (\ref{d62}) one can easily deduce the value of $re^{i\theta}$, which is again a complex number times $e^{3A}$. If we now define
\be
\Re \Om_{\Pi_3}\, \equiv\, - *_{\calm_6} \Im \Om_{\calb_3}
\label{ex2fiber}
\ee
it is easy to see that the extra constraint (\ref{d6eom}) is again automatically satisfied by our initial NS ansatz. Finally, the Bianchi identity for $F_2$ reads
\be
g_s\, \d F_2 \, = \, -  \nabla^2_{\T^3} e^{-4A} \d{\rm vol}_{\T^3}  - \a' (2\pi)^2 \sum_{a=1,2,3}  R_a^2\, \d\eta^a \wedge *_{\T^3} \d\eta^a
\label{ex2BIF3}
\ee

Quite similarly to (\ref{ex1defbg}), the background flux $F_2^{\text{bg}}$ satisfies the relation
\be
e^\Phi \left(F_2 - F_2^{\text{bg}}\right) \, =\, - *_{\calm_6} \frac{4}{3} \left(\d \Phi \wedge \Re \Om_{\Pi_3} \right)  
\label{defF2bg}
\ee
and this again simplifies the  computation of the supersymmetry variations. Indeed, the dilatino variation is now
\be\nonumber
\calo \, = \,  \slashed{\p}\Phi + \oh \slashed{H} \sig_3 - \frac{1}{4} e^{\Phi} \left(5 F_0  \sig_1 + 3 i \slashed{F}_2 \sig_2 \right)\, =\, - \frac{3}{4} e^{\Phi} \left( F_0  \sig_1 + i\slashed{F}_2^{\text{bg}} \sig_2\right) + \left( 2\slashed{\p}\Phi - e^\Phi i F_0  \sig_2\right) P_-^{\Pi_3} 
\label{ex2dilatino}
\ee
where we have defined the projectors
\be
P_\pm^{\Pi_3}\, =\, \oh \left(1 \pm \Re \slashed{\Om}_{\Pi_3} \g_{(6)} \sig_2 \right)
\label{ex2projD6}
\ee
Hence, if we consider a spinor $\eps$ satisfying $P_-^{\Pi_3} \eps = 0$, the corresponding dilatino variation is given by
\be
\calo \eps \, = \, - \frac{3}{4} \left(\slashed{f} + \slashed{H} \sig_3\right) \eps
\label{ex2dilatinob}
\ee
where we have used the identity $ie^{\Phi} \slashed{F}_2^{\text{bg}} =  \slashed{f} \Re \slashed{\Om}_{\Pi_3} \g_{(6)}$. The external gravitino operator is
\be\nonumber
\cald_\mu \, =\, \p_\mu + \frac{1}{6} \G_\mu \slashed{\p} \Phi  - \frac{1}{8} e^{\Phi} \G_\mu \left( F_0 \sig_1 + i \slashed{F}_2 \sig_2\right)\, =\,  \p_\mu - \frac{1}{8} e^{\Phi} \G_\mu \left( F_0 \sig_1 + i \slashed{F}_2^{\text{bg}} \sig_2\right) + \frac{1}{3}\G_\mu \slashed{\p} \Phi \, P_-^{\Pi_3}
\label{ex2gravext}
\ee
while the internal one
\be
\cald_m \, = \, \nabla_m + \frac{1}{4} \slashed{H}_m \sig_3 - \frac{1}{8} e^{\Phi} \left( F_0 \sig_1 +  i\slashed{F}_2 \sig_2\right) \G_m
\label{ex2gravint}
\ee
where $\nabla_m$ is given by (\ref{ex1spinc}), and $\Lam_{mn}$ is defined as in (\ref{ex1Lam}) but with $a \in \Pi_3$. By splitting the internal index $m$ into fiber and base indices, we obtain
\bseq
\begin{align}
\label{ex2gravintf}
\cald_{a}  & =  \p_{a}  - \frac{1}{8} \G_{a} \left(\slashed{f} +e^\Phi  F_0 \sig_1 \right) -  \Big(\frac{1}{4}\slashed{f} +  \frac{1}{3}\slashed{\p}\Phi \Big)\G_{a} P_-^{\Pi_3} \quad \quad a \in \Pi_3 \\
\label{ex2gravintb}
\cald_{j}  & =   \Big( \p_{j} - \frac{1}{6} \p_{j} \Phi\Big) + \frac{1}{8} \Lam_j{}^m\G_m  \Big(\slashed{f} + \slashed{H} \sig_3\Big) +\nonumber\\ &\qquad\qquad\Big[\Big(\frac{1}{4} \slashed{f}+ \frac{1}{3}\slashed{\p}\Phi \Big) \Lam_j{}^m\G_m +\frac{1}{4}\G_j \slashed{H} \sig_3 \Big]P_-^{\Pi_3} \quad\quad j \in \calb_3
\end{align}
\eseq
So by considering a spinor $\eps = e^{A/2} \eps'$ such that $\eps'$ is constant and satisfies $P_+^{\Pi_3} \eps' = 0$, the supersymmetry variations amount to
\bseq
\begin{align}
\calo\eps & =  - \frac{3}{4} \left(\slashed{f} + \slashed{H} \sig_3\right) \eps \\
\cald_\mu\eps & =   - \frac{1}{8}\G_{\mu} \left(\slashed{f} + \slashed{H} \sig_3\right) \eps   \quad \quad \quad \quad \quad \quad  \mu \in \reals^{1,3} \\
\cald_{a} \eps & =    \frac{1}{8} \Lam_{a}{}^n \G_n  \left(\slashed{f} + \slashed{H} \sig_3\right) \eps \quad \quad \quad \quad  \quad \ a \in \Pi_3 \\
\cald_{j} \eps & =  \frac{1}{8} \Lam_{j}{}^n \G_n \left(\slashed{f} + \slashed{H} \sig_3\right) \eps \quad \quad \quad \quad \quad \ j \in \calb_3
\end{align}
\eseq
obtaining a SUSY-breaking pattern of the form
\be
\begin{array}{ccc}
\calv_1 \, = \, - \frac{1}{4} \left(\slashed{f} + \slashed{H} \right)  \eta_1 & \quad & \calu^1_m \, =\, \frac{1}{8} \Lam_{mn} \G^n \left(\slashed{f} + \slashed{H} \right) \eta_1 \\
\calv_2 \, = \, - \frac{1}{4} \left(\slashed{f} - \slashed{H} \right)  \eta_2 & \quad & \calu^2_m \, =\, \frac{1}{8} \Lam_{mn} \G^n \left(\slashed{f} - \slashed{H} \right) \eta_2
\end{array}
\label{ex2susyparam}
\ee
Finally, the modified dilatino variation reads
\be
\Delta\eps\, =\, \left(\G^M\cald_M - \calo\right)\eps\, =\, \frac{1}{8} \left( 2 +  \Lambda_{mn} \G^m \G^n\right)\, \left(\slashed{f} + \slashed{H} \sig_3\right) \eps \, =\, \frac{1}{4} \left(\slashed{f} + \slashed{H} \sig_3\right)  \eps
\label{ex2dilat}
\ee
And so we again recover the ansatz (\ref{restr}), with
\be
r\eta_1^*\, =\, -\frac{1}{4} \left(\slashed{f} + \slashed{H}\right)\eta_1\quad \quad \text{and} \quad \quad r\eta_2^*\, =\, -\frac{1}{4} \left(\slashed{f} - \slashed{H}\right)\eta_2
\label{ex2r}
\ee
which imply that
\be
r\, =\, -\frac{1}{4} \Omega\cdot(f+H)
\label{ex2rb}
\ee
in agreement with (\ref{ex2relDWSB}).

%%%%%%%%%%%%%%%%%%%%%%%%%%%%%%

\subsection{Type IIB on a simple static $SU(2)$-structure vacuum}

Let us now construct a simple example of static $SU(2)$-structure background, in order to illustrate the discussion of section \ref{staticsec}. Instead of a twisted six-torus, we will now consider a standard factorizable six-torus with a non-trivial warp factor
\be
\d s^2_{\calm_6}\, =\, (2\pi)^2\alpha^\prime\Big\{e^{2A}\big[ R_1^2 (\d y^1)^2+R_2^2 (\d y^2)^2\big] +e^{-2A}\sum^6_{j=3}R_j^2(\d y^j)^2\Big\}
\ee  
and we fix the fibration describing our SUSY-breaking by choosing the 2-torus spanned by $y^1,y^2$ as the fiber $\Pi_2$ and the 4-torus  spanned by $y^3,y^4,y^5,y^6$ as the base $\calb_4$. Furthermore there are 16 O5-planes wrapping $\Pi_2$ at the fixed points of the involution $y^{3,4,5,6}\rightarrow - y^{3,4,5,6}$ on the base, and we allow for possible D5-branes wrapping $\Pi_2$ at arbitrary points in $\calb_4$. We also assume that all fields vary only along $\calb_4$. 

The static $SU(2)$-structure as defined in subsection \ref{staticsec}, is defined by the tensors
\bea\label{su2ex}
\mathpzc{j}&=&-(2\pi)^2\alpha^\prime\, (R_1R_4\d y^1\wedge \d y^4+R_2R_5\d y^2\wedge \d y^5)\cr
\mathpzc{w}&=&(2\pi)^2\alpha^\prime(e^AR_1\d y^1+i e^{-A}R_4\d y^4)\wedge (e^AR_2\d y^2+ie^{-A}R_5 \d y^5) \cr
\theta &=&2\pi\sqrt{\alpha^\prime}e^{-A}(R_3\d y^3+iR_6\d y^6)
\eea
so from (\ref{hol1form}) we obtain that we must set
\be
e^\Phi\,=\,g_s\,e^{2A}
\ee

From (\ref{su2ex}) we see that $\d \mathpzc{j}=0$ and thus the first condition in (\ref{jH}) is automatically satisfied while the second is fulfilled if we take
\be
H\, =\, (2\pi)^2\alpha^\prime N_{\rm NS}\, \d y^4\wedge \d y^5\wedge \d y^6
\ee
with $N_{\rm NS}\in\mathbb{Z}$. This corresponds to having $f=u=h=0$ and $g=i\pi \sqrt{\alpha^\prime} e^A N_{\rm NS}\d y^4\wedge \d y^5$ in (\ref{decjH}). Then, from (\ref{sbsu2}) we get the SUSY-breaking parameter
\be
r\,=\,\frac{i e^{3A} N_{\rm NS}}{8\pi\sqrt{\alpha^\prime} R_4R_5R_6}
\ee

One can also check that (\ref{su2half}) are satisfied, while (\ref{su2RR}) constrains the RR fields to be
\be
F_{1}\, =\, -\frac{R_3 N_{\rm NS}}{g_sR_4R_5R_6}\, \d y^3\quad \quad F_{3}\,=\,-\frac{1}{g_s}\hat*_{\calb_4} \,\d e^{-4A}\quad\quad F_{5}\,=\,0
\ee
where $\hat*_{\calb_4}$ is constructed from the (unwarped) flat metric $\d \hat s^2_{\calb_4}=(2\pi)^2\alpha^\prime\sum^6_{j=3}R_j^2(\d y^j)^2$ and restricted to $\calb_4$. Finally, (\ref{simpres}) and (\ref{einsu(2)}) are also easily checked to be satisfied. 

The RR-flux quantization implies that we must set
\be
\frac{R_3 N_{\rm NS}}{g_sR_4R_5R_6}\,=\,N_{\rm R}\,\in\,\mathbb{Z}
\ee
and so the RR-bianchi identities reduce to 
\be
-\hat\nabla_{\calb_4}^2 e^{-4A}\, =\, \frac{g_s}{(2\pi)^2\alpha^\prime\prod^6_{a=3} R_a}\big[ N_{\rm NS}N_{\rm R}+\!\!\!\!\sum_{i\in \text{D5's,O5's}}q_i\delta^4_{{\rm T}^4}(y_i)\big]
\ee
where $q_{\rm D5}=-q_{\rm O5}=1$. The corresponding tadpole condition requires that $ N_{\rm NS}N_{\rm R}+n_{\rm D5}=16$, which is a tadpole constraint quite similar to that obtained in toroidal GKP vacua. Indeed, it is easy to check that, by performing two T-dualities along $y^1,y^2$, one obtains the GKP background discussed at the beginning of the following subsection.

As in the previous examples, one can compute the dilatino and gravitino operators. The dilatino operator is given by
\be
\label{ex3dilatino}
\calo\, =\, \slashed{\p}\Phi + \oh \slashed{H} \sig_3 - e^{\Phi}\slashed{F}_1 i \sig_2  - \oh e^{\Phi} \slashed{F}_3\sig_1\, =\, -  \oh \slashed{H} \sig_3 + 2\left( \slashed{\p}\Phi +\slashed{H} \sig_3 \right) P_-^{\Pi_2}
\ee
where we have used the relations $e^{\Phi} \slashed{F}_3 = 2 i \slashed{\p} \Phi\, \Re \slashed{\mathpzc{w}}_{\Pi_2} \g_{(6)}$ and $e^{\Phi} \slashed{F}_1 = i \slashed{H}  \Re \slashed{\mathpzc{w}}_{\Pi_2} \g_{(6)}$, and defined the projectors
\be
P_\pm^{\Pi_2}\, =\, \oh \left(1 \pm \Re \slashed{\mathpzc{w}}_{\Pi_2} \g_{(6)}\sig_1 \right)
\label{ex3projD5}
\ee
The external gravitino variation is 
\be
\cald_\mu  = \p_\mu + \frac{1}{4} \G_\mu \slashed{\p} \Phi  - \frac{1}{8} \G_\mu e^{\Phi} \left( \slashed{F}_1 i\sig_2 + \slashed{F}_3 \sig_1\right) = \p_\mu - \frac{1}{4}\G_{\mu} \left[ \slashed{H} \sig_3  - \left(\slashed{H} \sig_3 + 2\slashed{\p}\Phi \right)P_-^{\Pi_2}\right]
\label{ex3gravext}
\ee
and the internal ones
\bseq
\begin{align}
\label{ex3gravintf}
\cald_{a}  & =  \p_{a}  - \frac{1}{8} \G_{a} \slashed{H}\sig_3  - \frac{1}{4} \left(\slashed{H}\sig_3 +  2\slashed{\p}\Phi \right)\G_{a} P_-^{\Pi_2} \quad \quad a \in \Pi_2 \\
\label{ex3gravintb}
\cald_{j}  & =   \left( \p_{j} - \frac{1}{4} \p_{j} \Phi\right) + \frac{1}{8} \G_j  \slashed{H}\sig_3  + \frac{1}{4} \left( \slashed{H}\sig_3+ 2\slashed{\p}\Phi \right) \G_j  P_-^{\Pi_2} \quad \quad j \in \calb_4
\end{align}
\eseq
where again we have separated the internal index into a fiber index $a$ and a base index $j$. Note  that we essentially obtain the same operators to those of subsection \ref{ex1}. Indeed, we only need to replace (\ref{ex1proj}) with (\ref{ex3projD5}), $\slashed{f}$ with $\slashed{H} \sig_3$, and take into account that $\Lam_{mn}$ is now diagonal, and the above expressions are obtained from (\ref{ex1dilatino}), (\ref{ex1gravext}), (\ref{ex1gravintf}) and (\ref{ex1gravintb}). Hence, a similar pattern of dilatino and gravitino variations, again inside our DWSB subansatz, follows.

%%%%%%%%%%%%%%%%%%%%%%%%%%%%%%
\subsection{Type IIB on a $\beta$-deformed background}
\label{betaexample}

To construct an explicit $\b$-deformed background, let us start from the following simple $\caln=0$ GKP background with toroidal internal manifold $\calm_6=\mathbf{T}^6\simeq \mathbf{T}^2\times \mathbf{T}^4$, modeled on the examples discussed in \cite{kachrutorus}. We take as internal unwarped (flat) metric
\be
\d s^2_{\mathbf{T}^6}\,=\, (2\pi)^2\alpha^\prime\sum^3_{i=1} R_i^2\d z^i\d\bar z^i
\ee
with $z^i=y^i+\lambda^i y^{i+3}$ (i.e., $\lambda^i$ are complex structure moduli). The axio-dilaton $\tau=C_{(0)}+i e^{-\Phi}$ is taken to be constant (with $e^{\Phi}\equiv g_s$) and the three-form fluxes are chosen in the following way
\bea
H&=&(2\pi)^2\alpha^\prime N_{\rm NS}\,\d y^4\wedge \d y^5\wedge \d y^6 \cr
F_{3} &=&(2\pi)^2\alpha^\prime N_{\rm R}\,\d y^1\wedge \d y^2\wedge \d y^3+(\Re\tau) H
\eea
where $N_{\rm NS}$ and $N_{\rm R}$ are even numbers.\footnote{We choose $N_{\rm NS}, N_{\rm R} \in 2\zet$ in order to avoid subtleties with the charge quantization condition \cite{fp02}.}
Then, the ISD condition (\ref{isd}) on $G_{(3)}=F_{3}+ie^{-\Phi}H$ is satisfied whenever (see e.g. Appendix A of \cite{ms04})
\bea
\label{nonholmoduli}
\lambda^1\lambda^2\bar\lambda^3=\lambda^1\bar\lambda^2\lambda^3=\bar\lambda^1\lambda^2\lambda^3=\bar\lambda^1\bar\lambda^2\bar\lambda^3=
\tau\, N_{\rm NS}/N_{\rm R}
\eea
For simplicity, we will assume that $\Re\lambda^i=\Re\tau=0$, so that our $\T^6$ is the direct product of 6 orthogonal ${\bf S}^1$'s. 

To complete the construction of this background, one needs to take into account $2^6$ O3-planes located at the fixed point of the $\mathbb{Z}_2$ orientifold action $z^i\rightarrow -z^i$. Both these O3-planes and D3-branes are sources of the internal RR five-form flux, which is related to the warping by $F_{5}=g^{-1}_s\hat*_6\d e^{-4A}$. Hence, the Bianchi identity (\ref{intBI}) reduces to
\be\label{T6BI}
-\hat\nabla^2 e^{-4A}\,=\,\frac{g_s}{(2\pi)^2\alpha^\prime\prod_i R^2_i\Im\lambda_i}\big[ N_{\rm NS}N_{\rm R}+\!\!\!\!\sum_{i\in \text{D3's,O3's}}q_i\delta^6(y_i)\big]\ ,
\ee
where $q_{\rm D3}=1$ and $q_{\rm O3}=-1/4$. If $n_{\rm D3}$ denotes the number of D3 branes, by integrating (\ref{T6BI}) over ${\bf T}^6/\mathbb{Z}_2$ one gets the tadpole condition
\be\label{tadpole}
 \frac12N_{\rm NS}N_{\rm R}+n_{\rm D3}\,=\,16
\ee
Finally, the pure spinors are given by
\bseq
\begin{align}
\Psi^{\rm GKP}_1&=\exp\big[2\pi^2\alpha^\prime e^{-2A}\sum_i R^2_i\d z^i\wedge \d \bar z^i \big]\\
\Psi^{\rm GKP}_2&=e^{-3A}(2\pi\sqrt{\alpha^\prime})^{3}(R_1 R_2 R_3)\d z^1\wedge \d z^2\wedge \d z^3
\end{align}
\eseq

Let us now apply the procedure described in subsection \ref{msb} by splitting ${\bf T}^6\simeq {\bf T}^2\times D$, where ${\bf T}^2$ is described by the coordinates $y^1,y^2$ and $D = {\bf T}^4$ is described by $y^3,\ldots,y^6$. Then, applying a $\beta$-deformation along ${\bf T}^2$ generated by the bi-vector $\beta=\gamma/[(2\pi)^2\alpha^\prime]\partial_{y^1}\wedge \partial_{y^2}$ we obtain the following NS background
\bea
\d s^2&=&(2\pi)^2\alpha^\prime e^{-2A}\Big\{ g_s^{-2} e^{2\Phi} \Big[R_1^2(\d y^1)^2+R_2^2(\d y^2)^2\Big] +\Big[ R^2_3(\d y^3)^2+\sum^3_{i=1} R_i^2\,\Im\lambda^i(\d y^{i+3})^2\Big] \Big\}\cr
B&=& (2\pi)^2\alpha^\prime\Big(N_{\rm NS}\,y^4\d y^5\wedge \d y^6-\gamma g_s^{-2} e^{2\Phi - 4A} R_1^2 R_2^2 \, \d y^1\wedge\d y^2\Big)\cr
e^\Phi&=&g_s(1+\gamma^2e^{-4A}R_1^2 R_2^2)^{-1/2}
\eea 
where $A$ and $\Phi$ depend only on the $\T^4$ coordinates. Similarly, by applying (\ref{betaRR}) one obtains the RR field-strengths 
\bea
F_{1}&=& -\gamma N_{\rm R}\,\d y^3 \cr
F_{3}&=&(2\pi)^2\alpha^\prime N_{\rm R}\, g_s^{-2} e^{2\Phi} \, \d y^1\wedge \d y^2\wedge \d y^3-\gamma g_s^{-1} R_1 R_2\,\hat*_{{\bf T}^4}\d e^{-4A}\cr
F_{5} &=&(2\pi)^2\alpha^\prime R_1 R_2\, g_s^{-3} e^{2\Phi} \, \d y^1\wedge \d y^2\wedge \hat*_{{\bf T}^4}\d e^{-4A}
\eea

As discussed in subsection \ref{msb},  if $\gamma=m/n$ then $\beta$-deformation maps $n$ D3-branes to $m$ D5-branes. The relation $F_1=-\gamma N_{\rm R}$ found above is the flux counterpart of this result and indeed flux-quantization imposes that $\gamma N_{\rm R}\in\mathbb{Z}$. The Bianchi identity (\ref{intBI}) reduces to
\be\label{T4BI}
-\hat\nabla_{{\bf T}^4}^2 e^{-4A}\, =\, \frac{g_s}{\gamma(2\pi)^2\alpha^\prime\prod_i R^2_i\Im\lambda_i}\big[ \gamma N_{\rm NS}N_{\rm R}+\!\!\!\!\sum_{i\in \text{D5's,O5's}}q_i\delta^4_{{\rm T}^4}(y_i)\big]
\ee
with $q_{\rm D5}=-q_{\rm O5}=1$. 
If we insist in imposing the identification $y^{3,4,5,6}\simeq y^{3,4,5,6}+1$, we would be forced to introduce exotic orientifolds with non-vanishing $B|_{O5}=-(2\pi)^2\alpha^\prime/\gamma\,\d y^1\wedge \d y^2$ on them.  If $\gamma=1/m$, this would be appropriately quantized to be equivalent to zero. In any case, the fulfillment of the corresponding projection conditions for the background fields seems non-trivial and, to be conservative, one can just consider one or more of the above coordinates as non-compact.

Finally, the pure-spinors of the $\b$-deformed background are given by
\bseq
\begin{align}
\Psi_1&=g_s^{-1} e^{\Phi} \exp\Big[(2\pi)^2\alpha^\prime\gamma R_1^2R_2^2 e^{-4A} \Big( g_s^{-2} e^{2\Phi} \d y^1\wedge \d y^2 - \Im\lambda^1\Im\lambda^2\d y^4\wedge\d y^5\Big)\Big]\wedge\nonumber\\ &\qquad\qquad  \exp\Big(2\pi^2\alpha^\prime e^{-2A}\sum_i R_i\d z^i\wedge \d \bar z^i\Big)\\
\Psi_2&=-2\pi \sqrt{\alpha^\prime} \g g_s^{-1} e^{\Phi-3A} R_1R_2 R_3\,\d z^3\wedge\nn\\
&\qquad\quad\exp\Big[(2\pi)^2\alpha^\prime\Big(-\frac{\d z^1\wedge \d z^2}{\gamma} + \gamma  R_1^2 R_2^2 g_s^{-2} e^{2\Phi-4A} \d y^1\wedge \d y^2\Big)\Big]
\end{align}
\eseq
and the DWSB ansatz has the form (\ref{finansatz2}) with $\Pi={\bf T}^2$ and
\be
r\,=\, \frac{i\,e^{3A}N_{\rm NS}}{8\pi\sqrt{\alpha^\prime}\prod_i R_i\Im\lambda^i}\quad\quad\quad R\,=\,\g^{-1} g_s^{-2} e^{2\Phi}\, \d y^1\wedge \d y^2
\ee
Note that, as claimed earlier, $R\rightarrow 0$ on top of the sources.

%%%%%%%%%%%%%%%%%%%%%%%%%%%%%%
%%%%%%%%%%%%%%%%%%%%%%%%%%%%%%

\section{DWSB and soft terms on D-branes}\label{sec:fsoft}

While in Section \ref{sec:4dint} we discussed the 4D structure of DWSB vacua, we mainly focused on the closed string sector of the theory. An important issue in $\caln=0$ compactifications, however, is how supersymmetry breaking is felt by D-branes, which contain the gauge sector and chiral matter of the theory. Following common wisdom, one expects that spontaneous breaking of SUSY in the bulk is communicated via gravitational effects to the D-branes, which as a result develop a moduli mediated soft-term pattern \cite{modulim}.\footnote{Additional contributions  such as  those induced by anomaly mediation will of course be present and, as pointed out in \cite{chno05}, they may be comparable to the moduli mediated ones. In the following we will focus on the tree-level computation of moduli mediated soft-term masses, leaving the analysis of other contributions for future work.} Such a scenario has been worked out for GKP vacua in \cite{ciu03,ggjl03,lrs04,ciu04,lmrs05}, and more recently for other no-scale vacua in \cite{cg07}. 

The purpose of the present section is to generalize the computation of moduli-mediated soft terms for setups beyond GKP, and in particular for the no-scale DWSB vacua of section \ref{1param}. In the spirit of the rest of the paper, our approach will be based on the analysis of the higher-dimensional D-brane action, as in \cite{ciu03,ciu04}, rather than relying on a 4D effective action. The final results, nevertheless, should be understandable within the 4D context of moduli mediation and, in particular, compatible with the 4D structure of F-terms obtained in Section \ref{sec:4dint}. In addition, if we restrict to the subcase of no-scale DWSB vacua with $SU(3)$-structure (see section \ref{subcases}), we are led to those backgrounds analyzed in \cite{cg07}, where soft terms were computed via an effective 4D approach. While the 10D versus 4D soft term computation may differ via some effects such as those of warping, in general one expects that they should agree qualitatively. We will show below that, at least for simple examples, this is indeed the case.

%%%%%%%%%%%%%%%%%%%%%%%%%%%%%%

\subsection{Soft gaugino masses}

Let us first consider the computation of the D-brane gaugino masses which, as we now show, can be carried out generally for no-scale DWSB vacua. As advertised in Section \ref{sec:dwsb}, the key quantity of the DWSB ansatz (\ref{finansatz}) for the structure of D-brane soft terms is the rotation matrix $\Lam$, and this is particularly transparent in the case of soft gaugino masses.

The starting point of our analysis will be the D-brane fermionic action, quadratic in the fermions, computed in \cite{mms,dirac}. Specialized to space-filling D-branes wrapping an internal cycle $\Sigma \subset \calm_6$, with worldvolume flux $\calf$, it gives the following four-dimensional Lagrangian density 
\be\label{faction}
\call_F\,=\, i\pi\int_\Sigma\d\sigma e^{4A-\Phi}\sqrt{\det(g|_\Sigma+\calf)}\,\bar\theta[1-\Gamma(\calf)]\left(\Gamma^\mu \cald_\mu+\tilde\calm^{\alpha\beta}\Gamma_\alpha \cald_\beta-\frac12\calo\right)\theta
\ee
where $\alpha,\beta,\ldots$ are world-volume indices on the internal cycle $\Sigma$, and $\cald_\mu$, $\cald_\alpha$ and $\calo$ are the pull-back of the operators appearing in the dilatino, external and internal gravitino, respectively, which are defined in (\ref{backsusy}). In addition, $\theta$ is the (doubled) GS-spinor living on the D-brane, $\Gamma(\calf)$ is the $\kappa$-symmetry operator and $\tilde\calm^{\alpha\beta}$ denotes the inverse of $\tilde\calm:= g|_\Sigma+\sigma_3\calf$. Finally, we are again setting $2\pi \sqrt{\a'}=1$.

In order to analyze fermionic masses, we first of all remove the pure gauge fermionic degrees of freedom by imposing the $\kappa$-fixing
\be\label{kfix}
\bar\theta\Gamma(\calf)\,=\,-\bar\theta
\ee
Then, in order to extract the gaugino bilinear from (\ref{faction}), we use the approximate supersymmetry generators $\epsilon=(\epsilon_1,\epsilon_2)$, specified by $\eta_1$ and $\eta_2$ as in (\ref{fermsplit}). As we are considering BPS D-branes they must be calibrated by $\om^{\rm (sf)}$, which means that $\bar\epsilon\Gamma(\calf)=\bar\epsilon$ even if the background is non-supersymmetric. Using this and the $\kappa$-fixing (\ref{kfix}), we are led to identify the gaugino $\lambda$ as the fermionic mode $\theta=(\theta_1,\theta_2)$  such that
\be\label{gino}
\theta_1\,=\,\frac1{4\pi} e^{-2A}\lambda\otimes \eta_1 +\ \text{c.c.}\quad\quad\quad \theta_2\,=\,-\frac{1}{4\pi}e^{-2A}\lambda\otimes\eta_2+\ \text{c.c.}
\ee
As a check of this decomposition, it is easy to see that from the general supersymmetry transformations found in \cite{mms,dirac} one obtains the standard four-dimensional supersymmetry transformations relating the gauge field to $\lambda$.

Plugging (\ref{gino}) into (\ref{faction}) and using (\ref{restr}), one gets the following effective four-dimensional terms 
\be\label{gaction}
\call_\lambda\, =\, \frac{i}{2}\Re f\, \bar\lambda\slashed{\partial}\lambda +\frac12\,m_\lambda\, \lambda^{T}\hat\gamma^{0}\lambda \  +\ \text{c.c.}
\ee
where
\be\label{gcoupl}
f\,=\,\frac{1}{2\pi}\int_\Sigma \calt|_\Sigma\wedge e^\calf
\ee
gives the gauge kinetic function as a function of the chiral field $\calt$ defined in (\ref{calt}), and the gaugino mass is given by
\be\label{gmass}
m_\lambda\, =\,\frac{i}{8\pi}\int_\Sigma \mathfrak{m}_{3/2}\big[\Re\calt|_\Sigma\wedge e^\calf\big]_{\rm top}\big[ 3-\frac12\tr \big(\Lambda \cdot \Lambda^{-1}_{(\Sigma,\calf)}\big)\big]
\ee
where $\mathfrak{m}_{3/2}$ is the gravitino mass density  (\ref{grm1}) which is related to the susy-breaking parameter $r$ by (\ref{rmid}), and $\Lambda_{(\Sigma,\calf)}$ is defined as in (\ref{lambda}) but with $(\Sigma,\calf)$ instead of $(\Pi,R)$.

As in Section \ref{sec:dwsb} we call a calibrated D-brane aligned if it  wraps a leaf $(\Pi,R)$ of the generalized foliation that defines our DWSB background. Again, these are special D-branes that are  selected by the polarization matrix $\Lam$ of our background. From (\ref{gmass}) it is immediate to see that they can also be characterized by the vanishing of their gaugino mass, since
\be\label{vmcond}
m_\lambda\, =\,  0\quad\quad\text{for aligned D-branes}
\ee
while non-aligned calibrated D-branes will always have non-vanishing gaugino mass.

Let us now check that (\ref{gmass}) fits the usual 4D supergravity formula for gaugino masses, which has the schematic form
\be\label{4dgaugino}
m^{\rm 4D}_\lambda\, =\, -\frac{i}{4}G^{\phi\bar\phi}F_\phi\overline{(\partial_\phi f)}
\ee
where $F_\phi$ is the F-term associated with the chiral field $\phi$ that $f$ depends on, and $G^{\phi\bar\phi}$ is the inverse of the K\"ahler metric for the chiral field. To see the relation between (\ref{gmass})  and (\ref{4dgaugino}), let us take (\ref{gcoupl}) for a calibrated D-brane and consider its variation under the variation of $\calt$. We first identify the field $\phi$ that enters the gauge kinetic function (\ref{gcoupl}) as
\be
\delta_\phi\calt\, =\, \phi\,\tilde*_6 j_{(\Sigma,\calf)}
\ee
since deformations orthogonal to this one do not affect (\ref{gcoupl}). Then, we can write (\ref{gmass}) as
\be\label{ggm}
m_\lambda\, =\,-\frac{i}{4}\int_{\calm_6} G^{\phi\bar\phi}\cald_\phi W\overline{(\delta_\phi f(\calt))}
\ee
where we have introduced the inverse metric density 
\be
G^{\phi\bar\phi}:=\frac{2}{\pi e^{2A}\langle\tilde*_6j_{(\Sigma,\calf)},j_{(\Sigma,\calf)}\rangle}
\ee
and the densities
\be
\cald_\phi W\,=\,\langle \tilde*_6 j_{(\Sigma,\calf)}, \cald_{\calt_0} W\rangle\quad\quad\quad \delta_\phi f(\calt)\,=\, \langle \tilde*_6 j_{(\Sigma,\calf)},\delta_\calt f(\calt)\rangle
\ee
where $\cald_{\calt_0} W$ is the F-term density given in (\ref{tft}). 

This rewriting of the gaugino mass provides a 4D interpretation of (\ref{vmcond}). Indeed, for aligned D-branes we have $\delta_\phi\calt = \delta_\alpha\calt$, with $\delta_\alpha\calt$ defined in (\ref{fluct}). Because of (\ref{falpha}) $\cald_\alpha W$ vanishes, and from (\ref{ggm}) so does $m_\lam$. For instance, in the GKP case we have $\Lambda=\bbone$ and $(\Pi,R)$ is the trivial foliation whose leaves are points of $\calm_6$. The D-branes sitting on such leaves (i.e., the aligned D-branes) are D3-branes, for which it is well-known that $m_\lam^{D3} =0$ in the presence of ISD $\calg_3$ fluxes \cite{granaferm,ciu03}. From the 4D viewpoint above this happens because, on the one hand, $f_{D3} \sim \tau$ (that is, the D3-brane gauge kinetic function only depends on the axio-dilaton $\tau$, which is nothing but $\delta_\alpha\calt$ in the GKP case). On the other hand, the GVW superpotential depends on $\tau$, and so the no-scale structure implies that $F_\tau$ must vanish on-shell. In general, we would expect exactly the same 4D situation for any aligned D-brane. Indeed, the gauge kinetic function of a general aligned D-brane will depend on the field $\a$ defined by (\ref{fluct}), and because of (\ref{poteffns}) and (\ref{vanft}), the corresponding F-term (density) $\cald_\alpha W$ needs to vanish on-shell.

Finally, we can test our computation for the other BPS D-branes in GKP vacua, namely D7-branes. These are not aligned and so their gaugini will get a soft mass. From the general formula (\ref{gmass}) we obtain
\be\label{gmassd7}
m^{{\rm D}7}_\lambda\, =\, \frac{i}{8\pi}\int_\Sigma e^{-\Phi}\mathfrak{m}_{3/2}\,[1+\frac14 \tr(g|_\Sigma+\calf)^{-1}(g|_\Sigma-\calf)]\,\big[\calf\wedge\calf-(J\wedge J)|_\Sigma\big]
\ee
which matches and extends previous results \cite{ciu04,lmrs05}.

%%%%%%%%%%%%%%%%%%%%%%%%%%%%%%

\subsection{Further soft terms}

Using a similar strategy, one can in principle compute the full spectrum of moduli mediated soft terms for a D-brane in DWSB vacua. In particular, (\ref{faction}) contains the information of all fermion masses that a D-brane with a U(1) gauge group develops in such backgrounds. Beyond the case of the gaugino, however, the dilatino and gravitino variations (\ref{restr}) do not contain  enough information to compute the fermion mass, and one should compute the full fermionic operators $\cald_M$ and $\calo$. In the following we would like to illustrate how such fermion mass computations work in some simple examples, leaving a more detailed study for future work.

Let us first consider the case of type IIB on a warped Calabi-Yau with constant dilaton and ISD $\calg_3 = F_3 + ie^{-\Phi_0} H$. The fermionic operators in this case are given by
\be
\begin{array}{lcl}
\calo & = & \slashed{H}\sig_3 P_+^{O3}\\
\cald_\mu & = & \p_\mu + \G_\mu \slashed{\p} A\, P_+^{O3}  - \frac{i}{8}\G_\mu \slashed{H}\g_{(6)}\sig_1 \\
\cald_m & = & \nabla^{CY}_m  - \frac{1}{2}\p_m A  +   \left(\slashed{\p}A +\frac{1}{4} \slashed{H}\sig_3\right) \G_m P_+^{O3} + \frac{1}{8}\G_m\slashed{H}\sig_3
\end{array}
\label{fopwcy}
\ee
where $\mu$ runs over the coordinates of $\reals^{1,3}$,  $\nabla^{CY}_m$ is the covariant derivative in the unwarped Calabi-Yau, and we have defined the projectors
\be
%P_\pm^{D3}\, =\, \oh \left(1 \pm \g_{(4)} \sig_2\right)\\
P_\pm^{O3}\, =\, \oh \left(1 \pm \g_{(6)} \sig_2\right)
\label{projwcy}
\ee
Then, if we consider a D7-brane in this background wrapping a 4-cycle $\cals_4$ with $\calf = 0$ we find that the fermionic operator appearing in (\ref{faction}) is
\be
\G^\mu \cald_\mu + \G^\alpha\cald_\alpha - \oh \calo\, =\, \slashed{\p}_{\reals^{1,3}} + \slashed{\nabla}^{CY} -  \slashed{\p} A \Big(\frac{1}{2} - 2 P_+^{O3}\Big) -\oh \left( \slashed{H} - g^{\alpha\beta}\G_\alpha\slashed{H}_\beta\right)\sig_3 P_+^{O3}
\label{D7faction}
\ee
where $\alpha, \beta$ are indices pulled-back onto $\cals_4$. Note  that the last term in (\ref{D7faction}) vanishes if $H$ has only one index on $\cals_4$, but not otherwise. In addition, an $H$ with three components on $\cals_4$ is not compatible with $\calf = 0$. So we are left with the fermionic Lagrangian density
\be
\bar{\th} P_-^{D7} \Big[\slashed{\p}_{\reals^{1,3}} + \slashed{\nabla}^{CY} -  \slashed{\p} A \Big(\frac{1}{2} - 2 P_+^{O3}\Big) + \oh \slashed{H}_{(2)} \sig_3 P_+^{O3}\Big] \th
\label{D7faction2}
\ee
where $P_-^{D7}$ is the D7-brane projector enforcing the $\kappa$-fixing (\ref{kfix}), and ${H}_{(2)}$ stands for those components of $H$ with two indices on the D7-brane worldvolume: the case analyzed in \cite{ciu04} in flat space. 

All flux-induced fermion masses arise then from the term
\be
\oh \bar{\th}\, \slashed{H}_{(2)} \sig_3 P_+^{O3} P_+^{D7}\, \th
\label{D7faction3}
\ee
and are non-vanishing only for those spinors that satisfy $P_+^{D7} \th = P_+^{O3} \th = \th$. This is indeed the case for the gaugino (\ref{gino}) and so, by using (\ref{rmid}) and (\ref{gkpparam}), we recover the gaugino mass (\ref{gmassd7}) for the case $\calf = 0$.

There are, of course, further fermionic modes $\th$ that also satisfy $P_+^{D7} \th = P_+^{O3} \th = \th$. Using the results in \cite{jl05}, it is easy to see that they correspond to the fermionic partners of the geometric deformations of $\cals_4$. That all such bosonic modes pick up a mass proportional to $H$ was seen in \cite{gmm05} using a DBI analysis. The above formula shows the analogous statement for the corresponding fermionic modes, since for generic $H$ all possible mass terms for these would-be geometric modulini will be non-zero. Finally, there are fermionic modes that satisfy $P_+^{D7} \th = \th$ but $P_+^{O3} \th = 0$, and so do not get any mass term. The bosonic partners are nothing but the D7-brane Wilson lines, which are indeed moduli that do not get lifted by background fluxes, even if the latter break supersymmetry \cite{ciu04,gmm05}. 

Similarly, one can compute the fermion mass structure of different D-branes in other DWSB backgrounds. For instance, let us consider a D9-brane in the twisted torus background of section \ref{ex1}. The Dirac action contains the operator
\be\nonumber
\G^M\cald_M - \oh \calo\, =\, \slashed{\p} -\frac{1}{4} \slashed{\p} \Phi + \oh \left(\slashed{f}_m + 2 \p_m \Phi\right) \Lam^m{}_n \G^n P_-^{\Pi_2}\, =\  \slashed{\p}_{10} - \slashed{\p} A\Big(\oh - 2 P_-^{\Pi_2}\Big) +\oh \slashed{f} P_-^{\Pi_2}
\label{D9faction}
\ee
where we have taken into account that $ \Lam^m{}_n$ is block diagonal. Hence, we again obtain a fermion mass structure of the form
\be
\oh \bar{\th}\, \slashed{f} P_-^{\Pi_2} P_+^{D9}\, \th
\label{D9faction2}
\ee
in full analogy with (\ref{D7faction3}). Following a similar reasoning as for the D7-brane case, we find that from the four fermion modes that satisfy $P_+^{D9} \th = \th$ and have vanishing KK mass, only two (those satisfying $P_-^{\Pi_2} \th = \th$) may obtain a mass induced by $f$, while the other two ($P_-^{\Pi_2} \th = 0$) will not get a mass at this classical level, even if supersymmetry is broken. From the two fermions with $f$-induced masses one is, of course, the gaugino, while the other one is a would-be Wilson line along the $\Pi_2$ fiber coordinates. Note  that, if we set $A = const.$,  this reproduces the D9-brane $\mu$-term computation of \cite{cg07}, table 3, obtained via a 4D effective approach. Similarly, one can compute the $\mu$-terms for the D5-branes in this background, matching again the results of \cite{cg07}. Finally, a similar computation can be performed for the fermion masses of D6-branes in the type IIA background of section \ref{ex2}, matching the results of \cite{D6torsion,cg07}.

%%%%%%%%%%%%%%%%%%%%%%%%%%%%%%

\subsection{Vanishing soft terms and fermionic projectors}\label{cond}

We have seen above that an aligned D-brane in a DWSB background does not develop a gaugino mass via moduli mediation. In fact, from the prototypical example of an aligned D-brane, namely a D3-brane in a GKP background, we know that a stronger statement may hold. Indeed, such a D3-brane does not develop any soft mass term at the classical level \cite{ciu03}. One may then wonder if such a statement is true for any aligned D-brane in a more general DWSB background. 

In light of the fermion mass computations above, it is easy to understand why D3-branes do not develop fermion masses. Indeed, in general the fermionic Lagrangian density in (\ref{faction}) has the form
\be
\bar{\th}\, P_-^{Dp}\, \Big( \vec{\G}^{\,T} \cdot \tilde\calm^{-1} \cdot \vec{\cald} - \oh \calo \Big)\th
\label{fdens}
\ee
where $\vec{\G}$ is defined in the tangent space of the D-brane and we have defined the projectors
\be
P_\pm^{Dp} \, =\, \oh\left(1 \pm  \G(\calf)\right)
\label{fproj}
\ee
$\G(\calf)$ being the $\kappa$-symmetry operator. In the case of a D3-brane in a constant dilaton background we have that $P_\pm^{D3} = (1 \pm \g_{(4)} \sig_2)/2$ and that the quantity in brackets reads
\be\label{D3faction}
\G^\mu \cald_\mu - \oh \calo \, =\, \slashed{\p}_{\reals^{1,3}} + P_-^{O3} \Big(4\slashed{\p} A\, -  \frac{1}{2}  \slashed{H} \sig_3\Big) \, \simeq\, \slashed{\p}_{\reals^{1,3}} + P_+^{D3}\Big(4\slashed{\p} A\, -  \frac{1}{2}  \slashed{H} \sig_3\Big)
\ee
where we have taken into account that $\bar{\th} P_-^{O3} \simeq \bar{\th} P_+^{D3}$ for a  type IIB spinor. Since we have to multiply this quantity with $P_-^{D3}$ to the right, only the 4D derivative term $\slashed{\p}_{\reals^{1,3}}$ survives. Note  that in the above discussion we are not making any assumption about the fermionic modes $\th$, and so all of the D3-brane fermionic modes are free of flux-induced soft term masses.

A similar statement holds for aligned D-branes in the twisted torus backgrounds of section \ref{sec:examples}. Indeed, for a D5-brane wrapping the fiber $\Pi_2$ in the background of section \ref{ex2} we have that the Dirac operator reduces to
\be
\G^\mu \cald_\mu + \G^{a} \cald_{a} - \oh \calo \, =\, \slashed{\p}_6 -\frac{1}{4} \left( e^\Phi \slashed{F}_3^{bg} \sig_1 +\slashed{f} \right) \, =\,  \slashed{\p}_6 - \oh P_+^{\Pi_2} \slashed{f},
\quad \quad a \in \Pi_2
\label{D5faction}
\ee
where $\slashed{\p}_6 = \slashed{\p}_{\reals^{1,3} \times \Pi_2}$. Since now $P_+^{\Pi_2} \simeq P_+^{D5}$, we again find that only the derivative term survives. Finally, the same result is obtained for a D6-brane wrapping $\Pi_3$ in the type IIA background of section \ref{ex2}.

It is then easy to see when all fermionic soft masses will vanish for a D-brane. Namely when the Dirac operator in (\ref{fdens}) reduces to
\be
\vec{\G}^{\, T} \cdot \calm^{-1} \cdot \vec{\cald} - \oh \calo\, =\, \{\text{derivatives}\} + P_+^{Dp} \{\text{something}\}
\label{cond2}
\ee
It would be interesting to see if this statement generalizes to aligned D-branes in more complicated DWSB backgrounds beyond twisted tori.

%%%%%%%%%%%%%%%%%%%%%%%%%%%%%%
%%%%%%%%%%%%%%%%%%%%%%%%%%%%%%

\section{Anti-D-branes in DWSB vacua}
\label{sec:anti}

The bulk of this paper is devoted to non-supersymmetric compactifications where the source of SUSY-breaking is certain background fluxes in the internal space. In particular, we have focused on backgrounds that can be seen as deformations of supersymmetric ones. In this case, the internal fluxes generate a mass for the 4D gravitino of an underlying 4D $\caln=1$ theory, and the supersymmetry may be seen as spontaneously broken. 

On the other hand, in variations of the GKP construction, like the KKL(MM)T  scenario \cite{kklt,kklmmt}, there is an additional source of supersymmetry breaking that comes from probe {\em anti}-D-branes  (or $\bar{{\rm D}}$-branes) which explicitly break the (approximate) $\caln=1$ supersymmetry selected by the background fluxes, D-branes and orientifolds. In these models, the effect of the $\bar{{\rm D}}$-branes on the four-dimensional physics is usually described by some effective 4D potential, extracted by combining 4D and 10D arguments. 

Here we would like to study some aspects of the 4D physics of $\bar{{\rm D}}$-branes in our generalized setup, again trying to keep a 10D approach. A first step will be to realize that the existence of the calibration $\omega^{\rm (sf)}$ for space-filling branes can be useful for $\bar{{\rm D}}$-branes too. Furthermore, this will also allow us to give a 10D derivation of the super-Higgs mechanism induced by $\bar{{\rm D}}$-branes, that cannot be described starting from a 4D effective $\caln=1$ theory. %This generates a four-dimensional gravitino mass that should be compared with the gravitino mass generated by the background fluxes, in order to establish which of the two effects is the leading one and determines the possible effective four-dimensional  low-energy description.

%%%%%%%%%%%%%%%%%%%%%%%%%%%%%%

\subsection{Effective potential for pseudo-calibrated $\bar{{\rm D}}$-branes}

In Section \ref{sec:vacua} we have characterized our backgrounds by the existence of the calibrations $\omega^{\rm (sf)}$ and $\omega^{\rm (string)}$ for space-filling and string-like D-branes, while due to the breaking of supersymmetry the calibration $\omega^{\rm (DW)}$ is not integrable because of (\ref{nonsusypure}). In this case, from (\ref{cal}) we see that we can choose $C^{\rm el}=\omega^{\rm (sf)}$ as electric RR-potentials and this amounts to setting
\be
\cale(\Sigma,\calf)_{\rm BPS}\,=\, \cale_{DBI}(\Sigma,\calf)_{\rm BPS}+\cale_{CS}(\Sigma,\calf)_{\rm BPS}=0
\ee

If we consider space-filling $\bar{{\rm D}}$-branes (D and $\bar{\rm D}$ strings are physically equivalent), $\omega^{\rm (sf)}$ provides the following local lower bound for the associated energy density\footnote{We set the orientation by imposing $\int_\Sigma \omega^{\rm(sf)}|_\Sigma\wedge e^\calf\geq 0$ and change the sign of the CS term.}
\bea\label{localower}
\cale_{\bar{\rm D}\text{-brane}}(\Sigma,\calf)&=&\cale_{DBI}(\Sigma,\calf)-\cale_{CS}(\Sigma,\calf)\cr
&=& e^{4A-\Phi}\sqrt{\det(g|_\Sigma+\calf)}\,\d\sigma+\big[\omega^{\rm(sf)}|_\Sigma\wedge e^\calf\big]_{\rm top}\cr 
&\geq&  2\big[\omega^{\rm(sf)}|_\Sigma\wedge e^\calf\big]_{\rm top}
\eea
Stable configurations must minimize
\be\label{nsbps}
\calv_{\bar{\rm D}\text{-brane}}(\Sigma,\calf)\,=\,\int_\Sigma\cale_{\bar{\rm D}\text{-brane}}(\Sigma,\calf)\,\geq \,  2\int_\Sigma \omega^{\rm(sf)}|_\Sigma\wedge e^\calf
\ee
and the latter inequality implies that stable configurations can be obtained by configurations that  minimize the simplified potential
\be\label{antieffpot}
\calv^{\text{(eff)}}_{\bar{\rm D}\text{-brane}}(\Sigma,\calf)\,:=\,  2\int_\Sigma \omega^{\rm(sf)}|_\Sigma\wedge e^\calf
\ee
and are furthermore {\em pseudo}-calibrated, i.e. satisfy (\ref{BPScond}) but nevertheless are not calibrated because of the wrong orientation. Indeed, if $(\Sigma,\calf)$ is a global (local) pseudo-calibrated minimum of $\calv^{\text{(eff)}}_{\bar{\rm D}\text{-brane}}$, then it is a global (local)  minimum of $\calv_{\bar{\rm D}\text{-brane}}$ since for any (small) deformation to $(\Sigma^\prime,\calf^\prime)$ 
\be
\calv_{\bar{\rm D}\text{-brane}}(\Sigma^\prime,\calf^\prime)\, \geq\,  \calv^{\text{(eff)}}_{\bar{\rm D}\text{-brane}}(\Sigma^\prime,\calf^\prime)\, \geq\, \calv^{\text{(eff)}}_{\bar{\rm D}\text{-brane}}(\Sigma,\calf)\, =\,\calv_{\bar{\rm D}\text{-brane}}(\Sigma,\calf)
\ee
Note that, in particular, if $(\Sigma,\calf)$ is an isolated calibrated configuration\footnote{The deformations of calibrated configurations have been studied in \cite{lucapaul1}.}, then the $\bar{{\rm D}}$-brane is automatically at a (locally) stable configuration. 

Moreover, by restricting ourselves to pseudo-calibrated $\bar{{\rm D}}$-branes, $\calv_{\bar{\rm D}\text{-brane}}$ reduces to $\calv^{\text{(eff)}}_{\bar{\rm D}\text{-brane}}$. The latter may be used as an effective (off-shell) potential where we have integrated out fluctuations  deforming the $\bar{{\rm D}}$-brane away from the  pseudo-calibration condition, that are indeed massive because of (\ref{localower}).\footnote{Here we are making  the (often reasonable) assumption that the masses of the fluctuations preserving the pseudo-calibration conditions are much lower then the masses of the `KK-modes' violating it.} Furthermore, we expect pseudo-calibrated D-branes to be described by a finite-dimensional (pseudo)-moduli space (as it happens in the supersymmetric case \cite{lucapaul1}) and so the resulting $\calv^{\text{(eff)}}_{\bar{\rm D}\text{-brane}}$ should in fact depend on a finite number of four-dimensional fields.

The simplest example is provided by $\overline{{\rm D}3}$-branes in GKP vacua, whose $\omega^{\rm (sf)}$ is given in (\ref{wcalib}).  In this case $\overline{{\rm D}3}$-branes are automatically pseudo-calibrated and 
\be
\calv^{\text{(eff)}}_{\overline{{\rm D}3}}\,=\,2e^{4A(y)-\Phi(y)}
\ee
Thus, in e.g. warped CY's with constant dilaton, $\overline{{\rm D}3}$-branes tend to move towards the points where the warping is minimal. The typical example is the case of the warped deformed conifold solution found in \cite{ks}, where $\overline{{\rm D}3}$-branes fall to the tip of the cone.

To summarize, we have seen how in studying $\bar{{\rm D}}$-branes it is natural to restrict to pseudo-calibrated ones and use the effective potential (\ref{antieffpot}). In the next section we turn our attention to the fermionic degrees of freedom associated with these pseudo-calibrated $\bar{{\rm D}}$-branes.

%%%%%%%%%%%%%%%%%%%%%%%%%%%%%%

\subsection{Super-Higgs effect: the pseudo-gaugino is the (almost-) goldstino}\label{almost}

$\bar{{\rm D}}$-branes have opposite CS terms and correspondingly we must make the sign change
\be\label{signflip}
1-\Gamma(\calf)\quad\rightarrow\quad 1+\Gamma(\calf)
\ee
in the $\kappa$-symmetry projector appearing in the fermionic action (\ref{faction}). Thus, we must also change the sign in the $\kappa$-fixing condition (\ref{kfix}) and, for pseudo-calibrated $\bar{{\rm D}}$-branes, this leads to identifying the pseudo-gaugino $\lambda$ inside $\theta=(\theta_1,\theta_2)$ in the following way 
\be\label{antigino}
\theta_1\, =\, \frac1{4\pi} e^{-2A}\lambda\otimes \eta_1+\ \text{c.c.}\quad,\quad \theta_2\,=\,\frac{i}{4\pi}e^{-2A}\lambda\otimes\eta_2+\ \text{c.c.}
\ee
We will call $\lambda$  the pseudo-gaugino  since it transforms non-linearly under the (almost-)\ supersymmetry preserved by the flux-vacuum, the latter being explicitly broken by the $\bar{{\rm D}}$-brane. Such explicit supersymmetry breaking is similar in spirit to a D-term supersymmetry breaking and indeed $\lambda$ can be considered  an (almost-)goldstino. This can be seen by plugging  the expansion (\ref{antigino})  in  (\ref{faction}) and taking into account the change (\ref{signflip}) and the pseudo-calibration condition. In this way, one obtains again terms of the form (\ref{gaction}), with $f$ as in (\ref{gcoupl}) but with mass now given by
\bea\label{gmass2}
m^{\bar{\rm D}}_\lambda&=&-\frac{i}{8\pi}\int_\Sigma \mathfrak{m}_{3/2}\big[\Re\calt|_\Sigma\wedge e^\calf\big]_{\rm top}\big\{ 
1+\frac12\tr[\Lambda^{-1}\Lambda_{(\Sigma,\calf)}]\big\}
\eea
Note  that in the supersymmetric case $\mathfrak{m}_{3/2}=0$ and thus $m^{\bar{\rm D}}_\lambda=0$ for all anti-D-branes, consistently with the fact that $\lambda$ is the goldstino associated with breaking of the background supersymmetry.\footnote{In the fluxless CY case, $\lambda$ would be the true gaugino associated with the $\caln =1 \subset \caln = 2$ bulk supersymmetry preserved by the $\bar{{\rm D}}$-brane. In the presence of fluxes such supersymmetry is no longer there.} On the other hand, in the non-supersymmetric case even aligned anti-D-branes (that have $\Lambda_{(\Sigma,\calf)}=\Lambda$) have non-vanishing gaugino mass term.

Consider for example the GKP case. In this case the pseudo-calibrated $\bar{\rm D}$-branes are ${\overline{{\rm D}3}}$ and ${\overline{{\rm D}7}}$ branes.  For ${\overline{{\rm D}3}}$-branes we have
\be\label{gantid3}
m^{\overline{{\rm D}3}}_\lambda\,=\,-\frac{i}{2\pi}e^{-\Phi}\mathfrak{m}_{3/2}
\ee
while for $\overline{{\rm D}7}$-branes we have
\be\label{gsntid7}
m^{\overline{{\rm D}7}}_\lambda\,=\,\frac{i}{8\pi}\int_\Sigma e^{-\Phi}\mathfrak{m}_{3/2}\big\{1-\frac14\tr(g|_\Sigma+\calf)^{-1}(g|_\Sigma-\calf)\big\}\big[(J\wedge J)|_\Sigma-\calf\wedge\calf\big]
\ee
Note  that if $\calf=0$ then $m^{\overline{{\rm D}7}}_\lambda=0$, consistently with the results of \cite{ciu04}. Nevertheless, as  for other $\bar{\rm D}$-branes, the $\overline{{\rm D}7}$ gaugino acquires a mass anyway by super-Higgs effect, once gravity is taken into account. We will discuss this point in the next subsection.

%%%%%%%%%%%%%%%%%%%%%%%%%%%%%%
 
\subsection{Super-Higgs effect from $\bar{{\rm D}}$-branes}

In the previous subsection we have seen that, for a supersymmetric background, $m^{\bar{{\rm D}}}_\lambda=0$ and it was explained how this result comes from the identification of $\lambda$ with the goldstino created by the complete breaking of the background supersymmetry by the $\bar{{\rm D}}$-brane. Thus, we expect $\lambda$ to acquire a mass, together with the 4D gravitino (and other fermions), by some kind of super-Higgs effect. 

To see concretely how this works, we need the quadratic interaction terms between background and world-volume fermions. These terms can be obtained by extending the procedure followed in \cite{mms,dirac} to include background fermionic fields.\footnote{Alternatively, it should be possible to obtain these couplings by direct expansion of the background superfields in the superspace D-brane actions \cite{bt}, see e.g. \cite{dimitrios}.} Without repeating all the details, since they are analogous to the case of pure bosonic backgrounds, we quote just the final result.  For a general D-brane configuration $(\Gamma,\calf)$ on an arbitrary background, the fermionic interaction term is given by
\be\label{faction2}
S^{\rm mixed}_F\,=\,-2\pi i\int_\Gamma\d\xi e^{-\Phi}\sqrt{\det(g|_\Sigma+\calf)}\,\bar\theta[1-\Gamma(\calf)](\tilde\calm^{ab}\Gamma_a \psi_b-\frac12\chi)
\ee
where $\xi^a$ are world-volume coordinates and $\psi_a$ refers to the pull-back  of the  background string-frame (doubled) gravitino $\psi_M$. Furthermore, differently from the rest of the paper, in (\ref{faction2}) we denote the dilatino by $\chi$, to distinguish it from the world-volume $\lambda$ field. 

We now restrict to our class of backgrounds. We are interested in the coupling between the worldvolume field $\lambda$ and the 4D gravitino $\psi^{\rm 4D}_\mu$, that was identified in section \ref{sec:4dint} and therein denoted by $\psi_{(0)_\mu}^{\rm 4D}$ (while $\psi^{\rm 4D}_\mu$ referred to the 4D gravitino density) -- here we will omit the subscript ${}_{(0)}$. 

Let us start by first considering space-filling BPS D-branes. In (\ref{faction2}) we must take $\Gamma=\mathbb{R}^{1,3}\times \Sigma$, $\calf$ completely internal and $\xi^a=(x^\mu,\sigma^\alpha)$. Plugging the decompositions (\ref{gino}) and (\ref{diag10d})-(\ref{4Dgravitino}) into (\ref{faction2}), it is not difficult to see that the $\lambda\psi^{\rm 4D}$ term vanishes, as expected. We then turn to $\bar{{\rm D}}$-branes. In practice this amounts to making the sign change (\ref{signflip}) in (\ref{faction2}) and using the decomposition (\ref{antigino}). In this case, one obtains the following non trivial interacting term
\be
\call_{\rm int}\,=\,-\frac{i}2\,\rho\,\lambda^T\hat\gamma^0\hat\gamma^\mu\psi^{\rm 4D}_\mu +\ {\rm c.c.}
\ee
with
\be\label{intf}
\rho\, =\, 2\int_\Sigma e^{2A}\Re\calt|_\Sigma\wedge e^\calf
\ee
Comparing with the 4D supergravity \cite{toineconf}, the above term leads to the identification of $\rho$ as a D-term. This identification is also supported by rewriting (\ref{antieffpot}) as
\be
\calv^{\rm (eff)}_{\bar{\rm D}\text{-brane}}\,=\,2\int_\Sigma e^{4A}\Re\calt|_\Sigma\wedge e^{\calf}
\ee
that has indeed the structure $ f^{-1}(\text{D-term})^2/2$ of a D-term potential, if we think in terms of densities, by removing the integrals in (\ref{gcoupl}) and (\ref{intf}) to get $f$ and the D-term respectively, and then integrating the resulting density to get the potential.  Extrapolating the known results about the 4D super-Higgs effect \cite{toine83} to our context, we are lead to
\be
m^{\bar{{\rm D}}}_\lambda\, \sim\, \frac{\calv^{\rm (eff)}_{\bar{\rm D}\text{-brane}}}{M^2_{\rm P}m_{3/2}}
\ee
where $m_{3/2}$ is the gravitino mass (\ref{gmi}). Note however that the proper evaluation of $m_{3/2}$ should depend on the backreaction of the  $\bar{{\rm D}}$-brane on the background. For example, assuming that $e^{\calk_{\rm E}}|\calw_{\rm E}|^2\sim M^2_{\rm P}\calv^{\rm (eff)}_{\bar{\rm D}\text{-brane}}$, we get $m^{\bar{{\rm D}}}_\lambda\sim\sqrt{\calv^{\rm (eff)}_{\bar{\rm D}\text{-brane}}}/M_{\rm P}$.

%%%%%%%%%%%%%%%%%%%%%%%%%%%%%%
%%%%%%%%%%%%%%%%%%%%%%%%%%%%%%

\section{DWSB AdS$_4$ vacua}\label{sec:ads4}

In our quest to generalize the GKP construction, we have mainly focused on obtaining 4D Minkowski and no-scale stable vacua. With this goal in mind, the DWSB ansatz that we have taken in sections  \ref{sec:vacua} and \ref{sec:dwsb} seems the most natural one. From the broader perspective of constructing 4D $\caln=0$ supergravity vacua we could have relaxed several assumptions taken in section \ref{extension} and, in particular, the fact that our 10D space is $X_{10}  = X_4 \times_{\om} \calm_6$ with $X_4 = \reals^{1,3}$. In the following, we would like to extend the analysis of these sections to $\caln=0$ compactifications where $X_4 = \text{AdS}_4$. 

Indeed, while from a phenomenological viewpoint $\text{AdS}_4$ vacua may a priori seem not too attractive, it has been shown that, in terms of moduli stabilization via fluxes, they possess  much nicer properties than Minkowski vacua \cite{dkpz04,vz05,dgkt05,camara2005}. In addition, following the ideas in \cite{kklt}, one may consider uplifting such an AdS$_4$ vacuum to a de Sitter one by including anti-D-branes. While in the original proposal of \cite{kklt} these uplifting ingredients were anti-D3-branes, we have seen in the previous section that in a generalized setup one could consider similar objects, namely pseudo-calibrated anti-D-branes,  that could also do the job. Finally, note that the key point to arrive at the DWSB ansatz (\ref{susypure}) and (\ref{nonsusypure}) was the understanding of $\caln =1$ Minkowski vacua in terms of D-brane calibrations \cite{luca}. As this understanding has been extended to $\caln=1$ AdS$_4$ vacua in \cite{lucapaul3}, it is natural to apply the same philosophy to the construction of $\caln=0$ AdS$_4$ backgrounds.

We can thus proceed as in section \ref{extension} and consider a 10D spacetime of the form $X_{10}=X_4\times_\omega \calm_6$, with metric
\be
\d s^2_{10}\, =\, e^{2A}\d s^2_{X_4}+g_{mn}\d y^m\d y^n
\label{10dansatz2}
\ee
and RR fields of the form (\ref{RRsplit}), but where now  $X_4$ is an AdS$_4$ space of radius $R_{\text{AdS}}$. Again, we will assume that $\calm_6$ is endowed with an $SU(3) \times SU(3)$ structure, corresponding to an approximate 4D supersymmetry in this background. As before, such an $SU(3) \times SU(3)$ structure is equivalent to the presence of the internal pure spinors $\Psi_1$ and $\Psi_2$, which define in turn the real polyforms (\ref{bcal}) playing the role of D-brane calibrations. Finally, the supersymmetry conditions for this background can again be expressed in terms of these calibrations as 
in (\ref{susygeneral}), where the only new ingredient is the complex constant $w_0$ defined by (\ref{defw0}) and related to the AdS$_4$ radius by $R_{\text{AdS}} = 1/|w_0|$.

Now, in our extension to AdS$_4$ we would like to keep an essential property of our DWSB backgrounds, which is that calibrated space-filling D-branes do not develop tachyons. This amounts to imposing from the very beginning the gauge BPSness condition, that in the present context reads
\boxedeq{\d_H(e^{4A-\Phi}\Re\Psi_1)\,=\, e^{4A}\tilde*_6 F+ 3(-)^{|\Psi_2|}e^{3A-\Phi}\Re(\bar w_0\Psi_2)\quad\quad \text{gauge BPSness} \label{adscal}}
where again $|\Psi_2|$ is the degree mod 2 of the polyform $\Psi_2$. Just like in the Minkowski case, (\ref{adscal}) can be rephrased in terms of the selfduality properties of the polyform (\ref{calg}). We now have that
\be
(\tilde*_6+i)\calg\, =\, 3(-)^{|\Psi_1|}e^{-A-\Phi}w_0\bar\Psi_2
\ee
reproducing (\ref{iasd}) for $w_0 = 0$.

Having imposed (\ref{adscal}), we now relax the other two BPSness conditions. First, note that, for $w_0 \neq 0$,   (\ref{adscal}) automatically implies that $\d_H[e^{3A-\Phi}\Re(\bar w_0\Psi_2)]=0$ so that `half' of the DW BPSness is  satisfied (see \cite{lucapaul3} for an interpretation of this). Thus, for $w_0 \neq 0$, the SUSY-breaking pattern will be encoded in the following {\em real} DW (non)BPSness condition
\boxedeq{\label{adsdwsb}
\d_H[\Im(\bar w_0 e^{3A-\Phi}\Psi_2)]-2(-)^{|\Psi_2|}|w_0|^2e^{2A-\Phi}\Im\Psi_1 = \{\text{DWSB}\}\  \hspace{.75cm}  \text{DW (non)BPSness}}
while the string (non)BPSness condition will be a consequence of the above. Indeed, by looking at (\ref{susygeneral}) one can check that it amounts to
\boxedeq{|w_0|^2\, \d_H(e^{2A-\Phi}\Im \Psi_1)\,\propto\, \d_H \{\text{DWSB}\} \quad\quad \text{ D-string (non)BPSness} \label{adsstsb}}
and so it is fixed by the DWSB ansatz in (\ref{adsdwsb}). In particular, if we impose DW BPSness and $w_0 \neq 0$, the D-string BPSness condition is automatically satisfied. Note  that, since (\ref{adsstsb}) encodes the 4D bulk D-flatness, the above is equivalent to the familiar statement that in AdS$_4$ vacua F-flatness implies D-flatness \cite{lucapaul2}. On the other hand, in compactifications to flat space we may impose DW BPSness and still have non-vanishing D-terms.

In analogy with the procedure of Section \ref{sec:dwsb}, we can translate the DWSB pattern into constraints for the SUSY-breaking spinors $\calv_{1,2}$, $\calu^{1,2}_m$ and $\cals_{1,2}$. As derived in Appendix \ref{ap:sb}, we obtain that (\ref{adscal}) imposes the following constraints
\bea\label{adsgauge}
\left\{\begin{array}{l} r_1+r_2+t_1+t_2=0 \\
s^1_m+u^1_m+\frac12(1+iJ_1)^n{}_m(p^2_n)^*=0\\
s^2_m+u^2_m+\frac12(1+iJ_2)^n{}_m(p^1_n)^*=0\\
(1-iJ_2)^k{}_m q^1_{kn}=(1-iJ_1)^k{}_nq^2_{km}
\end{array}\right. \quad\quad\quad\quad\text{gauge BPSness}
\eea 
on the set of parameters defined in (\ref{fexp}). By restricting to backgrounds where the SUSY-breaking vectors vanish, i.e., by setting
\be
s^{1,2}_m\, =\, u^{1,2}_m\, =\, p^{1,2}_m\, =\, 0
\ee
we obtain a SUSY-breaking pattern of the form
\bea\label{calgenads}
\calv_{1} =r_1\eta^*_{1}\quad & &\quad \calv_{2} =r_2\eta^*_{2} \cr
 \cals_{1}=t_1\eta^*_{1}\quad & &\quad \cals_{2}=t_2\eta^*_{2} \cr
 \calu^{1}_m=q^{1}_{mn}\gamma^n\eta_{1}^*\quad & &\quad \calu^{2}_m=q^{2}_{mn}\gamma^n\eta_{2}^*\quad 
\eea
where these SUSY-breaking parameters are, in addition, restricted by (\ref{adsgauge}).  From (\ref{psgen}) and (\ref{psexp}), we see that the gauge BPSness violation has the form
\bea\label{adsps2}
\d_H(e^{3A-\Phi}\Psi_2)&=&2i(-)^{|\Psi_2|}w_0e^{2A-\Phi}\Im\Psi_1 +
\frac12(-)^{|\Psi_1|}e^{3A-\Phi}(t_2\Psi_1-t_1\Psi_1^\star)\\
&& + \frac12e^{3A-\Phi}(q^1_{mn}\gamma^n\Psi_1^\star\gamma^m-q^2_{mn}\gamma^m\Psi_1\gamma^n)\hspace{0.5cm} \text{DW (non)BPSness}\nonumber
\eea
while the string BPSness violation looks like
\bea\label{adsps}
\d_H(e^{2A-\Phi}\Im \Psi_1)&=&\frac12 (-)^{|\Psi_1|}e^{2A-\Phi}\Im[(t_2-t_1)^\star\Psi_2]\\
&&+e^{2A-\Phi}\Im[(q^1_{mn})^\star\gamma^n\Psi_2\gamma^m] \hspace{1cm} \text{string (non)BPSness}
\nonumber
\eea
In the last expression we have already taken into account the following relations between the scalar SUSY-breaking parameters 
\be\label{adsparrel}
 r\, :=\, r_1\, =\, r_2\,=\, -\frac12(t_1+t_2)
\ee
required by the mutual consistency of  (\ref{adsps2}) and (\ref{adsps}).  Furthermore, if we write $w_0=e^{-i\chi}/R$, the restricted form (\ref{adsdwsb}) of the allowed DW (non)BPSness   requires that
\be\label{adsconst}
e^{i\chi}t_1\,=\,e^{-i\chi} t_2^*\quad\quad\text{and}\quad\quad e^{i\chi}(1+iJ_2)^k{}_mq^1_{kn}\,=\,e^{-i\chi}(1+iJ_1)^k{}_n(q^2_{km})^*
\ee
Finally, if we impose the string BPSness/D-flatness condition we obtain a further relation
\be\label{Fsb}
\begin{array}{c}\text{pure F-term}\\ \text{SUSY breaking}\end{array} \qquad\Leftrightarrow \qquad \begin{array}{c} t_1=t_2\quad\quad\text{and} \\ (1-iJ_2)^k{}_m q^1_{kn}=(1-iJ_1)^k{}_nq^2_{km}=0\end{array}
\ee

Given this AdS$_4$ DWSB framework, one could in principle pursue the philosophy of section \ref{sec:dwsb}, and define a one-parameter set of backgrounds, interpret them in terms of foliated or other kind of geometries, etc. We will not attempt to construct AdS$_4$ vacua in such way, but rather turn to a quite different, although complementary, approach to find $\caln=0$, AdS$_4$ supergravity vacua: that based on integrability.

%%%%%%%%%%%%%%%%%%%%%%%%%%%%%%
%%%%%%%%%%%%%%%%%%%%%%%%%%%%%%

\section{Integrability of $\caln =0$ vacua}\label{sec:inte}

In the previous sections we have discussed extensively a class of  type II flux compactifications to four-dimensions. One of their key ingredients is the existence of background (generalized) calibrations of the kind introduced in \cite{paul,luca} and  the assumption that the (fully-backreacting) localized sources are calibrated by them. This fact allowed to simplify considerably not only the open-string equations of motion, but also the closed-string ones,  even in absence of supersymmetry, that would have been otherwise extremely complicated by the presence of the sources.

In this section we will explain how this remarkable property has more general validity, not necessarily related to compactifications to four dimensions. In fact, as we will see, the same mechanism is at the origin of the integrability of supersymmetric (static) backgrounds with localized sources that was already studied in \cite{kt}, extending previous results valid in the sourceless case \cite{lt,gauntlett}. Supersymmetry implies the existence of well-defined calibrations -- i.e. satisfying a certain differential condition -- for the sources \cite{lucapaul3}.  In \cite{kt} it was shown that, under certain mild assumptions, the inclusion of supersymmetric (and thus calibrated)  sources does indeed guarantee that the (first-order)  Killing-spinor conditions imply that the appropriately source-modified (second-order) Einstein, dilaton and $B$-field equations of motion are automatically satisfied, once the Bianchi identities are taken into account. As we will see in the following, this result can be though of as a corollary of a more general integrability theorem, valid also for non-supersymmetric backgrounds. 

In the general non-supersymmetric case considered here we will assume the D-branes and orientifolds to be  calibrated with respect to a well-defined calibration constructed from an underlying globally-defined spinor $\epsilon$. This will allow us to rewrite the second-order bosonic equations of motion as spinorial equations involving the product of two  first-order operators, where the contribution from the localized sources has disappeared by using the (generalized) Bianchi identities\footnote{Recall that in the democratic formalism the Bianchi identities of the dual RR fields correspond, after implementing the self-duality condition,  to the equations of motion of the original fields.}. 
%This factorization highlights the origin of the integrability of the supersymmetric case discussed in \cite{kt} and, as we will see, can be used to simplify the solution of the field equations for  non-supersymmetric configurations as well.
Schematically, let $\epsilon$ be the ten-dimensional supersymmetry generator, so that $A\epsilon=\psi$, where $\psi$ parameterizes the supersymmetry-breaking and $A$ is a first-order differential operator. The integrability result presented here, amounts to identifying a first-order differential operator $B$ such that, provided the (generalized) Bianchi identities are satisfied, $\epsilon^TB\psi$ is a linear combination of the (second-order) equations of motion. Note that by taking $\epsilon$ to be a Killing spinor, so that $\psi$ vanishes, one reproduces as a corollary the 
integrability results of \cite{lt,gauntlett,kt} for supersymmetric backgrounds. 
It follows from the above that an alternative strategy for the construction of general ten-dimensional non-supersymmetric vacua would be to search for backgrounds such that $\psi$ is non-vanishing but nevertheless lies in the kernel of $\epsilon^TB$.

 In the next subsection we will first discuss the general sourceless case, to later introduce calibrated sources in static spaces in subsection \ref{intloc}. In subsection \ref{int4+6} we then specialize back to compactifications to four dimensions, writing explicitly the spinorial equations in this case. These equations will be used in subsection \ref{4+6vacua}, where we construct new non-supersymmetric IIA AdS$_4$ vacua, as well as in appendix \ref{gkpvacua}, where we revisit the GKP vacua.

%%%%%%%%%%%%%%%%%%%%%%%%%%%%%%

\subsection{Spinorial factorization of sourceless equations of motion}

In this section we show how sourceless equations of motion and Bianchi identities can be combined in spinorial equations involving the product of two first-order differential operators.

First, we choose a (doubled) supersymmetry generator $\epsilon=(\epsilon_1,\epsilon_2)$ and introduce the associated supersymmetry breaking spinorial `parameters' $X_M$, $Y$ and $Z$ as follows
\beal\label{d3}
(\mathcal{D}_M\epsilon)_i&=:X_M^i\nn\\
(\mathcal{O}\epsilon)_i&=:Y^i\nn\\
(\Delta \epsilon)_i&= \Gamma^MX_M^i-Y^i=:Z^i
\end{align}
where $i=1,2$ and $\mathcal{D}_M$, $\mathcal{O}$ and $\Delta$ are the operators entering the fermionic  supersymmetry transformations and are defined by (\ref{backsusy}) and (\ref{modified}). Clearly, if the background is supersymmetric and $\epsilon$ is the associated Killing spinor,  then $X_M$, $Y$ and $Z$ must all vanish.  Furthermore, let us define the operator $\mathcal{P}$
\beal\label{oppdef}
(\mathcal{P}Z)^1&:= \big(\slashchar{\nabla}-\slashchar{\partial}\Phi+\frac14\underline{H}\big)Z^1
-\frac{1}{16}e^{\Phi}\underline{F}\cdot \Gamma_{(10)}Z^2\nn\\
(\mathcal{P}Z)^2&:= \big(\slashchar{\nabla}-\slashchar{\partial}\Phi-\frac14\underline{H}\big)Z^2
+\frac{1}{16}e^{\Phi}\underline{\sigma(F)}\cdot \Gamma_{(10)}Z^1
\end{align}
as well as the following tensors entering the modified Einstein, $B$-field and dilaton bosonic equations of motion (\ref{einstein2}), (\ref{Heom}) and (\ref{dilatoneom2})
\beal
E_{MN}&:=R_{MN}+2\nabla_{M}\nabla_N\Phi-\frac12H_M\cdot
H_N-\frac14e^{2\Phi}F_M\cdot F_N\nn\\
\delta H &:=e^{2\Phi}*_{10}\left[
\d(e^{-2\Phi}*_{10} H) -\frac12(*_{10} F\wedge F)_{8}
\right]\nn\\
D&:=2R-H^2+8\left(\nabla^2\Phi-(\partial\Phi)^2\right)~.\label{dlth}
\end{align}

The main result of this subsection is the following set of identities, whose derivation is discussed in appendix \ref{inte},  that express the double action of first order operators acting on $\epsilon$ in terms of the tensors $E_{MN}$, $\delta H$  and $D$ defined in (\ref{dlth})
\bseq\label{seconone}
\begin{align}
\Gamma^M\cdot &(\cd_{[N} X_{M]})^1-\frac12 (\nabla_N+\frac14 \underline{\iota_NH})\cdot Y^1
+\frac12(\mathcal{O}\cdot X_N)^1=\nn\\
&-\frac14 E_{NK}\Gamma^K\epsilon_1+\frac18\left(
\delta H_{NK}\Gamma^K+\underline{\iota_N\d H}\right)\epsilon_1-\frac{1}{16}e^{\Phi}\underline{\d_HF}\cdot\Gamma_N\Gamma_{(10)}\epsilon_2\label{iden}\\ 
\label{idenb}
\Gamma^M\cdot &(\cd_{[N} X_{M]})^2-\frac12 (\nabla_N-\frac14 \underline{\iota_NH})\cdot Y^2
+\frac12(\mathcal{O}\cdot X_N)^2=\nn\\
&-\frac14 E_{NK}\Gamma^K\epsilon_2-\frac18\left(
\delta H_{NK}\Gamma^K+\underline{\iota_N\d H}
\right)\epsilon_2
+\frac{1}{16}e^{\Phi}\Gamma_{(10)}\underline{\sigma(\d_{H}F)}\cdot\Gamma_N\epsilon_1\\
\label{idenc}
(\mathcal{P}Z)^1&
-\big( \nabla^M-2\partial^M\Phi +\frac{1}{4}g^{MN}\underline{\iota_NH}\big)\cdot X_M^1=-\frac{1}{8}D\epsilon_1+\frac14 \underline{\d H} \epsilon_1+\frac18 \underline{\d_H F}\epsilon_2\\
\label{idend}
(\mathcal{P}Z)^2&
-\big( \nabla^M-2\partial^M\Phi -\frac{1}{4}g^{MN}\underline{\iota_NH}\big)\cdot X_M^2=-\frac{1}{8}D\epsilon_2-\frac14 \underline{ \d H} \epsilon_2+\frac18\underline{ \sigma(\d_{H}F) }\epsilon_1
\end{align}
\eseq
From the above equations one can immediately derive the integrability property of sourceless supersymmetric vacua discussed in \cite{lt,gauntlett}. Indeed, in this case the left-hand side of the above equations vanishes and, once we impose the sourceless Bianchi identities $\d H=0$ and $\d_H F=0$,  they reduce to identities setting to zero $D$ and combinations of  $E_{MN}$ and $\delta H_{MN}$ acting on the spinors. For example, in the case of static spaces with vanishing mixed time-space components of $E_{MN}$, this automatically implies that the full set of equations is satisfied. We will consider the static case in more detail in the next subsection, where we will include localized sources in the discussion. As we will see, equations (\ref{seconone}) naturally encode the possibility to incorporate {\em calibrated} localized sources in the background.

%%%%%%%%%%%%%%%%%%%%%%%%%%%%%%

\subsection{Adding calibrated sources}
\label{intloc}

To treat the sources we will assume a general static space-time $X_{10}=\mathbb{R}\times \calm_9$ of metric
\beal
G_{MN}\d x^M\d x^N = e^{2 A} \d t^2 + {g}_{mn} \d x^m \d x^n
\end{align}
with $A$ and $g_{mn}$ depending only on the internal coordinates $x^m$.
We will include static D-branes and/or O-plane sources  corresponding to a piece $S^{\rm (loc)}$ in the total action which is the sum of two terms\footnote{For simplicity, we avoid to explicitly write down the higher-order corrections, which can be easily included in the present formalism.}
\be\label{source}
S^{\rm loc}\,=\,-2\pi\tau\int_\Gamma\sqrt{-\det(G|_\Gamma+\calf)}+2\pi\tau\int_\Gamma C|_\Gamma\wedge e^\calf\
\ee
In units of $2\pi\sqrt{\alpha^\prime}=1$,   $\tau$  is given by $\tau_{\text{D}p}=1$ for all D-branes and $\tau_{{\rm O}q}=-2^{q-5}$ for O$q$-planes, and in the latter case $\calf=0$. By introducing the total current $j_{\rm tot}$ defined by
\be
\int_{X_{10}}\langle \alpha, j_{\rm tot}\rangle\,=\,\sum_{i\in \text{loc. sources}}\!\!\!\!\tau_i\int_{\Gamma_i}\alpha|_{\Gamma_i}\wedge e^{\calf_i}
\ee 
for any polyform $\alpha$, we can write the Bianchi identities as
$\d_H F=-j_{\rm tot}$.
For static sources we have $\Gamma=\mathbb{R}\times \Sigma$, where $\mathbb{R}$ denotes the time direction and  $\Sigma\subset \calm_9$ is an internal cycle, and thus $j_{\rm tot}$ is defined on $\calm_9$. Furthermore, we can split the RR field-strength in electric and magnetic parts $F^{\rm el}$ and $F^{\rm mg}$ defined on $\calm_9$ as follows
\be
F\,=\, F^{\rm mg}+\d t\wedge F^{\rm el}
\ee
so that $F^{\rm el}=e^{-A}\sigma(*_9 F^{\rm mg})$, $\d_H F^{\rm el}=0$ and
\be\label{bi1+9}
\d_H F^{\rm mg}\,=\,-j_{\rm tot}
\ee

We can now use a globally defined spinor $\epsilon=(\epsilon_1,\epsilon_2)$ to construct a calibration as follows. Following \cite{kt}, we decompose first $\epsilon_{1,2}$ in 
\eq{
\epsilon_1 = \left(\begin{array}{c} 1 \\ 0 \end{array}\right) \otimes
\chi_1    \quad \quad
\epsilon_2 = \left(\begin{array}{c} 1 \\ 0 \end{array}\right) \otimes
\chi_2  \ \text{(IIB)}   \quad \quad
\epsilon_2 = \left(\begin{array}{c} 0 \\ 1 \end{array}\right) \otimes
\chi_2 \ \text{(IIA)} 
\label{tencom}}
where $\chi_{1,2}$ are real spinors on $\calm_9$. The gamma-matrices decompose accordingly as
\eq{
\label{epsdecomp}
\qquad \Gamma_{\ul{0}}=
(i \sigma_2) \otimes \bbone \, ,\qquad\Gamma_{{m}} = \sigma_1 \otimes {\gamma}_{{m}} \, ,  \qquad \Gamma_{(10)} =  \sigma_3
\otimes \bbone
}
with $\sigma_i$ the Pauli matrices, and ${\gamma}_{{m}}$ the
9-dimensional gamma-matrices. Using the internal spinors $\chi_{1,2}$ one can construct on $\calm_9$ the real polyform 
\bea
\omega:= \sum_{p\ \text{even/odd}} \frac{e^{A-\Phi}}{p!|a|^2}
\left(\chi_1^T {\gamma}_{m_1 \ldots m_p} \chi_2\right)
\; \d x^{m_1}\wedge \ldots\wedge \d x^{m_p} 
\label{calform}
\eea
where one has to sum over $p$ even/odd in IIA/IIB respectively. For any $\Sigma\subset \calm_9$ (with world-volume coordinates $\xi$), the polyform $\omega$ defined in (\ref{calform}) satisfies the algebraic inequality
\bea\label{loclb1+9}
[\omega|_\Sigma\wedge e^{\calf}]_{\rm top}\leq e^{A-\Phi}\sqrt{\det(g|_\Sigma+\calf)}\,\d\xi
\eea
Thus, if furthermore $\omega$ satisfies the differential condition
\bea\label{difgencal}
\d_H\omega=-F^{\rm el}
\eea
then $\omega$ is a proper (generalized) calibration. 

In the following we will assume that  the differential condition (\ref{difgencal}) is satisfied,\footnote{The condition (\ref{difgencal}) is automatically satisfied if $\epsilon$ is a Killing spinor \cite{lucapaul3}, up to some additional mild assumptions.} thus indirectly imposing a differential condition on $\epsilon$. We will also 
assume that all localized sources are calibrated by $\omega$, i.e. they saturate the local lower bound (\ref{loclb1+9}). These restrictions will allow us to greatly simplify the following discussion for two reasons. First, the open-string equations of motion are automatically satisfied, as can be seen by a simple extension  to this generalized setting of the stability argument (\ref{compustab}). Second, in deriving the NS closed-string equations of motion we can use the  simplified `effective' action
\beal
\label{actsrc}
S_{\rm eff}^{\text{loc}}
= -2\pi\int_{X_{10}}\langle \d t\wedge\omega, j_{\rm tot} \rangle
\end{align} 
Using (\ref{actsrc}), it is not difficult to see that the source-modified Einstein, dilaton and $H$-field equations of motion read:
\bseq\label{sourcecorr}
\beal
E_{MN}&=  - \frac12 e^{2\Phi} *_{10}\big[
G_{K(M} \langle\d x^{K} \wedge \iota_{N)} (\d t\wedge \omega),j_{\rm tot} \rangle
- \frac{1}{2} G_{MN}\langle \d t\wedge \omega, j_{\rm tot} \rangle
\big] \, ,\label{einst}\\
\delta H &=\frac12 e^{2\Phi}*_{10}\langle \d t\wedge \omega, j_{\rm tot} \rangle_8  \label{hfeom}\\
D&= e^{2\Phi}*_{10}\langle \d t\wedge \omega, j_{\rm tot} \rangle\label{dileom}
\end{align}
\eseq
Note that, as expected for static sources, the mixed time/space components on the
right-hand side of \eqref{einst} vanish and indeed, for the backgrounds we consider here, we automatically have  
\beal\label{mixedein}
E_{0m}=0
\end{align}

Consider now eq.~(\ref{iden}) for $N=n$, along the spatial directions.  Assuming that the contraction of the left-hand side 
with $\epsilon^T_1\Gamma_m$ vanishes and  taking eqs.~(\ref{tencom})-(\ref{mixedein}) and the $H$-field Bianchi identity $\d H=0$ into account, we obtain
\beal\label{presrc}
|a|^2\big( E_{mn}+\frac{1}{2}\delta H_{mn}\big)\mp\frac{1}{4}e^{\Phi}
(\chi^T_1\gamma_m\cdot \underline{\d_H F}^{\rm mg}\cdot\gamma_n\chi_2)  =0
\end{align}
for IIA/IIB.  
%In deriving the above, we have observed that there is no scalar in the  decomposition of the symmetric tensor product of two $Spin(9)$ spinors and hence:
%
%\beal
%(\hat{\epsilon}^i\gamma_{mn}\hat{\epsilon}^i)=0~,
%\end{align}
%
%for $i=1,2$. 
Substituting into (\ref{presrc}) the RR Bianchi identities (\ref{bi1+9}),  symmetrizing in $m,n$ and taking 
eq.~(\ref{plopa}) into account, we obtain the space/space components 
of the source-corrected Einstein equation (\ref{einst}).  
The time/time component 
of the source-corrected Einstein equation is obtained similarly 
by considering eq.~(\ref{iden}) for $N=0$, 
i.e. along the time direction, assuming the contraction of the left-hand side 
with $\epsilon^T_1\Gamma_0$ vanishes.  

Moreover, by antisymmetrizing eq.~(\ref{presrc}) in $m,n$ and taking 
eq.~(\ref{plopb}) into account, we obtain instead the 
source-corrected $H$-field equation-of-motion (note that $\delta H$ has no 
non-vanishing time components), eq.~(\ref{hfeom}).
Finally, in order to obtain the source-corrected dilaton equation-of-motion we assume that 
the  contraction of the left-hand side of  eq.~(\ref{idenc}) 
with $\epsilon^T_1$ vanishes. Taking (\ref{tencom}) into account and 
imposing the $H$-field Bianchi identity we thus obtain
\beal\label{d12}
|a|^2D=e^{\Phi}(\chi^T_1 \underline{\d_H F}^{\rm mg}\chi_2)
\end{align}
Imposing the RR Bianchi identities (\ref{bi1+9}) and using eq.~(\ref{plopc}), we arrive at the dilaton equation-of-motion (\ref{dileom}).  Note  that the same results can similarly be obtained starting from (\ref{idenb}) and (\ref{idend}) 
instead of (\ref{iden}) and (\ref{idenc}). 

{\it In summary: } if we consider static space-times with localized sources calibrated by a calibration $\omega$,  constructed from a doubled spinor $\epsilon=(\epsilon_1,\epsilon_2)$ as in (\ref{calform}), the vanishing of the following contraction of the left-hand side of { either} eq.~(\ref{iden}), or eq.~(\ref{idenb})\footnote{\label{foot12}In (\ref{inteq1}) and (\ref{inteq2}) the $\epsilon^i$'s are given by (\ref{tencom}). One could write 
the spinor contraction in a convention-independent form by replacing 
$\epsilon^i$ on the left with $\bar{\epsilon}^i\Gamma_0$. 
In addition, it has to be remembered that in this case the
${\epsilon}^i$'s are {\it commuting} ten-dimensional spinors.}
\boxedeq{\label{inteq1}\epsilon^T_i\Gamma_K\left\{
\Gamma^M\cdot (\cd_{[N} X_{M]})^i-\frac12 (\nabla_N-\frac14 (-)^i\underline{\iota_NH})\cdot Y^i
+\frac12(\mathcal{O}\cdot X_N)^i\right\}=0
}
either for $i=1$ or for $i=2$ and for $(K,N)$ space/space or time/time indices (but not mixed), is sufficient to guarantee that the source-corrected Einstein and $H$-field equation-of-motion are automatically satisfied, 
provided one imposes the (source-corrected) RR-field Bianchi identities (\ref{bi1+9}) and  
the $H$-field Bianchi identity.  Similarly, 
the vanishing of the following contraction of 
the left-hand side of { either} eq.~(\ref{idenc}), or eq.~(\ref{idend})
\boxedeq{\label{inteq2}\epsilon^T_i\left\{
(\mathcal{P}Z)^i
-\big( \nabla^M-2\partial^M\Phi -\frac14 (-)^i g^{MN}\underline{\iota_NH}\big)\cdot X_M^i
\right\}=0}
either for $i=1$ or for $i=2$, is sufficient to guarantee that the 
source-corrected dilaton equation-of-motion is automatically satisfied.

Finally, let us observe that if we consider space-times without sources, the equations (\ref{inteq1}) and (\ref{inteq2}) are valid 
% just assuming a reduction $SO(1,9)\rightarrow SO(9)$ of space-time structure group and the 
%condition (\ref{mixedein}), 
without any need to restrict to  static space-times or impose the condition (\ref{difgencal}) --  since the latter is only needed in order to ensure that the open-string equations of motion are satisfied.

%%%%%%%%%%%%%%%%%%%%%%%%%%%%%%

\subsection{Integrability conditions for flux compactifications}
\label{int4+6}

In this section, we give the general form of eqs.~(\ref{inteq1}) and (\ref{inteq2}) for compactifications to $\mathbb{R}^{1,3}$ or AdS$_4$. We decompose the ten-dimensional spinor as in eq.~(\ref{fermsplit}), with the four-dimensional spinor $\zeta$ satisfying the Killing equation (\ref{susygen3}), where the parameter $w_0$ is related to the AdS radius by $R=1/|w_0|$ and flat space is recovered by specializing to $w_0=0$. 

Direct substitution 
of definitions (\ref{sbf1ap}) into eqs.~(\ref{inteq1}, \ref{inteq2}), using the equations above, leads to the following conditions
\boxedeq{\label{4plus6b}\spl{0&=
(\slashed{\nabla}-\slashed{\partial}\phi+2\slashed{\partial}{A}+\frac{1}{4}\slashed{H})\calv^1-w_0({\calv}^1+{\cals}^1)^*\\
&+\slashed{\partial}A\cals^1
-2\partial^mA~\calu_m^1
+\frac{1}{4}e^{\phi}\slashed{F}\gamma_{(6)}\calv^2
+\frac{1}{4}e^{\phi}\gamma^m\slashed{F}\calu_m^2}}
and
\boxedeq{\label{4plus6a}\spl{0&=
2w_0\,{\calu}_n^{1*}-(\slashed{\nabla}-\slashed{\partial}\phi+2\slashed{\partial}{A}+\frac{1}{4}\slashed{H})\calu_n^1+\frac{1}{2}\slashed{H}_{n}{}^{m}\calu_m^1\\
&+(\nabla_n+\frac{1}{4}\slashed{H}_n)\cals^1+2\partial_nA\calv^1
+\frac{1}{8}e^{\phi}\gamma^m\slashed{F}\gamma_n\gamma_{(6)}\calu_m^2
+\frac{1}{4}e^{\phi}\slashed{F}\gamma_n\calv^2
~,
}}
coming from equation eq.~(\ref{inteq1}), as well as
\boxedeq{\label{4plus6c}\spl{0=\eta_1^{\dagger}\Big\{&
(\slashed{\nabla}-\slashed{\partial}\phi+2\slashed{\partial}{A}+\frac{1}{4}\slashed{H})\cals^1
-\frac{1}{8}e^{\phi}\slashed{F}\gamma_{(6)}\cals^2\\
&-(\nabla^m-2\partial^m\Phi+\frac{1}{4}\slashed{H}^m)\calu_m^1+\slashed{\partial}A\calv^1+w_0({\calv}^1-2{\cals}^1)^*
\Big\}}}
coming from eq.~(\ref{inteq2}). 

In deriving the above we have observed 
that the only nonvanishing spinor bilinears which can be constructed from the {\it commuting} (cf. footnote \ref{foot12}) four-dimensional spinor $\zeta$ are $\bar{\zeta}\hat\gamma_{\mu}\zeta$, $\bar{\zeta}^*\hat\gamma_{\mu\nu}\zeta$ and their complex conjugates.

Note that condition (\ref{4plus6c}) 
is expressed as the vanishing of a spinor contraction, i.e. it 
has the same form as the ten-dimensional integrability 
equation (\ref{inteq2}) from which it descends. 
On the other hand, 
the two conditions (\ref{4plus6a}) and (\ref{4plus6b}), coming from the ten-dimensional integrability 
equation (\ref{inteq2}), do not contain spinor contractions. This is because 
requiring that equation (\ref{inteq2}) be satisfied for all $(K,M)$ such that 
$K$, $M$ are either both spatial or both timelike, amounts to requiring that the right-hand sides  of (\ref{4plus6a}) and (\ref{4plus6b}) vanish upon contraction 
with both $\eta_1^{T}$ and $\eta_1^{\dagger}\gamma_m$ for any six-dimensional gamma matrix $\gamma_m$. For a nonzero six-dimensional Weyl spinor $\eta_1$, this is equivalent to requiring that the right-hand sides  vanish identically.

%%%%%%%%%%%%%%%%%%%%%%%%%%%%%%

\subsection{Non-supersymmetric AdS$_4$ vacua from integrability}
\label{4+6vacua}

We would now like to provide some examples where we can apply the results of subsection \ref{int4+6} to look for susy-breaking vacua. The strategy is to select an underlying $SU(3)\times SU(3)$ structure defined by two internal spinors $\eta_{{1}}$ and $\eta_2$, compatible with the possible localized space-filling sources that must be calibrated, and use equations (\ref{4plus6b}), (\ref{4plus6a}) and (\ref{4plus6c}) to investigate the restrictions imposed by the equations of motion. As we explain in appendix \ref{gkpvacua}, one can easily recover in this way the fact that the GKP vacua of section \ref{gkp} satisfy the equations of motion. Here we show how using the above results one can systematically investigate non-supersymmetric  vacua of IIA supergravity of the form AdS$_4\times \calm_6$, where $\calm_6$ is a six-dimensional nearly-K\"{a}hler manifold with constant dilaton and warp factor. To our knowledge, the first class of solutions presented here (eq.~(\ref{class1}) below) is new. The second class (eq.~(\ref{class2}) below) was first constructed by Romans in \cite{romans}.\footnote{Non-supersymmetric solutions of the form AdS$_4\times \calm_6$, where $\calm_6$ is a six-dimensional {\it complex} manifold, were also constructed in \cite{romans}. These are different from the  solutions presented here: the nearly-K\"{a}hler solutions  use an almost complex structure on $\calm_6$ which is {\it not} integrable. It would be interesting to examine whether the non-supersymmetric AdS$_4$ vacua of four-dimensional effective supergravity considered in \cite{camara2005} admit a ten-dimensional lift  to the solutions presented here.}
 Finally, supersymmetric vacua of this form (eq.~(\ref{class3}) below), which are a special case of the broader class of AdS$_4$ vacua of \cite{lt}, were first constructed in \cite{bc}.

A nearly-K\"{a}hler manifold is a special case of an $SU(3)$-structure six-dimensional manifold and, as such, it possesses a nowhere-vanishing spinor $\eta$ of positive chirality. Its only non-zero torsion class -- the torsion classes of  an $SU(3)$-structure are defined in (\ref{tc}) --  is ${W}_1$ (which can be taken to be constant and imaginary);  it is related to the nowhere-vanishing spinor through
\beal\label{triton1}
\nabla_m\eta=\frac{1}{4}{W}_1\gamma_m\eta^*
\end{align}
Acting with a covariant derivative on the left-hand side we obtain, after some standard manipulations:
\beal\label{triton2}
R_{mn}=\frac{5}{4}|W_1|^2g_{mn}
\end{align}
hence nearly-K\"{a}hler manifolds are Einstein spaces. In the following we will set ${W}_1=i\omega$, with $\omega$ a real constant. Finally let us note that in terms of the $SU(3)$-structure $(J,\Omega)$, equation (\ref{triton1}) can be written equivalently as
\beal\label{triton3}
\d J&=\frac{3}{2}\omega\mathrm{Re}\Omega\nn\\
\d\mathrm{Im}\Omega&=\omega J\wedge J
\end{align}

Let us now assume the following ansatz for the fluxes:
\beal\label{triton5}
F_{0}=\alpha;~~F_{2}=\beta J;~~F_{4}=\frac{1}{2}\gamma J^2;~~F_{6}=\frac{1}{6}\delta J^3;~~H=\varepsilon \mathrm{Re}\Omega
\end{align}
where $\alpha,\beta,\gamma,\delta,\varepsilon$ are real constants. Using (\ref{triton3}) it is easy to see that the Bianchi identities for the RR fluxes are equivalent to the conditions
\beal\label{triton9}
\beta=-\frac{2\varepsilon}{3\omega}\alpha;~~\gamma=\frac{2\varepsilon}{3\omega}\delta
\end{align}
while the H-field Bianchi identity is automatically satisfied. 

We are now ready to come to the integrability equations (\ref{4plus6b}-\ref{4plus6c}). Assuming a strict $SU(3)$ ansatz
\beal
\eta_1=\eta;~~\eta_2=e^{i\varphi}\eta^*
\end{align}
for some constant phase $\varphi$ and plugging the above ansatz into (\ref{4plus6b}) we obtain, after some algebra,
\beal\label{condi1}
0=3|w_0|^2-\frac{1}{16}(3|\lambda|^2+|\mu|^2)
\end{align}
where we have introduced:
\beal\label{triton7}
\lambda&:=\alpha- i \beta+\gamma-i \delta\nn\\
\mu&:=\alpha-3i \beta-3\gamma+i \delta
\end{align}
that satisfy the following equations
\beal\label{triton11}
\varepsilon(\mu^*+3\lambda)-\frac{3i}{2}\omega(\mu^*-\lambda)=0
\end{align}
by virtue of the RR Bianchi identities (\ref{triton9}). The second integrability condition, eq.~(\ref{4plus6a}), can similarly be seen to be equivalent to
\beal\label{condi2}
0=-\frac{5}{8}\omega^2+\frac{1}{2}\varepsilon^2+\frac{i}{2}\varepsilon\omega
-\frac{1}{16}(\lambda\mu^*+(\lambda^*)^2)
\end{align}
Separating real and imaginary parts, taking the definitions (\ref{triton7}) into account, eq.~(\ref{condi2}) can be seen to be  
equivalent to the following two conditions
\beal\label{triton10a}
0&=\varepsilon^2-\frac{5}{4}\omega^2-\frac{1}{4}(\alpha^2+\beta^2-\gamma^2-\delta^2)\\
0&=\varepsilon\omega-\frac{1}{2}(\alpha \beta+2 \beta \gamma+\gamma \delta)
\label{triton10b}
\end{align}
Finally the third integrability condition, eq.~(\ref{4plus6c}), can be seen to be equivalent to the following equation:
\beal\label{condi3}
0=3|w_0|^2-\frac{15}{8}\omega^2
+\frac{1}{2}\varepsilon^2
\end{align}
where again we have made use of (\ref{triton11}). Thus, according to the integrability prescription, the solution is determined by the coupled set of equations (\ref{triton9},\ref{condi1},\ref{triton10a},\ref{triton10b},\ref{condi3}). We find the following three classes of solutions

{\it First solution:} $\mathrm{sgn}(\omega\alpha)=-\mathrm{sgn}(\delta\varepsilon)$ and
\beal\label{class1}
|w_0|^2=\frac{1}{2}\omega^2\quad\quad
\alpha^2=\frac{3}{4}\omega^2\quad\quad
\delta^2=\frac{9}{4}\omega^2\quad\quad
\varepsilon^2=\frac{3}{4}\omega^2
\end{align}

{\it Second solution:}
\beal\label{class2}
|w_0|^2=\frac{5}{8}\omega^2\quad\quad
\alpha^2=\frac{5}{4}\omega^2\quad\quad
\delta^2=\frac{25}{4}\omega^2\quad\quad
\varepsilon=0
\end{align}

{\it Third solution:} $\mathrm{sgn}(\omega\alpha)=\mathrm{sgn}(\delta\varepsilon)$ and
\beal\label{class3}
|w_0|^2=\frac{3}{5}\omega^2\quad\quad
\alpha^2=\frac{15}{16}\omega^2\quad\quad
\delta^2=\frac{81}{16}\omega^2\quad\quad
\varepsilon^2=\frac{3}{20}\omega^2
\end{align}
The third solution (\ref{class3}) is the supersymmetric solution of \cite{bc}, while the first two are genuinely non-supersymmetric. In all three cases one can check directly that the supergravity equations of motion are satisfied, in agreement with the general discussion of subsections \ref{intloc} and \ref{int4+6}. 

Unfortunately, only the supersymmetric case satisfies the gauge BPSness conditions (\ref{adsgauge}), so that the above non-supersymmetric vacua do not naturally allow the introduction of D-branes or orientifolds. To show this, let us first observe that the SUSY-breaking has the form (\ref{calgenads}) with parameters 
\beal
& r_1=r_2=-w_0-\frac{1}{4}e^{i\varphi}\mu^* \ \ 
t_1=-2w_0+\frac{3i}{2}\omega+\varepsilon\ \ 
t_2=-2w_0+\frac{3i}{2}\omega e^{2i\varphi}+\varepsilon e^{2i\varphi}\\ \nn
& q^1_{mn}=\big(\frac{i}{8}\omega+\frac{1}{4}\varepsilon -\frac{1}{16}\lambda^*e^{i\varphi} \big)(1-iJ)_{mn}\ \  q^2_{mn}=\big(\frac{i}{8}\omega e^{2i\varphi}+\frac{1}{4}\varepsilon e^{2i\varphi} -\frac{1}{16}\lambda^*e^{i\varphi}\big)(1+iJ)_{mn}
\end{align}
where we have taken into account that $J_{mn}=i\eta^{\dagger}\gamma_{mn}\eta=(J_1)_{mn}=-(J_2)_{mn}$. 
From the third line of the gauge BPSness equation (\ref{adsgauge}), we arrive at the condition: $\varphi=0,\pi$; i.e. $\eta_1=\pm\eta^*_2$. It is then straightforward to verify that only the supersymmetric solution (\ref{class3}) satisfies the first line of (\ref{adsgauge}). In particular, $\varphi=0$ corresponds to the subcases $\mathrm{sgn}(\omega\delta)=1$, while $\varphi=\pi$ corresponds to $\mathrm{sgn}(\omega\delta)=-1$. Finally,  let us note that the second line in (\ref{adsgauge}) is trivially satisfied by all three classes of solutions.

%%%%%%%%%%%%%%%%%%%%%%%%%%%%%%
%%%%%%%%%%%%%%%%%%%%%%%%%%%%%%

\section{Conclusions and outlook}

In the present paper we have analyzed the structure of non-supersymmetric type II flux vacua from the vantage point of generalized complex geometry. While GCG techniques have mainly been applied to supersymmetric type II vacua, we have shown that they are equally useful for $\caln = 0$ backgrounds, as long as an approximate 4D supersymmetry survives. As a first application of this idea we have rephrased the well-known properties of $\caln=0$ warped CY/F-theory vacua in terms of generalized calibrations, that are the natural objects  describing the BPS properties of probe D-branes in general flux backgrounds. Roughly-speaking, while in an $\caln =1$ background the full set of D-branes must obey a BPS bound, only a subset of these bounds will survive in the absence of supersymmetry, and so we can classify $\caln=0$ backgrounds in terms of the D-brane BPS bounds/generalized calibrations they contain. In the case of GKP vacua, the D-branes whose BPSness is affected by SUSY-breaking look like 3D domain walls from the 4D viewpoint, and so we have named this SUSY-breaking pattern `Domain-Wall SUSY-breaking' (DWSB).

We have analyzed the structure of DWSB backgrounds from different perspectives, with the particular goal of finding those backgrounds that are most similar to the $\caln=0$ no-scale vacua of \cite{gkp}. In this quest we have selected in Section \ref{sec:dwsb} a simple DWSB subansatz which, analyzed from the 4D perspective in sections \ref{sec:inteflux} and \ref{sec:4dint}, can indeed reproduce the desired no-scale structure. From the geometric point of view, this subansatz is based on compactification manifolds $\calm_6$ that contain a calibrated generalized foliation, which is a rather strong restriction. However, the techniques used in sections \ref{sec:inteflux} and \ref{sec:4dint} are valid for general $\caln=0$ flux compactifications, and so they can be applied to DWSB backgrounds beyond the simple subansatz considered above. It would be interesting to see which new kinds of 6D geometries and 4D effective theories can be obtained in this way. In particular, it would be interesting to explore how likely the generalization of the KKLT \cite{kklt} and Large Volume \cite{LVM} scenarios is in this context, as well as whether new scenarios for constructing de Sitter vacua may naturally appear. 

The effective potential considered in Section \ref{sec:inteflux} also allows to address a basic issue of $\caln=0$ vacua, which is the presence of closed string tachyons. We have seen that even for the DWSB subansatz of section \ref{sec:dwsb} the absence of tachyons is not guaranteed, and that some mild assumptions on the off-shell gravitino and dilatino variations should be made. While these assumptions come naturally in the case of warped Calabi-Yau vacua,  their interpretation is less clear for more general backgrounds, and in particular for those beyond our DWSB subansatz. It would be very interesting to gain further understanding on such tachyon-free conditions for this kind of backgrounds.

Note that while the techniques of sections \ref{sec:inteflux} and \ref{sec:4dint} admit a 4D perspective, the approach used is fully ten-dimensional. Hence, it allows us to address issues that effective 4D-like approaches cannot deal with, such as warping effects. This is particularly manifest in the results of section \ref{sec:subcases}, when comparing some simple subcases of DWSB backgrounds with some no-scale vacua found in \cite{cg07}. Indeed, by comparing results we see that the 10D eom's derived in Section \ref{sec:inteflux} seem to impose further constraints beyond the background relations found in \cite{cg07}. As in these cases the extra conditions become trivial for constant warping, it is tempting to speculate that they arise from warping corrections to the 4D K\"ahler potential, along the lines of \cite{giddings,lucapaul2,bcdgmqs06,stud07}. Since the explicit examples provided in section \ref{sec:examples} satisfy these extra 10D constraints automatically, it would be very interesting to extend the set of examples to include some where this is not the case.

Another obvious extension of this work is to construct vacua which are not Minkowski. Indeed, as shown in Section \ref{sec:ads4} the general philosophy of this paper can be easily extended to compactifications to AdS$_4$. There we have described the minimal requirements to construct phenomenologically viable vacua, which amounts to requiring that 4D spacetime-filling  D-branes develop a BPS bound, while domain walls and D-strings may not have such a BPSness property. In this AdS$_4$ context, it is essential to consider uplifting mechanisms to de Sitter space, and so our discussion of anti-D-branes in  $\caln=0$ flux backgrounds in Section \ref{sec:anti} could be a key point for the construction of novel examples of metastable vacua.

Another important point that has been addressed in this paper is the structure of F-terms in the closed string sector and the flux-induced soft terms in the open string sector. We have found good agreement between both, in particular for the case of the gaugino mass, that we have computed in full generality. Other soft terms, like fermionic $\mu$-terms, have also been computed in particular cases, finding qualitative good agreement with the results of \cite{cg07}. It would be interesting to extend this computation to more general situations, including more backgrounds beyond twisted tori and compactifications to AdS$_4$, and to also compute the spectrum of scalar soft terms for these cases. 

Finally, we have seen that an alternative, complementary approach, based on integrability, can be used
 for the construction of  $\caln =0$ type II vacua. The method relies on the ability to factorize 
 second-order equations of motion into two first-order equations involving spinorial quantities. We 
have illustrated the procedure by constructing a new class of  $\caln =0$ vacua of the form AdS$_4\times \mathcal{M}_6$, where $\mathcal{M}_6$ can be any nearly-K\"{a}hler manifold. It would be  interesting to explore whether this approach can be used as a tool for the classification of general non-supersymmetric type II supergravity backgrounds. 

We hope that the ideas and techniques developed in this paper serve to understand the set of $\caln=0$ supergravity/string theory vacua from a different perspective, that allows to derive interesting results on the above and related issues. 

%%%%%%%%%%%%%%%%%%%%%%%%%%%%%%
%%%%%%%%%%%%%%%%%%%%%%%%%%%%%%

\bigskip

\bigskip

\centerline{\bf Acknowledgments}

\bigskip

It is a pleasure to thank C.~Angelantonj, P.\,G.~C\'amara,  G.\,L.~Cardoso, D.~Cassani, G.~Curio, M.~G\'omez-Reino, M.~Haack, L.\,E.~Ib\'a\~nez, P.~Koerber, M.~Petrini, A.~Tomasiello and  M. Zagermann for useful discussions. F.M. would like to thank the Arnold-Sommerfeld-Center for hospitality while part of this work was done. This work  is supported in part by the European Commission under the
  Project MRTN-CT-2004-005104, and by the Cluster of Excellence ``Origin and Structure of
  the Universe'' in M\"unchen, Germany.

\newpage

\appendix

%%%%%%%%%%%%%%%%%%%%%%%%%%%%%%
%%%%%%%%%%%%%%%%%%%%%%%%%%%%%%

\section{Supergravity conventions}\label{ap:conv}

\subsection{Bosonic sector}

Our bosonic conventions are identical to those of \cite{toine}, up to the sign changes $H\rightarrow -H$ in IIB and $C_{2n+1}\rightarrow (-)^n C_{2n+1}$ in IIA. This implies that the self-duality relations read  $F_{n}=(-)^{\frac{(n-1)(n-2)}{2}}*_{10}F_{10-n}$.\footnote{ The ten dimensional Hodge-star operator $*_{10}$ is defined by
\be\label{star10}
*_{10}\omega_{p}\, =\, -\frac{1}{p!(10-p)!}\sqrt{-g}\,\epsilon_{M_1\ldots M_{10}}\omega^{M_{11-p}\ldots M_{10}} \d x^{M_1}\wedge \ldots\wedge \d x^{M_{10-p}}\nonumber
\ee
where $\epsilon^{01\ldots 9}=1$.} By introducing the operator $\sigma$ which reverses the order of the indices of a $p$-form, the self-duality condition can be written as
\be\label{10dself}
F\, =\, *_{10}\sigma(F)
\ee
The pseudo-action (i.e. a mnemonic tool to obtain the e.o.m's) of the democratic formulation is
\be
S\, =\, \frac{1}{2\kappa^2_{10}}\int\d^{10}x\sqrt{-g}\Big\{e^{-2\Phi} \big[R+4(\d\Phi)^2-\frac12 H^2] -\frac14 F^2 \Big\}+S^{\rm (loc)}
\ee
where $2\kappa^2_{10}=(2\pi)^7(\alpha^\prime)^4$ and for any form $\omega$ we define $\omega^2=\omega\cdot\omega$, with $\cdot$  given by
\be
\omega_{p}\cdot\chi_{p}\, =\, \frac{1}{p!}\omega_{M_1\ldots M_p}\chi^{M_1\ldots M_p}
\label{cdot}
\ee
If $\omega$ is complex, we also define $|\omega|^2=\omega\cdot \bar\omega$.\footnote{On non-spinorial quantities, we use both $\overline{({\ldots})}$ and $(\ldots)^*$ interchangeably to indicate the ordinary complex conjugation. On the other hand, on ordinary spinors $\overline{({\ldots})}$ denotes the Dirac conjugate.}
In addition to the RR e.o.m./BI's 
\be
\d_H F\, =\, -j_{\text{source}}
\ee
we get:

\bigskip

\noindent The {\em dilaton} e.o.m.:
\be
2\kappa^2_{10}\frac{\delta S}{\delta\Phi}=-8e^{-2\Phi}\big[\nabla^2\Phi-(\d\Phi)^2+\frac14 R-\frac18 H^2\big]+\frac{2\kappa^2_{10}}{\sqrt{-g}}\frac{\delta S^{\rm (loc)}}{\delta\Phi}\,=\, 0
\ee
and thus
\be\label{dilatoneom2}
\nabla^2\Phi-(\d\Phi)^2+\frac14 R-\frac18 H^2-\frac14\frac{\kappa^2_{10}e^{2\Phi}}{\sqrt{-g}}\frac{\delta S^{\rm (loc)}}{\delta\Phi}\, =\, 0
\ee

\noindent  The {\em $B$-field} e.o.m.: 
\bea\label{Heom}
2\kappa^2_{10}\frac{\delta S}{\delta B}&=& -\d(e^{-2\Phi}*_{10}H)-\frac12 [F\wedge\sigma(F)]_{8}+2\kappa^2_{10}\frac{\delta S^{\rm (loc)}}{\delta B}\cr
&=&  -\d(e^{-2\Phi}*_{10}H)+\frac12 [*_{10}F\wedge F]_{8}+2\kappa^2_{10}\frac{\delta S^{\rm (loc)}}{\delta B}=0
\eea

\noindent The {\em Einstein} e.o.m.:
\bea\label{einsteinstring}
\frac{2\kappa^2_{10}}{{\sqrt{-\det g}}}\frac{\delta S}{\delta g^{MN}}&=&e^{-2\Phi}\big[G_{MN}+2g_{MN}\d\Phi\cdot\d\Phi-2g_{NM}\nabla^2\Phi\cr && +2\nabla_{M}\nabla_{N}\Phi -\frac12\iota_M H\cdot\iota_NH+\frac14 g_{MN}H\cdot H\big]\cr
&&-\frac14\iota_M F\cdot \iota_N F-\kappa^2_{10}T_{MN}^{\rm (loc)}=0
\eea
where 
\be
T_{MN}^{\rm (loc)}\, =\, -\frac{2}{{\sqrt{-\det g}}}\frac{\delta S^{\rm (loc)}}{\delta g^{MN}}
\ee
By combining (\ref{einsteinstring}) with the dilaton equation  we get the {\em modified Einstein}  e.o.m.
\be\label{einstein2}
R_{MN}+2\nabla_{M}\nabla_{N}\Phi-\frac12\iota_MH\cdot\iota_NH-\frac14e^{2\Phi}\iota_MF\cdot\iota_N F- \kappa^2_{10}e^{2\Phi}\Big( T_{MN}^{\rm (loc)}  +\frac{g_{MN}}{2\sqrt{-g}}\frac{\delta S^{\rm (loc)}}{\delta \Phi}\Big)\, =\, 0
\ee

Finally, let us recall that the Mukai pairing on a certain space of dimension $n$ is defined by 
\be
\langle\omega,\chi\rangle\, =\, \omega\wedge\sigma({\chi})|_{n}
\label{mukain}
\ee
 for any pair of polyforms $\omega$ and $\chi$, where $\sigma$ is the operator that reverses the order of the indices of a form. More generically, we can also define
 \be
 \langle\omega,\chi\rangle_k\, =\, \omega\wedge\sigma({\chi})|_{k}
 \ee
 for $k < n$.

%%%%%%%%%%%%%%%%%%%%%%%%%%%%%%

\subsection{Fermionic sector}

We can use a representation in which the 10D gamma matrices $\Gamma_{M}$ are real. Underlying the flat indices, the 10D chiral operator is given by
\be
\Gamma_{(10)}\, =\, \Gamma^{\ul{01\ldots 9}}
\ee
For any form $\omega$, with denote  by both $\slashed{\omega}$ and $\underline{\omega}$  its image under Clifford map. More explicitly, for a $p$-form
\be
\slashed{\omega}_{p}\equiv\underline{\omega_p}\, :=\, \frac{1}{p!}\omega_{M_1\ldots M_p}\Gamma^{M_1\ldots M_p}
\ee
Then the self-duality condition (\ref{10dself}) can be written as
\be
\Gamma_{(10)}\slashed{F}\,=\,\slashed{F}
\ee

The type II supersymmetry transformations are parameterized by two MW spinors,  $\epsilon_1$ and $\epsilon_2$, which in our representation are real and satisfy $\Gamma_{(10)}\epsilon_1=\epsilon_1$ and $\Gamma_{(10)}\epsilon_2=\mp\epsilon_2$ in IIA/IIB. 
In our conventions, the type II supersymmetry transformations of \cite{toine} can be written as follows
\bea\label{backsusy}
\delta\psi^{(1)}_M\, =\,(\cald_M\epsilon)_1&\equiv & (\nabla_M+\frac14\slashchar{H}_M)\epsilon_1+\frac{1}{16}e^\Phi
\slashchar{F}\Gamma_M\Gamma_{(10)}\epsilon_2\cr
\delta\psi^{(2)}_M\, =\, (\cald_M\epsilon)_2 & \equiv & (\nabla_M-\frac14\slashchar{H}_M)\epsilon_2-\frac{1}{16}e^\Phi
\sigma(\slashchar{F})\Gamma_M\Gamma_{(10)}\epsilon_1\cr
\delta\lambda^{(1)}\, = \,(\calo\epsilon)_1& \equiv & (\slashchar{\partial}\Phi+\frac12\slashchar{H})\epsilon_1+\frac{1}{16}e^{\Phi}\Gamma^M\slashchar{F}\Gamma_M\Gamma_{(10)}\epsilon_2\cr
\delta\lambda^{(2)}\, =\, (\calo\epsilon)_2 & \equiv & (\slashchar{\partial}\Phi-\frac12\slashchar{H})\epsilon_2-\frac{1}{16}e^{\Phi}\Gamma^M\sigma(\slashchar{F})\Gamma_M\Gamma_{(10)}\epsilon_1
\eea
Note  also that one has the following modified dilatino equations 
\bea\label{modified}
\Gamma^M\delta\psi^{(1)}_M-\delta\lambda^{(1)}&=&\Delta\epsilon_1\, \equiv\, (\slashchar{\nabla}-\slashchar{\partial}\Phi+\frac14\slashchar{H})\epsilon_1\cr
\Gamma^M\delta\psi^{(2)}_M-\delta\lambda^{(2)}&=&\Delta\epsilon_2\, \equiv\, (\slashchar{\nabla}-\slashchar{\partial}\Phi-\frac14\slashchar{H})\epsilon_2
\eea
In double spinor notation, the equations (\ref{backsusy}) we can be written as
\bea
\delta\psi_M & = &\left[\nabla_M+\frac14\slashchar{H}_M\sigma_3+\frac{1}{16}e^\Phi
\left(\begin{array}{cc} 0 & \slashchar{F} \\ -\sigma(\slashchar{F})& 0 \end{array}\right)\Gamma_M\Gamma_{(10)}\right]\epsilon \cr
\delta\lambda &=&\left[ \slashchar{\partial}\Phi+\frac12\slashchar{H}\sigma_3+\frac{1}{16}e^{\Phi}\Gamma^M  \left(\begin{array}{cc} 0 & \slashchar{F} \\ -\sigma(\slashchar{F})& 0 \end{array}\right) \Gamma_M\Gamma_{(10)}\right] \epsilon
\eea

%%%%%%%%%%%%%%%%%%%%%%%%%%%%%%

\subsection{Splitting to 4+6 dimensions and pure spinors} 
\label{4dstructure}

Let us now consider a ten-dimensional spacetime of the form $X_{10}=X_4\times_\omega \calm_6$, with $X_4$ either AdS$_4$ or $\mathbb{R}^{1,3}$, and split the coordinates accordingly $x^M\rightarrow (x^\mu,y^m)$. We assume that the ten-dimensional metric (in the string frame) has the form
\be
\d s^2_{10}\, =\, e^{2A}\d s^2_{X_4}+g_{mn}\d y^m\d y^n
\label{10dansatzap}
\ee
and that the $H$-field has only internal legs. In addition, we assume that the ten-dimensional RR field-strengths, denoted here by $F^{\rm tot}$, split as follows
\be
F^{\rm tot}\, =\, F+e^{4A}\d{\rm Vol}_4\wedge \tilde*_6 F
\ee 
where $F$ has only internal legs, $\d{\rm Vol}_4$ is the volume form of $\d s^2_{X_4}$ and we have defined\footnote{The six-dimensional Hodge-star is defined as 
\be\label{star6}
*_{6}\omega_{p}\, =\, \frac{\sqrt{g}}{p!(6-p)!}\epsilon_{m_1\ldots m_{6}}\omega^{m_{7-p}\ldots m_{6}} dy^{m_1}\wedge \ldots\wedge dy^{6-p}\nonumber
\ee}
\be
\tilde*_6\, =\, *_6\circ \sigma
\label{circstar6}
\ee
so that we have $\tilde*_6^2=-1$. Finally, all fields are independent of the external coordinates $x^\mu$.

The ten-dimensional gamma matrices $\Gamma^M$ can also be split  in terms of four- and six-dimensional gamma matrices $\hat\gamma^\mu$ (associated with the unwarped $X_4$ metric) and $\gamma^m$ in the following way
\be
\Gamma^{\mu}\, =\, e^{-A}\hat\gamma^{\mu}\otimes \bbone\quad \quad \quad\Gamma^{{m}}\, =\,\gamma_{(4)}\otimes \gamma^{{m}} 
\ee
where $\gamma_{(4)}=i\hat\gamma^{\ul{0123}}$ is the standard four-dimensional chiral operator. The six-dimensional chiral operator is in turn $\gamma_{(6)}=-i\gamma^{\ul{123456}}$ and so we have that $\Gamma_{(10)}=\gamma_{(4)}\otimes\gamma_{(6)}$. The ten-dimensional type II supersymmetry generators can accordingly be decomposed as
\bea\label{fermsplit}
\epsilon_1\, =\, \zeta\otimes\eta_1+\ \text{c.c.}\quad\quad\quad\quad \epsilon_2\, =\, \zeta\otimes \eta_2+\ \text{c.c.}
\eea
where $\zeta=\gamma_{(4)}\zeta$ is the generic Killing spinor of $X_4$. Furthermore, $\gamma_{(6)}\eta_1=\eta_1$ for both IIA and IIB, while $\gamma_{(6)}\eta_2=-\eta_2$ in IIA and $\gamma_{(6)}\eta_2=\eta_2$ in IIB. We also assume that $\eta_1^\dagger\eta_1=\eta_2^\dagger\eta_2=|a|^2$, since this is a necessary condition in order to have calibrations for static D-branes \cite{luca,lucapaul2}. The internal spinors $\eta_1$ and $\eta_2$ define the $SU(3)\times SU(3)$-structure of the configuration, that can be alternatively characterized in terms of the pure spinors $\Psi_1$ and $\Psi_2$ defined by
\bea
\slashed{\Psi}_1=-\frac{8i}{|a|^2}\eta_1\otimes\eta^\dagger_2\quad\quad,\quad\quad \slashed{\Psi}_2=-\frac{8i}{|a|^2}\eta_1\otimes\eta^T_2\ ,
\eea
where the overall normalization is chosen for later convenience. As polyforms of definite parity, $\Psi_1$ is even/odd in IIB/IIA and  $\Psi_2$ is even/odd in IIA/IIB.

Let us now consider the case where $\calm_6$ is an $SU(3)$-structure manifold in some detail. For this to be true in IIA supergravity, the condition $\eta_1=ie^{-i\theta}\eta^*_2$ has to be satisfied for some (possibly point-dependent) phase $e^{i\theta}$. Similarly, in type IIB we should require that $\eta_1=ie^{i\theta}\eta_2$.  We can thus introduce the normalized spinor $\chi=\eta_1/|a|$ and use it to construct the following tensors on $\calm_6$
\be\label{jomega}
J_{mn}\, =\, i\chi^\dagger\gamma_{mn}\chi\quad\quad\quad\quad\Omega_{mnp}\, =\, \chi^T\gamma_{mnp}\chi
\ee
$J$ is the two-form associated with the almost complex structure $J^m{}_n$, with respect to which $\Omega$ is a $(3,0)$-form. Together $J$ and $\Omega$ provide an alternative definition of the $SU(3)$ structure of the configuration. They are normalized so that 
\be
(1/3!)J\wedge J\wedge J\, =\, -(i/8)\Omega\wedge\bar\Omega\, =\, \d\text{Vol}_6\
\ee
In this case, the pure spinors $\Psi_1$ and $\Psi_2$ take the form
\bea\label{su3ps}
\Psi_1\, =\, e^{i\theta}\Omega\qquad & &\qquad \Psi_2\, =\,e^{-i\theta}e^{iJ}\qquad\, \text{in IIA}\cr
\Psi_1\, =\, e^{i\theta}e^{iJ}\qquad& &\qquad \Psi_2\,=\,e^{-i\theta}\Omega\qquad\ \ \text{in IIB}
\eea
the particular type IIB case of \cite{gp1,gkp,gp2} being obtained by setting $e^{i\theta}=1$.

Going back to a generic $SU(3)\times SU(3)$-structure, one can write the background supersymmetry conditions in terms of the pure spinors \cite{gmpt}.  In our conventions, such equations have the form
\bseq\label{susygeneral}
\begin{align}
\d_H(e^{4A-\Phi}\Re\Psi_1)&=\, e^{4A}\tilde*_6 F + 3 (-)^{|\Psi_2|}e^{3A-\Phi}\Re(\bar w_0\Psi_2)\label{susygen1}\\
\d_H(e^{2A-\Phi}\Im\Psi_1)&=\, 0\label{susygen2}\\
\d_H(e^{3A-\Phi}\Psi_2)&=\, 2i (-)^{|\Psi_2|}w_0e^{2A-\Phi}\Im \Psi_1\label{susygen3}
\end{align}
\eseq
where $w_0$ is the constant entering the AdS$_4$ Killing spinor equation
\be
\label{defw0}
\nabla_\mu\zeta\, =\, \frac12\bar w_0\hat\gamma_\mu\zeta^*
\ee
and is related to the AdS$_4$ radius by $R=1/|w_0|$. Hence, it vanishes in the case of compactification to flat space. Note that in the AdS$_4$ case $w_0 \neq 0$, (\ref{susygen2}) does not contain any new information, since it is a consequence of (\ref{susygen3}). In addition to (\ref{susygeneral}), one needs to impose the supersymmetry condition $\d\log |a|^2=\d A$ relating the norm of the internal spinors to the warp factor.

%%%%%%%%%%%%%%%%%%%%%%%%%%%%%%
%%%%%%%%%%%%%%%%%%%%%%%%%%%%%%

\section{SUSY-breaking and pure spinors}\label{ap:sb}

Let us consider a ten-dimensional ansatz of the form (\ref{10dansatzap}), supporting an $SU(3)\times SU(3)$ structure background. Then the ten dimensional bispinor $\epsilon=(\epsilon_1,\epsilon_2)^T$ is specified by two internal chiral spinors $\eta_1$ and $\eta_2$ as in (\ref{fermsplit}). Since in addition we are assuming that supersymmetry is broken, we generically have
\bea\label{sbf1ap}
(\cald_\mu\epsilon)_1=\frac12e^{A}\hat{\g}_{\mu}\zeta\otimes \calv_1+\ \text{c.c.}\quad& &\quad (\cald_\mu\epsilon)_2=\frac12e^{A}\hat{\g}_{\mu}\zeta\otimes \calv_2+\ \text{c.c.} \cr
(\cald_m\epsilon)_1=\zeta\otimes \calu^1_m+\ \text{c.c.}\quad& &\quad (\cald_m\epsilon)_2=\zeta\otimes \calu^2_m+\ \text{c.c.}\cr
\Delta\epsilon_1=\zeta\otimes \cals_1+\ \text{c.c.}\quad& &\quad \Delta\epsilon_2=\zeta\otimes \cals_2+\ \text{c.c.}
\eea
where $\calv_{1,2}$, $\calu^{1,2}_m$ and $\cals_{1,2}$ are internal spinors parametrizing the supersymmetry breaking. Their explicit form in terms of $\eta_1$ and $\eta_2$ is
\bea\label{spinsb}
\calv_1 &\equiv& \slashed\partial A\eta_1+\frac14 e^{\Phi}\gamma_{(6)}\slashed{F}\eta_2+e^{-A}w_0\eta_1^* \cr
\calv_2 &\equiv& \slashed\partial A\eta_2-\frac14 e^{\Phi}\gamma_{(6)}\slashed{F}^\dagger\eta_1+e^{-A}w_0\eta_2^* \cr
\cals_1&\equiv& (\slashed{\nabla}-\slashed\partial\Phi+2\slashed\partial A+\frac14\slashed{H})\eta_1+2e^{-A}w_0\eta_1^* \cr
\cals_2&\equiv& (\slashed{\nabla}-\slashed\partial\Phi+2\slashed\partial A-\frac14\slashed{H})\eta_2+2e^{-A}w_0\eta_2^* \cr
\calu_m^1&\equiv& (\nabla_m+\frac14\slashed{H}_m)\eta_1+\frac18 e^\Phi\slashed{F}\gamma_m\gamma_{(6)}\eta_2 \cr
\calu_m^2&\equiv& (\nabla_m-\frac14\slashed{H}_m)\eta_2-\frac18 e^\Phi\slashed{F}^\dagger\gamma_m\gamma_{(6)}\eta_1
\eea
where we have allowed for a non-trivial AdS$_4$ parameter $w_0$ for $X_4$. We choose the norm of the internal spinors $|a|^2=\eta_1^\dagger\eta_1=\eta_2^\dagger\eta_2$ to be related to the warp-factor by $\d|a|^2=|a|^2\d A$ as in the supersymmetric case, which gives the following constraint
\be\label{snorm}
\Re\big(\eta_1^\dagger{\cal U}^1_m-\frac12 \eta^\dagger_2\gamma_m{\cal V}_2\big)\, =\,
\Re\big(\eta_2^\dagger{\cal U}^2_m-\frac12 \eta^\dagger_1\gamma_m{\cal V}_1\big)\, =\, 0
\ee

Let us now expand the spinorial supersymmetry parameters $\calu^{1,2}_m,\ \cals_{1,2}$ and $\calv_{1,2}$ in terms of tensorial susy-breaking parameters in the following way
\bea\label{fexp}
\calv_{1} =r_{1}\eta^*_{1}+s^{1}_m\gamma^m\eta_{1}\quad & &\quad \calv_{2} =r_{2}\eta^*_{2}+s^{2}_m\gamma^m\eta_{2} \cr
 \cals_{1}=t_{1}\eta^*_{1}+u^{1}_m\gamma^m\eta_{1}\quad & &\quad \cals_{2}=t_{2}\eta^*_{2}+u^{2}_m\gamma^m\eta_{2} \cr
 \calu^{1}_m=p^{1}_m\eta_{1}+q^{1}_{mn}\gamma^n\eta_{1}^*\quad & &\quad \calu^{2}_m=p^{2}_m\eta_{2}+q^{2}_{mn}\gamma^n\eta_{2}^*\quad 
\eea
Note  that, because of (\ref{snorm}), we must take $\Re p^1_m = \Re s^2_m$, $\Re p^2_m = \Re s^1_m$ and that by definition 
\bea
(1-iJ_1)^k{}_m u^1_k=0\quad & \quad (1-iJ_1)^k{}_m s^1_k=0\quad & \quad    (1+iJ_1)^k{}_n q^1_{mk}=0\cr
(1-iJ_2)^k{}_m u^2_k=0\quad & \quad (1-iJ_2)^k{}_m s^2_k=0\quad & \quad (1+iJ_2)^k{}_n q^2_{mk}=0
\eea
where the almost complex structures $J_{1,2}$ are defined as
\bea\label{almc}
(J_{1})^m{}_n=\frac{i}{|a|^2}\eta^\dagger_1\gamma^m{}_n\eta_1& \quad \quad& (J_{2})^m{}_n=\frac{i}{|a|^2}\eta^\dagger_2\gamma^m{}_n\eta_2\ .
\eea
It is important to note that the new susy-breaking parameters $r_{1,2},\ t_{1,2},\ s_m^{1,2},\ p^{1,2}_m,\ u^{1,2}_m,\ q^{1,2}_{mn}$ do not mix under T-duality.\footnote{Assuming an isometry along a coordinate $\phi$ and applying the T-duality formul\ae{} for spinors of \cite{hassan}, we find that the above susy-breaking parameters transform as follows under T-duality 
\be\label{tsb}
\begin{array}{lcl}
\tilde r_1=r_1\quad\quad \tilde r_2=r_2 & \quad & \tilde t_1=t_1\quad\quad \tilde t_2=t_2\\
\tilde s_m^1=(Q_+)^n{}_m s^1_n & \quad & \tilde s_m^2=(Q_-)^n{}_m s^2_n\\
\tilde u_m^1=(Q_+)^n{}_m u^1_n & \quad & \tilde u_m^2=(Q_-)^n{}_m u^2_n\\
\tilde p_m^1=(Q_+)^n{}_m p^1_n & \quad & \tilde p_m^2=(Q_-)^n{}_m p^2_n\\
\tilde q_{mn}^1=(Q_-)^k{}_m (Q_+)^l{}_n q^1_{k,l} &\quad & \tilde q_{mn}^2=(Q_+)^k{}_m (Q_-)^l{}_n q^2_{k,l}
\end{array}
\ee
where the tildes are used for the T-transformed quantities and 
\bea\label{qt}
Q_{+}=\left(\begin{array}{cc}  g^{-1}_{\phi\phi} & -g^{-1}_{\phi\phi}(g + B)_{\phi \hat m} \\ 0 &  \bbone_5 \end{array}\right)\quad \quad
Q_{-}=\left(\begin{array}{cc}  -g^{-1}_{\phi\phi} & -g^{-1}_{\phi\phi}(g - B)_{\phi \hat m} \\ 0 &  \bbone_5 \end{array}\right)
\eea
with $y^{\hat m}\neq \phi$.}

Then, we have the $SO(6,6)$ spinorial identities
\bea\label{psgen}
&&e^{-2A+\Phi}\d_H(e^{2A-\Phi}\Psi_1)+2\d A\wedge \Re\Psi_1-e^{\Phi}\tilde*_6 F-3(-)^{|\Psi_2|}e^{-A}\Re(\bar w_0\Psi_2)=\Upsilon \cr
&&e^{-3A+\Phi}\d_H(e^{3A-\Phi}\Psi_2)-2i(-)^{|\Psi_2|}w_0e^{-A}\Im\Psi_1=\Xi
\eea
where the polyforms $\Upsilon$ and $\Xi$ are given by
\bea\label{psexp}
\Upsilon &=&\frac12 (-)^{|\Psi_1|}(r_1^* + t^*_2)\Psi_2\, +\, \frac12(-)^{|\Psi_1|}(r_2 + t_1)\Psi_2^*\,  +\, \frac12(s^1_m)^*\gamma^m\Psi_1^*\, +\, \frac12 (-)^{|\Psi_1|}s^2_m \Psi_1^* \gamma^m
\cr &&+\frac12\big[ u^1_m+(p^2_m)^* \big]\gamma^m\Psi_1\, +\, \frac12(-)^{|\Psi_1|}
\big[ (u^2_m)^*+p^1_m  \big]\Psi_1\gamma^m\cr 
&& +\frac12(q^2_{mn})^*\gamma^m\Psi_2\gamma^n\, -\, \frac12 q^1_{mn}\gamma^n\Psi_2^*\gamma^m\cr
\Xi &=& \frac12(-)^{|\Psi_1|}t_2\Psi_1\, -\, \frac12 (-)^{|\Psi_1|}t_1\Psi_1^*\, +\,\frac12(u_m^1+p_m^2)\gamma^m\Psi_2\, +\, \frac12(-)^{|\Psi_2|}
(u_m^2+p_m^1)\Psi_2\gamma^m\cr && +\frac12 q^1_{mn}\gamma^n\Psi_1^*\gamma^m\, -\, \frac12 q^2_{mn} \gamma^m\Psi_1\gamma^n 
\eea
and where we are following the usual notation where the action of $\gamma_m$ on a form $\omega$ from the left and the right stands for
\bea
{\gamma}_m\omega=(\iota_m+g_{mn}\d y^n\wedge)\omega\quad\text{and}\quad \omega\gamma_m=(-)^{|\omega|+1}(\iota_m-g_{mn}\d y^n\wedge)\omega\ .
\eea
respectively.

Note  that each of the polyforms in the expansion of $\Upsilon$ and of $\Xi$ belong to a different element of the so-called pure Hodge diamond (see e.g. \cite{gmpt}), and as such are independent elements in the space of polyforms. As a consequence, requiring that $\Upsilon =0$ is equivalent to asking that all the coefficients  of its expansion (\ref{psexp}) vanish and the same applies to $\Xi$. With this observation,  one can easily check that $\Upsilon = \Xi = 0$ is equivalent to the vanishing of all the susy-breaking parameters in (\ref{fexp}), and hence to having a supersymmetric vacuum. This fact was first proved in \cite{gmpt} (see also \cite{gmpt2} for a more detailed derivation), so the above observation is an alternative, although rather elegant derivation of the same result. On the other hand, when dealing with $\caln=0$ vacua we may only impose one of the two conditions, say $\Upsilon = 0$, and so the above formalism becomes essential. In order to illustrate how the computations proceed in such cases, let us focus on the backgrounds of main interest in this paper.

%%%%%%%%%%%%%%%%%%%%%%%%%%%%%%

\subsection{Pure DWSB Minkowski backgrounds} 
\label{flat4dex}

Let us consider the particular case of compactifications to flat space ($w_0 = 0$) of pure DWSB backgrounds. That is, as in Section \ref{sec:dwsb} we will impose both eqs.(\ref{cal}) and (\ref{df}) or, equivalently, (\ref{susypure}). The background need thus only satisfy $\Upsilon\equiv 0$, and this is equivalent to imposing the following relations between the susy-breaking parameters appearing in (\ref{fexp})
\bea\label{fcalcond}
t_1=-r_2\quad & &\quad t_2=-r_1\ ,\cr
 u^1_m=-\frac12(1+iJ_1)^k{}_m (p^2_k)^*\quad & &\quad   u^2_m=-\frac12(1+iJ_2)^k{}_m (p^1_k)^*\ ,\cr
 s^1_m=0\quad & & \quad s^2_m=0\ ,\cr
 (1-iJ_2)^k{}_m q^1_{kn}=0\quad & & \quad (1-iJ_1)^k{}_m q^2_{kn}=0\ .
\eea

Once we impose these restrictions, the original fermionic susy-breaking parameters $\calv_{1,2},\ \cals_{1,2}$ and $\calu^{1,2}_m$ take the form
\bea
\calv_{1} =r_{1}\eta^*_{1}\quad & &\quad \calv_{2} =r_{2}\eta^*_{2}\ ,\cr
 \cals_{1}=-r_2\eta^*_{1}-(p^2_m)^*\gamma^m\eta_{1}\quad & &\quad \cals_{2}=-r_1\eta^*_{2}-(p^1_m)^*\gamma^m\eta_{2}\ ,\cr
 \calu^{1}_m=p^{1}_m\eta_{1}+q^{1}_{mn}\gamma^n\eta_{1}^*\quad & &\quad \calu^{2}_m=p^{2}_m\eta_{2}+q^{2}_{mn}\gamma^n\eta_{2}^*\quad .
\eea
and the susy-breaking equation (\ref{nonsusypure}) reads
\bea\label{dwbroken2}
e^{-3A+\Phi}\d_H(e^{3A-\Phi}\Psi_2) &=&    -\frac12 (-)^{|\Psi_1|}\big(r_1\Psi_1-r_2\Psi_1^*\big)+ \frac12 q^1_{mn}\gamma^n\Psi^*_1\gamma^m-\frac12 q^2_{mn}\gamma^m\Psi_1\gamma^n\cr &&+(p^1+p^2)_m\d y^m\wedge\Psi_2 +(p^2-p^1)^m\iota_m\Psi_2\ .
\eea 

Moreover, since $\d_H(e^{2A-\Phi}\Im\Psi_1)=0$ and $\langle \Im\Psi_1,\Psi_2\rangle_5=0$, by consistency we have
\bea
\frac12(-)^{|\Psi_1|}(r_1-r_2)\langle e^{2A-\Phi}\Im\Psi_1, e^{3A-\Phi}\Re\Psi_1 \rangle& =&-\langle e^{2A-\Phi}\Im\Psi_1, \d_H(e^{3A-\Phi}\Psi_2) \rangle\\
&=&-\langle \d_H(e^{2A-\Phi}\Im\Psi_1), e^{3A-\Phi}\Psi_2\rangle=0\nonumber
\eea
and thus
\bea\label{rcond}
r_1=r_2\equiv r
\eea

To summarize, the DWSB compactifications to flat space considered in the main text are characterized by the SUSY-breaking fermionic parameters 
\bea\label{calpresapp}
\calv_{1} =r\eta^*_{1}\quad & &\quad \calv_{2} =r\eta^*_{2}\ \cr
 \cals_{1}=-r\eta^*_{1}+p^2_m\gamma^m\eta_{1}\quad & &\quad \cals_{2}=-r\eta^*_{2}+p^1_m\gamma^m\eta_{2}\ \cr
 \calu^{1}_m=p^{1}_m\eta_{1}+q^{1}_{mn}\gamma^n\eta_{1}^*\quad & &\quad \calu^{2}_m=p^{2}_m\eta_{2}+q^{2}_{mn}\gamma^n\eta_{2}^*\quad 
\eea
with the following extra constraints
\bea\label{sbconstapp}
\Re p^1_m\,  = & 0 & =\, \Re p^2_m \cr
(1+iJ_1)^k{}_n q^1_{mk}=\, & 0 & =\, (1-iJ_2)^k{}_m q^1_{kn}\cr
(1+iJ_2)^k{}_n q^2_{mk}=\, & 0 & =\, (1-iJ_1)^k{}_m q^2_{kn}
\eea
as claimed in Section \ref{sec:dwsb}. From (\ref{tsb}), this subset of vacua is closed under T-duality.

%%%%%%%%%%%%%%%%%%%%%%%%%%%%%%
%%%%%%%%%%%%%%%%%%%%%%%%%%%%%%

\section{The scalar curvature from pure spinors}\label{ap:scalarR}

In this section we present a formula that expresses the combination $\calr-H^2/2$ of the six-dimensional scalar curvature $\calr$ and the $H$-field in terms of the internal pure spinors, extending the one given in eq.~(4.20) of \cite{cassani2} to cases where the restrictions imposed in that paper -- see eq.~(4.19) therein -- are not valid and the warp factor can be non-trivial. Our formula is practically contained in the derivation presented in that paper, although it needs to be adapted to our setting and completed with some further steps.\footnote{Note  that our conventions are different from the ones used in \cite{cassani2} and the appropriate changes have to be taken into account.}   

Indeed, going through  the derivation in \cite{cassani2}, we obtain the following very general equation
\bea\label{gencur}
\calr-\frac12 H^2&=&-\frac12 f^{-2}\left\{\frac{\langle\tilde*_6 \d_H(f\Psi_1),\d_H(f\bar\Psi_1)\rangle }{\d \text{Vol}_6}+\frac{\langle\tilde*_6 \d_H(f\Psi_2),\d_H(f\bar\Psi_2)\rangle }{\d \text{Vol}_6}\right\}\cr
&& +\frac14\left\{\left| \frac{\langle \Psi_1,\d_H\Psi_2\rangle }{\d\text{Vol}_6}\right|^2+\left| \frac{\langle \bar\Psi_1,\d_H\Psi_2\rangle }{\d\text{Vol}_6}\right|^2\right\}\cr
&& +4(\d\Phi-\frac52\d A+\frac12\d\log f+u_{\rm R}^1)^2+4(\d\Phi-\frac52\d A+\frac12\d\log f+u_{\rm R}^2)^2\cr
&&-4\nabla^2(\Phi-\frac52 A+\frac12\log f)-2\nabla^m(u_{\rm R}^1+u_{\rm R}^2)_m+2f^{-1}\nabla^2f
\eea  
where $f$ is an arbitrary positive definite real function and $u_{\rm R}^{1,2}:=u^{1,2}+u^{* 1,2}=(u_m^{1,2}+u_m^{* 1,2})\d y^m$ are the real extension of SUSY-breaking one-forms  introduced in (\ref{fexp}) - note that $u_{\rm R}^{1,2}$ and $u^{1,2}$ contain the same amount of information. One can express $u^{1,2}$  in terms of the pure spinors as follows 
\bea\label{ups}
u^1_m&=&\frac{i\langle\gamma_m\bar\Psi_1,\d_H(e^{2A-\Phi}\Im\Psi_1)\rangle}{e^{2A-\Phi}\langle\Psi_1,\bar\Psi_1\rangle}+\frac{\langle\gamma_m\bar\Psi_2,\d_H(e^{3A-\Phi}\Psi_2)\rangle}{2e^{3A-\Phi}\langle\Psi_2,\bar\Psi_2\rangle}\cr
u^2_m&=&\frac{i(-)^{|\Psi_2|}\langle\Psi_1\gamma_m,\d_H(e^{2A-\Phi}\Im\Psi_1)\rangle}{ e^{2A-\Phi}\langle\Psi_1,\bar\Psi_1\rangle}+\frac{(-)^{|\Psi_1|}\langle\bar\Psi_2\gamma_m,\d_H(e^{3A-\Phi}\Psi_2\rangle}{2 e^{3A-\Phi}\langle\Psi_2,\bar\Psi_2\rangle}
\eea

For example, the sub-case considered in \cite{cassani2} corresponds to the choice $A=0$ and $f=e^{-\Phi}$ and to imposing the conditions given in eq.~(4.19) of that paper, that in this case are equivalent to the conditions  $u^1=u^2=0$.  

On the other hand, for the purposes of this paper, the most convenient choice is $f=e^{3A-\Phi}$. In this case,  equation (\ref{gencur}) can be written in the form
\bea\label{gencur2}
\calr-\frac12 H^2&=&-\frac12 e^{2\Phi-8A}\frac{\langle\tilde*_6 \d_H(e^{4A-\Phi}\Re\Psi_1), \d_H(e^{4A-\Phi}\Re\Psi_1)\rangle }{\d \text{Vol}_6}\cr
&&-\frac12 e^{2\Phi-4A}\frac{\langle\tilde*_6 \d_H(e^{2A-\Phi}\Im\Psi_1), \d_H(e^{2A-\Phi}\Im\Psi_1)\rangle }{\d \text{Vol}_6}
\cr&& 
-\frac12 e^{2\Phi-6A}\frac{\langle\tilde*_6 \d_H(e^{3A-\Phi}\Psi_2), \d_H(e^{3A-\Phi}\Psi_2)\rangle }{\d \text{Vol}_6}\cr
&&+ \frac14e^{2\Phi-6A}\left\{\left| \frac{\langle \Psi_1,\d_H(e^{3A-\Phi}\Psi_2)\rangle }{\d\text{Vol}_6}\right|^2+\left| \frac{\langle \bar\Psi_1,\d_H(e^{3A-\Phi}\Psi_2)\rangle }{\d\text{Vol}_6}\right|^2\right\}\cr
&& +22(\d A)^2+4(\d\Phi)^2-20\d A\cdot\d\Phi+10\nabla^2A-4\nabla^2\Phi\cr
&& +4(\d\Phi-2\d A)\cdot(u_{\rm R}^1+u_{\rm R}^2)-2\nabla^m(u_{\rm R}^1+u^2_{\rm R})_m+4[(u_{\rm R}^1)^2+(u_{\rm R}^2)^2]
\eea

%%%%%%%%%%%%%%%%%%%%%%%%%%%%%%
%%%%%%%%%%%%%%%%%%%%%%%%%%%%%%

\section{Comments on non-geometric backgrounds} \label{ap:ng}

The analysis presented in this paper is mainly local and does not explicitly involve the global structure of the background. Thus, we could apply it to non-geometric compactifications, where local patches of the internal space are related by elements of the extended T-duality group  $O(6,6;\mathbb{Z})$ (for a review, see e.g. \cite{wecht}). Generalized geometry provides a natural framework to discuss these kinds of configurations. For example, if all the fields (pure spinors included) are invariant under $n$ $U(1)$ symmetries, the conditions we have obtained are naturally covariant under the associated  $O(n,n;\mathbb{Z})$ T-duality group. This is better described in the twisted picture, where we consider the pure spinors 
\bea
\Psi^{\rm tw}_{1,2}=\Psi_{1,2}\wedge e^B
\eea
 instead of $\Psi_{1,2}$, and all the twisted differentials $\d_H$ are substituted by the ordinary exterior derivative d. $\Psi^{\rm tw}_{1,2}$ contain the full information about metric and $B$-field  and transform as spinors under the $O(6,6;\mathbb{R})$ structure group of  the extension bundle $E$ defined by
\bea\label{extbundle}
T^*_{\calm_6}\ \rightarrow\  E\  \rightarrow\  T_{\calm_6}
\eea
where the transition functions are given by $B$-field gauge transformations \cite{hitchin2}. Now, the key point  is that, under T-duality, $e^{-\Phi}\Psi^{\rm tw}_{1,2}$ transform as follows:
\bea\label{pstduality}
e^{-\Phi}\Psi^{\rm tw}_{1,2}\rightarrow e^{-\tilde\Phi}\tilde\Psi^{\rm tw}_{1,2}=\calo\cdot(e^{-\Phi}\Psi^{\rm tw}_{1,2})
\eea 
where $\calo\cdot$ is exactly the spinorial representation of the element $\calo$ of $O(n,n;\mathbb{Z})$ seen as a subgroup of the local structure group $O(6,6;\mathbb{R})$ and the overall dilaton factor takes care of the normalization of the pure spinors, see e.g. \cite{wittcal,gmpt2,alessandrobeta,gmpw}. This can be seen also from (\ref{psgen})-(\ref{psexp}). Indeed, it is known that twisted RR-fields transform as $O(n,n;\mathbb{Z})$ spinors \cite{fukuma}) and since the transformation (anti-)commutes (if type II changes) with the ordinary exterior derivative \cite{hassan}, we see that $e^{-\Phi}\Psi^{\rm tw}_{1,2}$ must transform as  $F^{\rm tw}$, up to an overall different sign if the type II theory changes. 

Thus, T-duality maps DWSB vacua to DWSB vacua and in particular the associated twisted  smeared currents $\tilde\jmath^{\rm tw}_{(\Pi,R)}:=e^B\wedge\tilde\jmath_{(\Pi,R)}$  transform as $F^{\rm tw}$.\footnote{Our discussion is local and we completely  ignore global topological issues.} Equivalently, the transformation of $\Lambda$ under T-duality can be described by using the matrices $Q_+$ and $Q_-$ that enter the two possible transformations of the vielbein $e^a_m\rightarrow (\tilde e^a_{\pm})_m=e^a_n(Q_\pm)^n{}_m$, defined in \cite{hassan}\footnote{Note that $Q_\pm$ here corresponds to $Q_\mp^{-1}$ in \cite{hassan}.} - in the case of a single T-duality, $Q_\pm$ are given in (\ref{qt}).  Then, $\Lambda$ transforms as follows
\bea
\Lambda\quad\rightarrow\quad \tilde\Lambda=Q_+^{-1}\Lambda Q_-
\eea
This, together with the fact that the warp-factor $e^A$ and the susy-breaking parameter $r$ are invariant under T-duality, completes the description of the gluing rules that must be used to patch together  the local conditions (\ref{calgen}) and (\ref{finansatz})-(\ref{finansatz2}) in a globally  non-geometric configuration.

A standard way to obtain non-geometric vacua is by T-dualizing a geometric vacuum with non-trivial $H$-flux. For example, one can start from the simple non-supersymmetric  GKP vacuum of subsection \ref{betaexample} and T-dualize it along the 2-torus $Q_2$ spanned by $y^5,y^6$.  In fact, this gives a concrete warped non-supersymmetric realization of the usual toy model obtained by a double T-duality on a flat 3-torus with constant $H$-flux. For example, using natural units of $2\pi\sqrt{\alpha^\prime}$ and setting $R_i=\Im\lambda_i=1$ in the GKP vacuum of subsection \ref{betaexample} for simplicity, the T-dual NS fields are given by 
\bea
\label{fullmetric65}
\d s^2 & = & e^{2A} \d s^2_{X_4} +  e^{-2A} \left( \d s^2_{{\T^4}} + \d s^2_{Q_2} \right)\cr
ds^2_{Q_2} & = & g_s^{-1}e^{2\Phi} \big[(\d y^5)^2 + (\d y^6)^2 \big]\qquad\qquad \d s^2_{\T^4} =  \sum_{m=1}^4 (\d y^m)^2\cr
B  &= & -  g_s^{-1}\, e^{2\Phi}N_{\rm NS}\, y^4\, \d y^5 \wedge \d y^6 \qquad\qquad 
e^{\Phi} = \frac{g_s}{\sqrt{e^{-4A} + (N_{\rm NS}y^4)^2}}
\eea

The generalized geometry description of the corresponding toy-model has been discussed e.g. in \cite{pascal} and applies to our non-supersymmetric vacuum too. The key point is that in the T-dual GKP vacuum
the $B$-field can be written as $B_{\rm GKP}=N_{\rm NS}\,y^4\d y^5\wedge \d y^6$ and thus has monodromy $B_{\rm GKP}\rightarrow B_{\rm GKP}+N_{\rm NS}\,\d y^5\wedge \d y^6$ under $y^4\rightarrow y^4+1$. This transformation can be see as an element of the extended T-duality group $O(2,2;\mathbb{Z})$  along $Q_2$ and in spin representation it acts on the GKP twisted polyforms as $\calo_{b}\cdot=e^{b}\wedge$, with $b=N_{\rm NS}\,\d y^5\wedge \d y^6$. Now, all the twisted polyforms of the T-dual non-geometric vacuum - i.e. $e^{-\Phi}\Psi^{\rm tw}_{1,2}$, $F^{\rm tw}$ and $\tilde\jmath^{\rm tw}_{ (\Pi,R)}$ - are related to the GKP ones by a T-duality operator $\calo_{Q_2}$. Thus the monodromy in the non-geometric background is given by $\calo_{Q_2}\calo_{b}\calo^{-1}_{Q_2}$ that turns out to be  exactly a  beta-transformation $\calo_\beta$ of the kind discussed in subsection \ref{msb}, with $\beta=N_{\rm NS}\,\partial_{y^5}\wedge \partial_{y^6}$ \cite{pascal}. Note  that, since $\tilde\jmath^{\rm tw}_{\rm GKP}\sim \d y^1\wedge\ldots\wedge \d y^6$ then $\tilde\jmath^{\rm tw}_{ (\Pi,R)}\sim \d y^1\wedge \ldots\wedge\d y^4$ and thus $\tilde\jmath_{ (\Pi,R)}\sim e^{-B}\wedge  \d y^1\wedge\ldots\wedge \d y^4$. Comparing with (\ref{finansatz2}) we see that we can identify $(\Pi,R)$ with $(Q_2,B)$. Thus, this non-geometric vacuum gives another example of `magnetized' DWSB. Furthermore, note that the monodromy operator $\calo_\beta$ acts trivially on $\tilde\jmath^{\rm tw}_{ (\Pi,R)}$ and thus affects $(\Pi,R)=(Q_2,B)$ only through the transformation of $B$. Finally, the same arguments can be applied to deduce  that D3-branes and O3-planes are mapped to D5-branes and O5-planes in the non-geometric vacuum, with $\calf=B|_{\rm sources}=0$.

%%%%%%%%%%%%%%%%%%%%%%%%%%%%%%
%%%%%%%%%%%%%%%%%%%%%%%%%%%%%%

\section{10d integrability}\label{inte}

In this section we give some further details on the derivation of the
integrability
conditions of section \ref{sec:inte}. 
From the definitions (\ref{d3}) and  (\ref{backsusy}) it follows that\footnote{Here we give the details 
of the computation 
 for IIA. The proof for the IIB case is completely analogous.}:
\beal\label{a1}
(\cd_{[N} X_{M]})^1&=\left(
\frac{1}{8}R_{NMAB}\Gamma^{AB}+\frac14\nabla_{[N}\left(\underline{\iota_{M]}H}\right)
+\frac{1}{16}\underline{\iota_{[N}H}\cdot \underline{\iota_{M]}H}+\frac{1}{256}
e^{2\Phi}\underline{F}\cdot\Gamma_{[N}\cdot \underline{\sigma(F)}\cdot\Gamma_{M]}\right)
\epsilon_1\nn\\
&-\left(\frac{1}{16}\nabla_{[N}(e^{\Phi}\underline{F})\cdot\Gamma_{M]}
-\frac{1}{64} e^{\Phi} \underline{F}\cdot \Gamma_{[N}\cdot \underline{\iota_{M]}H}
+\frac{1}{64}e^{\Phi}\underline{\iota_{[N}H}\cdot \underline{F}\cdot \Gamma_{M]}\right)\epsilon_2
\end{align}
and
\beal\label{a2}
(\nabla_N&+\frac14 \underline{\iota_NH})Y^1=\Big(
\nabla_N\slashchar{\partial}\Phi+\frac12\nabla_N\underline{H}+\frac14 \left[ \underline{\iota_NH},\partial\Phi
\right]
+\frac18 \left[ \underline{\iota_NH}, \underline{H} \right]\nn\\
&
-\frac{1}{16}\Gamma^K\cdot\nabla_N(e^{\Phi}\underline{F})\cdot\Gamma_K
-\frac{1}{256}e^{2\Phi}\Gamma^K\cdot
\underline{F}\cdot\Gamma_K\cdot\underline{\sigma(F)}\cdot\Gamma_N
\Big)\epsilon_1\nn\\
&+\frac{1}{16}\Big(e^{\Phi}\slashchar{\partial}\Phi\cdot \underline{F}\cdot\Gamma_N
-\Gamma^K\cdot\nabla_N(e^{\Phi}\underline{F})\cdot\Gamma_K
+\frac{1}{2}e^{\Phi}\underline{H}\cdot \underline{F}\cdot \Gamma_N\nn\\
&-\frac{1}{4}e^{\Phi}\left\{ \underline{\iota_NH},\Gamma^K\cdot \underline{F}\cdot\Gamma_K\right\}
\Big)\epsilon_2 +(\mathcal{O}\cdot X_N)^1
\end{align}
Taking the above into account we are now ready to
compute the left-hand side of eq.~(\ref{iden}), by
straightforwardly  expanding all gamma-matrix
products\footnote{We have found the symbolic gamma-matrix algebra
program \cite{gamm} to be very useful.}
on the right-hand sides
of eqs.~(\ref{a1},\ref{a2}). We break down the computation to
the following parts.

\bigskip

\noindent {\it Terms proportional to $\epsilon_1$.} These are of the form:
\beal\label{e1}
&-\frac14\Big\{
R_{NK}+2\nabla_N\partial_K\Phi
-\frac12 \iota_NH\cdot \iota_KH-\frac12\delta
H_{NK}-\frac12(F_{NP}F_K{}^{P}+\frac{1}{3!}F_{NPQR}F_K{}^{PQR})\nn\\
& +\frac14g_{NK}(F^2+\frac{1}{2!}F_{MP}F^{MP}+\frac{1}{4!}F_{MPQR}F^{MPQR})
\Big\}\Gamma^K\epsilon_1
-\frac{1}{12}\nabla_{[K}H_{LMN]}\Gamma^{LMN}\epsilon_1
\end{align}
where
\beal
\delta H_{NK}:=
e^{2\Phi}\Big[\nabla^L(e^{-2\Phi}H_{LNK})+
F_{NK}F&+\frac12 F_{NKLM}F^{LM}\nn\\
&+\frac{1}{2(4!)^2}\varepsilon_{NKM_1\dots M_8}
F^{M_1\dots M_4}F^{M_5\dots M_8}
\Big]
\end{align}
can be recognized as  the component form of eq.~(\ref{dlth}).
In deriving (\ref{e1}) we have imposed the self-duality condition
on the RR fields
as well as the Hodge-duality relation on the gamma-matrices:
\beal\label{gself}
\Gamma^{(n)}=(-)^{\frac{1}{2}n(n-1)}*_{10}\Gamma^{(10-n)}\Gamma_{(10)}
\end{align}
where we recall that $\Gamma_{(10)}$ is the chirality matrix in ten dimensions. To
compare with
eq.~(\ref{iden})
note that the self-duality condition implies:
\bea
\iota_M F_{(10-n)}\cdot \iota_N F_{(10-n)}=\iota_M F_{(n)}\cdot \iota_N
F_{(n)}
-g_{MN}F_{(n)}\cdot F_{(n)}
\eea
from which it then follows that (for both IIA/IIB):
\bea
-\frac14\iota_M F\cdot \iota_N F+\frac18g_{MN}F\cdot F&=&\sum_{n<5}\left(
-\frac12\iota_M F_{(n)}\cdot \iota_N F_{(n)}+\frac14g_{MN}F_{(n)}\cdot
F_{(n)}
\right)\nn\\
&&-\frac14\iota_M F_{(5)}\cdot \iota_N F_{(5)}
\eea
We can now see that the terms in (\ref{e1}) exactly reproduce the
terms proportional to $\epsilon_1$ in eq.~(\ref{iden}), where we must
also note that
\beal
(\d H)_{MNKL}&=4\nabla_{[M}H_{NKL]}
\end{align}

\bigskip

\noindent {\it Terms proportional to $\epsilon_2$.} These can be assembled
in the form given in eq.~(\ref{iden}), by using the following identities:
\beal\label{tyita}
\frac{1}{6!}\Gamma^{M_1\dots M_6}(\d_H F)_{NM_1\dots M_6}&=
-\frac{1}{6}\Gamma_N{}^{MPQ}\nabla^LF_{LMPQ}+\frac{7}{144}\Gamma^{M_1\dots
M_6}H_{[N M_1 M_2}F_{M_3\dots M_6]}\nn\\
\frac{1}{7!}\Gamma_N{}^{M_1\dots M_7}(\d_H F)_{M_1\dots M_7}&=
-\frac12\Gamma^{MP}\nabla^LF_{LNMP}
+\frac{1}{144}\Gamma_N{}^{M_1\dots M_7}H_{[M_1\dots M_3}F_{M_4\dots M_7]}
~
\end{align}
and
\beal\label{tyitb}
\frac{1}{8!}\Gamma^{M_1\dots M_8}(\d_H F)_{NM_1\dots M_8}&=
-\Gamma_N{}^{M}\nabla^LF_{LM}+\frac{1}{6}\Gamma_N{}^{M}H^{PQR}F_{PQRM}\nn\\
\frac{1}{9!}\Gamma_N{}^{M_1\dots \calm_9}(\d_H F)_{M_1\dots \calm_9}&=
-\nabla^LF_{LN}
+\frac{1}{6}H^{PQR}F_{PQRN}
\end{align}
where in components we have:
\beal
(\d_H F)_{M_1\dots M_n}&=
n\nabla_{[M_1}F_{M_2\dots M_n]}+\frac{n!}{3!(n-3)!}H_{[M_1\dots
M_3}F_{M_4\dots M_n]}
\end{align}
These identities can be shown by
use of the self-duality of the RR fields as well as the Hodge-duality
relation (\ref{gself}).

The easiest way to derive eq.~(\ref{idenb}) is to observe that the
left-hand side of
that equation can be obtained from the left-hand side of
eq.~(\ref{iden}), upon substituting:
 $\epsilon_1\leftrightarrow\epsilon_2$, $H\leftrightarrow -H$ and
 $F\leftrightarrow \pm\sigma(F)$, for IIA/IIB.

Equations (\ref{idenc},\ref{idend}) can be shown by similar manipulations. The proof 
amounts to expanding all products of gamma-matrices on the left-hand side, taking 
the following identities into account:
\beal
\Gamma^{MN}\Gamma^{PQ}R_{MN,PQ}=-2R
\end{align}
which can be shown by expanding the products of gamma-matrices and taking the Bianchi identities 
of the Riemann tensor into account, and:
\beal
\frac{1}{7!}\Gamma^{M_1\dots M_7}(\d_H F)_{M_1\dots M_7}&=
-\frac16\Gamma^{NMP}\nabla^LF_{LNMP}
+\frac{1}{144}\Gamma^{M_1\dots M_7}H_{[M_1\dots M_3}F_{M_4\dots M_7]}\nn\\
\frac{1}{9!}\Gamma^{M_1\dots \calm_9}(\d_H F)_{M_1\dots \calm_9}&=
\left(-\nabla^LF_{LN}
+\frac{1}{6}H^{PQR}F_{PQRN}\right)\Gamma^N 
\end{align}
which can be shown by contracting (\ref{tyita},\ref{tyitb}) with $\Gamma^N$.

To treat the sources, let us first note the following gamma-matrix
identities:
\beal\label{idga}
\gamma_{(m}\cdot \underline{j^{(p)}}\cdot\gamma_{n)}&=\frac{(-)^p}{p!}\left\{
g_{mn}\gamma^{q_1\dots q_{p}}j_{q_1\dots
q_{p}}-2p~\gamma_{(m}{}^{q_1\dots q_{p-1}}j_{n)q_1\dots q_{p-1}}
\right\}\\
\gamma_{[m}\cdot\underline{j^{(p)}}\cdot\gamma_{n]}&=\frac{(-)^p}{p!}\left\{
\gamma_{mn}{}^{q_1\dots q_{p}}j_{q_1\dots q_{p}}+p(p-1)
\gamma^{q_1\dots q_{p-2}}j_{mnq_1\dots q_{p-2}}
\right\}
\end{align}
where $\underline{j^{(p)}}:= 1/p!\gamma^{q_1\dots q_{p}}j_{q_1\dots q_{p}}$. Using
the definition of
the calibration $\omega$ given in eq.~(\ref{calform}) as well as the
nine-dimensional
gamma-matrix Hodge duality
\beal
\gamma^{(p)}=-(-)^{\frac{1}{2}p(p-1)}*_9\gamma^{(9-p)}
\end{align}
it is straightforward to prove that:
\beal\label{plopa}
\chi^T_1\gamma_{(m}\cdot\underline{j}\cdot\gamma_{n)}\chi_2&= \pm |a|^2e^{\Phi}
*_{10}\big(2g_{k(m} \langle \d x^{k} \wedge \iota_{n)}(\d t\wedge \omega),j \rangle
-  g_{mn}\langle \d t\wedge \omega, j \rangle \big)\, ,\\
\label{plopb}
\chi^T_1\gamma_{[m}\cdot\underline{j}\cdot\gamma_{n]}\chi_2&=\mp |a|^2e^{\Phi}*_{10}\big(
\langle \d t\wedge \omega, j \rangle_8
+\langle\d t\wedge \omega, *_9\sigma(j) \rangle_8 \Big)_{mn}
\end{align}
for IIA/IIB. 
It can be shown that the second term on the right-hand side of
eq.~(\ref{plopb}) above
vanishes identically for static calibrated sources.
 Let us also note the following relation:
\beal\label{plopc}
(\chi^T_1\underline{j}\chi_2)&= -|a|^2e^{\Phi}
*_{10}\langle \d t\wedge \omega, j \rangle 
\end{align}
which can be arrived at either directly, or by contracting (\ref{plopa}) with $g^{mn}$.

%%%%%%%%%%%%%%%%%%%%%%%%%%%%%%
%%%%%%%%%%%%%%%%%%%%%%%%%%%%%%

\section{Integrability of GKP vacua}\label{gkpvacua}

Let us first note some useful relations. The ISD condition (\ref{isd}) together 
with the six dimensional Hodge-duality of the gamma matrices
\beal
*_6\gamma^{(n)}=-i(-)^{\frac{1}{2}n(n-1)}*_6\gamma^{(6-n)}\gamma_{(6)}
\end{align}
where $\gamma_{(6)}$ is the six-dimensional chiral operator defined in appendix \ref{4dstructure},
imply that
\beal
\slashed{F}_{3}\lambda_{\pm}=\pm ie^{-\Phi}\slashed{H}\lambda_{\pm}
\end{align}
for any six-dimensional  Weyl spinor $\lambda_{\pm}$, where the subscript denotes the chirality. 
Similarly from (\ref{warpgkp}) and the definition of $\tau$ we obtain
\beal
\frac{i}{4}e^{\Phi}\left( \slashed{F}_{1}+\slashed{F}_{5}\right)\lambda_{+}
&=\left( \frac{i}{4}e^{\Phi}\slashed{\partial}\tau-\slashed{\partial}A\right)\lambda_{+}\nn\\
\frac{i}{4}e^{\Phi}\left( \slashed{F}_{1}+\slashed{F}_{5}\right)\lambda_{-}
&=\left( \frac{i}{4}e^{\Phi}\slashed{\partial}\bar{\tau}+\slashed{\partial}A\right)\lambda_{-}
\end{align}
In the GKP background the six-dimensional spinors $\eta_{1,2}$ are related 
via $\eta:=\eta_1=i\eta_2$. Moreover $\eta$ is related to the unimodular spinor $\hat{\eta}$ of the underlying six-dimensional manifold through $\eta=e^{\frac{A}{2}}\hat{\eta}$, where $A$ is the 
warp factor appearing in (\ref{alph1}). The holomorphy condition (\ref{holt}) implies
\beal\label{kji}
\slashed{\partial}\tau\eta&=0\nn\\
\slashed{\partial}\bar{\tau}\eta&=2ie^{-\Phi}\slashed{\partial}\Phi\eta
\end{align}
where we have taken into account that $\gamma_m\eta$ is holomorphic\footnote{The reader may find it useful to consult appendix 
B of \cite{lt} for some relevant formul\ae.} 
with respect to the almost complex structure constructed out of $\hat{\eta}$ and the (unwarped) metric $\hat{g}$ of (\ref{alph1}). It is also useful to note that 
${\eta}$ obeys the Killing equation
\beal
\hat{\nabla}_m\eta=\left(\frac{1}{2}\partial_mA+\frac{i}{4}e^{\Phi}F_m\right)\eta
\end{align}
from which it follows that 
\beal
{\nabla}_m\eta=\frac{1}{4}\left\{
\left(2\slashed{\partial}A-\slashed{\partial}\Phi\right)\gamma_m+{i}e^{\Phi}\partial_m\tau\right\}\eta
\end{align}
where $\hat{\nabla}$ and $\nabla$ are associated with the metrics $\hat{g}$ and $g$,  respectively.
The $SU(3)$ structure corresponding to $(\hat{\eta},\hat{g})$ can be specified equivalently by the data  $(\hat{J},\hat{\Omega})$, where the associated K\"{a}hler form is given in (\ref{alph2}) and the (3,0) form $\hat{\Omega}$ is related to the normalized holomorphic form $\Omega_0$ of (\ref{alph2}) through $\hat{\Omega}=e^{-\frac{\Phi}{2}}\Omega_0$. 

Taking the above as well as eqs.~(\ref{wcy}) into account, a direct computation involving some standard gamma-matrix algebra reveals that 
 the right-hand side of (\ref{4plus6b}) is equal to 
\beal
\left(r\slashed{\partial}\Phi-3r\slashed{\partial}A+\slashed{\partial}r\right)\eta^*
\label{koupi}
\end{align}
where $r$ is defined through
\beal
\frac{1}{4}\slashed{H}\eta=-r\eta^*
\end{align}
Similarly, the right-hand side of eq.~(\ref{4plus6a}) can be evaluated to give
\beal
-\frac{1}{2}\gamma_n\left(r\slashed{\partial}\Phi-3r\slashed{\partial}A+\slashed{\partial}r\right)\eta^*
\label{koupib}
\end{align}
On the other hand, using the fact that 
$|\hat{\Omega}|^2=8$, where the measure is computed in the unwarped metric $\hat{g}$, the above equation can be written equivalently as
\beal
H^{0,3}=-\frac{1}{2}re^{\Phi-3A}\Omega_0^{*}
\end{align}
The Bianchi identity for the NS three-form and 
the fact that $\Omega_0$ is closed imply that
\beal
\partial^+(re^{\Phi-3A})\wedge\Omega^*_0=0
\end{align}
where $\partial^+$ is the projection of the exterior differential to its holomorphic part. 
It then follows from the above equation and the fact that $\gamma_m\eta^*$ is antiholomorphic, that the expressions (\ref{koupi},\ref{koupib}) vanish. 

Finally, the expression in the curly brackets on the right-hand side of eq.~(\ref{4plus6c}) can be seen to be equal to 
\beal
-2r\slashed{\partial}A\eta^*
\end{align}
The contraction with $\eta^{\dagger}$, and therefore the right-hand side of (\ref{4plus6c}), then vanishes by virtue of the 
fact that the bilinear $\eta^{\dagger}\gamma_m\eta^*$ vanishes for any six-dimensional gamma-matrix $\gamma_m$. This concludes the proof that the GKP background satisfies the integrability conditions (\ref{4plus6b}-\ref{4plus6c}), or equivalently, conditions~(\ref{inteq1}, \ref{inteq2}).

%%%%%%%%%%%%%%%%%%%%%%%%%%%%%%
%%%%%%%%%%%%%%%%%%%%%%%%%%%%%%

\newpage

%bibi

\end{document}